\documentclass[12pt,nofootinbib,tightenlines,preprintnumbers,superscriptaddress,floatfix]{revtex4-1}
\usepackage[T1]{fontenc}
\usepackage[utf8]{inputenc}
\usepackage{lmodern}
\usepackage[english]{babel}
\usepackage{graphicx}
\usepackage{amsmath,mathtools,amssymb}
\usepackage{hyperref}
\usepackage{tabularx}
\usepackage{hhline}
\usepackage{booktabs}
\usepackage[dvipsnames]{xcolor}
\usepackage{tikz}
\usepackage{bm}
\usepackage{pgf-pie}
\usepackage{placeins}

\usetikzlibrary{decorations.pathreplacing,decorations.pathmorphing}

\usepackage{siunitx}
\usepackage{soul}

\newcommand{\md}{\mathrm d}
\newcommand{\beq}{\begin{equation}}
\newcommand{\eeq}{\end{equation}}
\newcommand{\bit}{\begin{itemize}}
\newcommand{\eit}{\end{itemize}}
\newcommand{\ben}{\begin{enumerate}}
\newcommand{\een}{\end{enumerate}}

\newcommand{\be}{\begin{equation}}
\newcommand{\ee}{\end{equation}}
\newcommand{\ba}{\begin{eqnarray}}
\newcommand{\ea}{\end{eqnarray}}
\newcommand{\la}{\label}
\newcommand{\<}{\langle}
\renewcommand{\>}{\rangle}

\renewcommand{\vec}[1]{\boldsymbol{#1}}

\usepackage{physics}
\renewcommand\bold{\mathbf}

\DeclareFontFamily{U}{mathx}{\hyphenchar\font45}
\DeclareFontShape{U}{mathx}{m}{n}{<-> mathx10}{}
\DeclareSymbolFont{mathx}{U}{mathx}{m}{n}
\DeclareMathAccent{\widebar}{0}{mathx}{"73}

\begin{document}

\preprint{MITP-24-055}

\title{Hot QCD matter around the chiral crossover:\\ a lattice study with $\mathcal{O}(a)$-improved Wilson fermions}


\author{Ardit Krasniqi} 
\affiliation{PRISMA$^+$ Cluster of Excellence \& Institut f\"ur Kernphysik,
Johannes Gutenberg-Universit\"at Mainz,
D-55099 Mainz, Germany}

\author{Marco C{\`e}}
\affiliation{Dipartimento di Fisica, Universit\'a di Milano-Bicocca and INFN, Sezione di Milano-Bicocca, \\ Piazza della Scienza 3, 20126 Milano, Italy}


\author{Renwick~J.~\!Hudspith} 
\affiliation{GSI Helmholtzzentrum für Schwerionenforschung,\\ 64291 Darmstadt, Germany}

\author{Harvey~B.~\!Meyer} 
\affiliation{PRISMA$^+$ Cluster of Excellence \& Institut f\"ur Kernphysik,
Johannes Gutenberg-Universit\"at Mainz,
D-55099 Mainz, Germany}
\affiliation{Helmholtz~Institut~Mainz,
Johannes Gutenberg-Universit\"at Mainz,\\
D-55099 Mainz, Germany}


\begin{abstract}
  Using lattice QCD simulations with $\mathcal{O}(a)$-improved Wilson quarks and physical up, down and strange quark masses,
  we investigate the properties of thermal QCD matter at the temperatures $T=\{128,154,192\}$\,MeV
  with a fixed lattice spacing $a=0.064$\,fm and volume $V=(6.1\,\text{fm})^3$.
  We find that the pion quasiparticle, defined as the low-energy pole in the two-point function
  of the axial charge, becomes lighter as the temperature increases and give an argument based on hydrodynamics
 as to why the pole becomes purely diffusive above the chiral crossover.
  We study the thermal modification of the isovector vector spectral function using the Backus-Gilbert method,
  finding an enhancement at low energies and a depletion at energies around 1\,GeV.
  The analogous study of the axial-vector channel reveals a larger enhancement at energies below 1\,GeV,
  and we show that these findings are consistent with rigorous spectral sum rules.
  The difference between vector and axial-vector correlators, an order parameter for chiral symmetry,
  turns out to be overall suppressed by more than an order of magnitude at the crossover.
\end{abstract}

\maketitle

\section{Introduction}
\label{sec:intro}
Quark matter under extreme conditions of temperature and
density is interesting both from the experimental and theoretical
point of view. In the early universe, on a time scale of
microseconds, the strongly-interacting constituents (quarks
and gluons) were in a hot and dense phase called the Quark-Gluon
Plasma (QGP).  Heavy ion colliders, like the Large Hadron Collider
(LHC) at CERN and the Relativistic Heavy Ion Collider (RHIC) at BNL
enable similar conditions to be reached in the lab.
As a result of its expansion, the universe gradually cooled down,
undergoing a transition to a hadronic phase in which we now find
ourselves. Note that at physical quark masses the transition is
actually a crossover~\cite{Aoki:2006we} characterized by a
pseudocritical temperature $T_{pc}=156.5(1.5)$\,MeV\,\cite{HotQCD:2018};
see also the earlier results of Ref.\,\cite{Borsanyi:2010bp}.

In the limit of massless quarks, the QCD Lagrangian has a global
$SU(N_f)_L \times SU(N_f)_R$ symmetry corresponding to independent
rotations of the left- and right-handed components of the Dirac
fields. This symmetry is spontaneously broken to $SU(N_f)_V$ at low
temperatures and becomes restored in the high-temperature phase --
chiral symmetry restoration. A non-vanishing value of the scalar
density $\langle\bar{\psi}\psi\rangle(T)$ in the chiral limit
characterizes the low-temperature phase
($0\leq T\leq T_c \approx132\,\text{MeV}$\,\cite{HotQCD:2019}).
By contrast,
$\langle\bar{\psi}\psi\rangle(T) = 0$ for $T>T_c$, indicating that
chiral symmetry is restored. In other words, the quark condensate
$\langle\bar{\psi}\psi\rangle $ is an order parameter for chiral
symmetry breaking.

The most immediate consequence of a broken chiral symmetry
is the existence of light pions, as implied by Goldstone's theorem.
Thus, starting from the QCD vacuum, increasing the temperature initially
leads to a dilute gas of pions. As the temperature is further
increased, other hadron species also begin to contribute. At the same
time, one expects the excitations of the medium to be quasiparticles
with somewhat modified properties as compared to the standard hadrons,
which are excitations of the vacuum.  A natural starting point in the
investigation of the medium's quasiparticles is to examine the
properties of the pion in the thermal environment~\cite{Shuryak:1990ie,Goity:1989gs}. The pion mass and decay constant
have been studied to one loop in a thermal chiral perturbation theory
(ChPT) approach ~\cite{Gasser:1987ah}. Additionally, the energy
density, the pressure and the quark condensate have been investigated
up to $\mathcal{O}(p^8)$ in a chiral expansion below the phase
transition~\cite{Gerber:1988tt}. In Ref.\,\cite{Schenk:1993ru} the
shift in the pion pole was calculated as a function of temperature up
to second order in the density. Toublan\,\cite{Toublan:1997rr}
calculated also the pion decay constant within thermal ChPT to two
loops and additionally examined the validity of the
Gell-Mann--Oakes--Renner relation at finite
temperature. It is not \emph{a priori} clear how far up in the
temperature this expansion is applicable, since the partition function
is certainly no longer dominated by the pions for $T \gtrsim
100$\,MeV~\cite{Shuryak:1990ie,Gerber:1988tt}. However, the
Goldstone-boson nature of pions guarantees the existence of a
divergent static correlation length for vanishing quark
masses~\cite{Pisarski:1996mt}.

In standard thermal ChPT, the quark mass as well as the temperature
are treated as small parameters, resulting in an expansion around
$m_q=0$ and $T=0$. In Refs.\,\cite{Son:2001ff,Son:2002ci}, however,
Son and Stephanov investigated perturbations only around $m_q=0$,
keeping the temperature $T$ fixed to any value in the chirally broken
phase. Although an explicit relation of parameters like the quark
condensate and pion decay constant to their $T=0$ counterparts is no
longer possible in this framework, the validity of their results is
extended to a regime where neither ChPT nor perturbative QCD is
usable.  Since lattice simulations rely on the imaginary-time
formalism, extracting real-time observables such as ‘pole masses’ out
of lattice quantities (e.g. Euclidean correlators) is in general a
challenging task. Nevertheless, at small $u,d$ quark masses the real
part of the dispersion relation of soft pions can be obtained from
static Euclidean correlators, with corrections that vanish in the
chiral limit. What happens to the pion pole as the system crosses over
into the chirally restored phase is an aspect that we investigate both
numerically and with guidance from hydrodynamics.

More generally, changes in the spectrum of QCD excitations are an
important aspect of the phenomenology of heavy ion collisions.
In particular, the strongly interacting matter can produce lepton pairs via timelike photons,
and the invariant-mass spectrum of the latter gives access
to the vector spectral function of the thermal medium~\cite{McLerran:1984ay} via a Kubo-Martin-Schwinger relation.
In the hadronic phase, dileptons are primarily produced through the decay of vector
mesons\,\cite{Braun-Munzinger:2015hba,Rapp:1999ej}.
At sufficiently  high temperatures, one expects to be able to compute the vector spectral function
using QCD weak-coupling methods~\cite{Laine:2013vma}.
Presently, it is however not certain in what range of temperatures
mesonic excitations analogous to the $\rho,\omega$ and $\phi$ 
disappear from the medium. Here we address this question 
in the isovector vector channel.

We note that, via a timelike $Z$ boson, dileptons can in principle also be produced
through the axial current~\cite{Laine:2013vma}, though at a very low rate for $T\lesssim 1\,$GeV,
which would probe the axial-vector spectral function of the thermal medium.
A restored chiral symmetry implies the vanishing of the difference of
the isovector vector and axial-vector spectral functions, $V-A$.
Determining in what temperature window this happens for physical quark masses
is one goal of this paper. In fact, we observe larger thermal effects
in the axial-vector spectral function than in the vector case.
There are a number of other probes of chiral symmetry restoration that can be studied.
In Refs.\,\cite{Aarts:2015mma, Aarts:2017rrl} consequences of chiral
symmetry restoration have been investigated in the baryonic sector in
the context of parity doubling.

In the low-temperature phase, the dispersion relation of the pion quasiparticle can be described at low momenta in terms of a renormalization-group invariant (RGI) parameter $u(T)$. This temperature-dependent 
parameter $u(T)$ is directly accessible on the lattice from screening quantities and fulfills a dual role: not only does this parameter predict the modified dispersion relation, but it also relates the \textit{static} pion screening mass to the (\textit{dynamic}) pole mass. 
This has been done in Refs.\,\cite{Brandt:2014qqa,Brandt:2015sxa} and Refs.\,\cite{Ce:2022dax,Krasniqi:2022djb} for the $N_{\mathrm{f}}=2$ and $N_{\mathrm{f}}=2+1$ case, respectively. Here we extend these studies to physical quark masses and higher statistical precision.

The paper is structured as follows: In Sec.\,\ref{sec:prelim} we start with the introduction of some basic definitions which play a key role in the description of the pion quasiparticle. We continue with the implementation of the lattice correlators, followed by a brief description of the numerical setup.  Our results on the pseudoscalar sector, divided into subsections, are presented in Sec.\,\ref{sec:results}. First we extract the mass and decay constant of the screening pion (\ref{sec:PCAC}-\ref{sec:extraction_screening_quant}).
Next, we determine the pion velocity $u$ and examine to what extent its extraction depends on a finite pion thermal width $\Gamma(T)$ (\ref{sec:velocity}-\ref{sec:thermal_width}). Subsequently, we reconstruct a smeared version of the spectral function associated with the two-point function
of the axial charge and compare our results with the literature (\ref{sec:Backus}-\ref{sec:comparison_lit}).
We provide the static pseudoscalar, vector and axial-vector screening masses along our temperature scan in Sec.\,\ref{sec:comp_HotQCD}.
Thereafter, we compare our lattice estimate for the isovector quark number susceptibility with the prediction from the hadron resonance gas model (HRG) in Sec.\,\ref{sec:suscep}. 
We look at several order parameters for chiral symmetry restoration in Sec.\,\ref{sec:order_param}, including the $V-A$ difference.
Sec.\,\ref{sec: vector_and_axial_vector_spectral_functions} contains our results on the thermal modification
of the vector and axial-vector spectral functions.
Finally, we collect our conclusions in Sec.\,\ref{sec:conclusion}.

\section{Preliminaries}
\label{sec:prelim}
In this section we introduce the notation and some basic definitions as well as the key quantities characterizing the pion quasiparticle that we will use throughout the paper. Furthermore, the lattice implementation of the correlators and the numerical setup are described briefly.

\subsection{Definition of operators and correlation functions}
\label{sec:basic definitions}

The notation and conventions used in this work are adapted from Ref.\,\cite{Ce:2022dax}. Our framework is the up, down and strange-quark sector of Euclideanized QCD on the space $S^1 \times \mathbb{R}^3$, $S^1$ denoting the Matsubara cycle of length $\beta \equiv 1/T$. We define the (isovector) pseudoscalar density, the vector current and the axial-vector current as
\begin{align}
    \label{eq:ps. density, vector curr, ax. curr}
    P^a(x) = \bar{\psi}(x)\gamma_5\frac{\tau^a}{2}\psi(x)\,,\ \ \ V^a_{\mu}(x) = \bar{\psi}(x)\gamma_{\mu}\frac{\tau^a}{2}\psi(x)\,,\ \ A^a_{\mu}(x) = \bar{\psi}(x)\gamma_{\mu}\gamma_5\frac{\tau^a}{2}\psi(x)\,,
\end{align}
where $a \in \{1,2,3\}$ is an adjoint $SU(2)_{\text{isospin}}$ index, $\tau^a$ is a Pauli matrix and $\bar\psi(x)=(\bar u(x)~ \bar d(x))$ is a Dirac field flavor doublet. The partially-conserved axial current (PCAC) relation is an operator identity that holds in Euclidean space when inserted in expectation values. It relates the divergence of the axial vector current $A^a_{\mu}(x)$ to the pseudoscalar density $P^a(x)$,
\begin{align}
    \label{eq:PCAC relation}
    \partial_{\mu}A_{\mu}^a(x) = 2m_{\text{PCAC}}\,P^a(x)\,.
\end{align}
In the path integral formulation, this relation results from performing a chiral rotation $\delta_A^a\psi(x)=\frac{\tau^a}{2}\gamma_5\psi(x)$ of the fields (see Ref.\,\cite{Luscher:1998pe}). Applying the pseudoscalar density operator on both sides and taking the expectation value one can obtain the bare PCAC quark mass,
\begin{align}
    \label{eq:PCAC bare mass}
    2 m_{\text{PCAC}}{\langle P^b(x)P^b(0)\rangle} = {\partial_{\mu}\langle A_{\mu}^a(x)P^a(0)\rangle}\,.
\end{align}
It is convenient to integrate over three space-time directions, leaving only one direction in which the derivative acts on the axial current.
Since the PCAC relation is an operator identity, we are free to choose this direction. 
 In our thermal system, the spatial direction is at least four times larger than the temporal one. As a consequence, measuring along the spatial direction results in a longer plateau and thus, smaller errors.
 Therefore, we will extract the PCAC quark mass from the relation
 \begin{align}
   \label{eq:mPCAClat}
    m_{\text{PCAC}}(x_3) = \frac{1}{2}\frac{\int\mathrm{d}x_0\mathrm{d}^2x_{\perp}\,\langle\widetilde\partial_3 A_3^{a}(x)P^a(0)\rangle}{\int\mathrm{d}x_0\mathrm{d}^2x_{\perp}\,\langle P^b(x)P^b(0)\rangle}\, ,\ \ \ \ \ \ x_{\perp} = (x_1,x_2)\, .
\end{align}

We introduce the static screening axial correlator,  given by
\begin{align}
    \label{eq:asymp}
    \delta^{ab} G_A^s(x_3,T) = \int \, \md x_0 \md ^2x_{\perp} \langle A_3^{a}(x)A_3^{b}(0) \rangle \overset{\abs{x_3}\rightarrow \infty}{=} \delta^{ab} \frac{f_{\pi}^2m_{\pi}}{2} e^{-m_{\pi} \abs{x_3}}\, ,
\end{align}
where we have specified the asymptotic form of the correlator,
which defines the pion screening mass $m_\pi$ and decay constant $f_\pi$.
Analogously, we define\footnote{Please note that in the case of identical operators at source and sink, we do not use double labels for the correlators, e.g. $G_P^s(x_3,T)$, instead of $G_{PP}^s(x_3,T)$ } the following static screening correlators:
\begin{align}
    \label{eq:def_static_PP_corr}
    \delta^{ab} G_P^s(x_3,T) &= \int \, \md x_0 \md ^2x_{\perp} \langle P^a(x)P^b(0) \rangle \\
    \label{eq:def_static_AP_corr}
    \delta^{ab} G_{AP}^s(x_3,T) &= \int \, \md x_0 \md ^2x_{\perp} \langle A_3^{a}(x)P^b(0) \rangle\,.
\end{align}
In order to probe the dynamical properties of the thermal system,
we define time-dependent correlators, projected to a definite spatial momentum,
\begin{align}
  \delta^{ab} G_{A_0}(x_0,T) &= \int \, \md^3x\, \langle A_0^{a}(x)A_0^{b}(0) \rangle \\
    \delta^{ab} G_{P}(x_0,T) &= \int \, \md^3x\, \langle P^a(x)P^b(0) \rangle \\
    \label{eq:A0P}
    \delta^{ab} G_{PA_0}(x_0,T) &= \int \, \md^3x\, \langle P^a(x)A_0^{b}(0) \rangle =- \int \, \md^3x\, \langle A_0^{a}(x)P^b(0) \rangle  \\
    \label{eq:def_time_dep_corr}
  \delta^{ab} G_{A}(x_0,{\bf p},T) &= -\frac{1}{3} \sum_{i=1}^3 \int \, \md^3x\, e^{-i{\bf p}\cdot {\bf x}}\,\langle A_i^{a}(x)A_i^{b}(0) \rangle \\
  \delta^{ab} G_{V}(x_0,{\bf p},T ) &= -\frac{1}{3} \sum_{i=1}^3 \int \, \md^3x\, e^{-i{\bf p}\cdot {\bf x}}\,\langle V_i^{a}(x)V_i^{b}(0) \rangle\,.
\end{align}
The time-dependent correlators are related to the corresponding spectral function  (see e.g.\ the review~\cite{Meyer:2011gj}) via
\begin{equation}
  \label{eq:axial_spec_func}
  G_X(x_0,\vec p,T) = \int_0^{\infty}\,\mathrm{d}\omega\, \rho_X(\omega, {\bf p},T)\,\frac{\text{cosh}(\omega(\beta/2-x_0))}{\text{sinh}(\omega\beta/2)}\,, \quad X=A_0,P,A,V.
\end{equation}
In Sec.\,\ref{sec:Backus} we will analyze the axial spectral function using the Backus-Gilbert method.
The temperature argument of the correlators and spectral functions will not always be shown explicitly.

\subsection{Pion properties at finite temperature}
It has been established within several frameworks~\cite{Son:2001ff,Son:2002ci} that at temperatures below the chiral phase transition a pion quasiparticle persists, with the real part of the dispersion relation of sufficiently soft pions given by
\begin{align}
    \label{eq:dispersion}
    \omega_{\bold{p}} = u(T) \sqrt{m_{\pi}^2 + \bold{p}^2} \, , \ \ \ \text{for any} \  T \lesssim T_c\, .
\end{align}
In the chiral limit, the parameter $u(T)$ can be interpreted as the group velocity of a massless pion excitation.
While the quasiparticle mass $\omega_{\bold{0}}$ is the real-part of a pole of the retarded correlator $G^R_P(\omega, {\bf p}=0, T)$
of the pseudoscalar density in the frequency variable, the screening mass $m_\pi$ is a pole of
$G^R_P(\omega = 0, \bold{p},T)$ in the spatial momentum $\abs{\bold{p}}$ and represents an inverse spatial correlation length.
A simple interpretation of the dispersion relation (\ref{eq:dispersion}) was given in Ref.\,\cite{Brandt:2015sxa} in terms of the poles
of the screening and the time-dependent correlators. Son and Stephanov~\cite{Son:2001ff,Son:2002ci}
showed that the pion velocity $u$ in the chiral limit is given by the ratio of $f_\pi^2$ to the axial charge susceptibility.
The latter, however, contains an ultraviolet divergence at any non-vanishing quark mass~\cite{Brandt:2014qqa} and is therefore not practical for lattice calculations. As an alternative, in Refs.\,\cite{Brandt:2014qqa,Brandt:2015sxa} the parameter $u$ was determined using lattice correlation functions at vanishing spatial momentum via two independent estimators,
\begin{align}
    \label{eq:u_m}
    u_m &= \left[\left.-\frac{4m_q^2}{m_{\pi}^2}\,\frac{G_P(x_0,T)}{G_{A_0}(x_0,T)}\right\vert_{x_0=\beta/2}\right]^{1/2}\, , \\
    \label{eq:u_f}
    u_f &= \frac{f_{\pi}^2 m_{\pi}}{2G_{A_0}(\beta/2,T)\,\text{sinh}(u_fm_{\pi}\beta/2)}\, ,
\end{align}
that agree in the chiral limit. We adopt them for the present study.
In doing so, for the estimator $u_m$, the parametric dominance of the pion in the time-dependent Euclidean axial as well as the pseudoscalar density correlator at small quark masses is exploited. The estimator $u_f$ only exploits the pion dominance in the axial correlator; on the other hand, it relies on the residue determined from the static screening correlator.
The pion contribution to the spectral function $\rho_{A_0}$ is expected to take the form of a sharp peak, 
\begin{equation}
\label{app:eq:spec}
    \rho_{A_0}(\omega, T) = \text{sgn}(\omega)\text{Res}(\omega_{\bold{0}})\delta(\omega^2-\omega_{\bold{0}}^2)+\dots\,,
\end{equation}
where in Ref.\,\cite{Son:2002ci} (see also Ref.\,\cite{Brandt:2015sxa}) the residue was predicted to have the form 
\begin{equation}
    \label{eq:def_res}
 \text{Res}(\omega_{\bold{0}}) \equiv (f_{\pi}^t)^2\omega_0^2 = f_{\pi}^2m_{\pi}^2\,,
\end{equation}
such that we can access the quasiparticle decay constant via $f_{\pi}^t = f_{\pi}/u_m$.

We note that the equations of this subsection unambiguously define the observables $\omega_{\vec 0}$, $u_m$, $u_f$, $m_\pi$ and $f_\pi$,
even for temperatures above the chiral crossover, even though the notation reflects their interpretation in the chirally broken phase.
In the next subsection, we elaborate on the expected pole structure at low energies in the axial-charge correlator.

\subsection{On the pole in the axial-charge correlator above the chiral crossover}

From the static pseudoscalar correlator, we define the overlap $G_s$ onto the (screening) ground state,
\be
\int_0^\beta \mathrm{d}x_0\; \<P^a(x)\; P^b(0)\>  \stackrel{|\vec x|\to\infty}{=} - \delta^{ab}\,G_s^2 \; \frac{e^{-m_s|\vec x|}}{4\pi |\vec x|}.
\ee
Note that the engineering dimension of $G_s$ is GeV$^2$.
In this and the following two paragraphs,
we use Minkowski-space notation with  mostly minus metric $\eta^{\mu\nu}$ and $\{\gamma^\mu,\gamma^\nu\} = 2\eta^{\mu\nu}$.
In Ref.\,\cite{Son:2002ci}, the spectral functions of the operators $A^{0,a}$ and $P^a$ were determined in the chirally broken phase.
Both are hydrodynamic fields at small quark masses, $A^0$ because it is the density of the almost conserved axial charge 
and $P$ because $\varphi^a = 2i \frac{P^a}{\<\bar \psi \psi\>}$ corresponds to the phase of the condensate~[see e.g. Ref.\,\cite{Sharpe:2006pu}, Sec.\,2.2], which as a Goldstone mode
should be included in the list of hydrodynamic fields~\cite{Son:2002ci}.

Here we are concerned with the chirally symmetric phase. Our main goal is to determine
the form of the spectral function of the axial charge density, which
is a hydrodynamic field, at low frequencies and momenta.
The corresponding current, up to higher-order terms,  can be expressed via Fick's law,
\be
\vec A^{a} = -D_A \nabla A^0 + \dots
\ee
In the chirally restored phase, the pseudoscalar density is no longer a hydrodynamic field.
Thus, it should be expressed via the axial charge density and its spatial gradients.
To leading order, we have the constitutive equation
\be
\phi^a \equiv iP^a = -\frac{\bar\lambda}{2} m_q G_s^2 A^{0,a} + \dots,
\ee
where the suppression by the quark mass reflects the fact that
$P$ and $A^0$ belong to different chiral multiplets and  as such decouple in the chiral limit at the linear level.
Note that $\bar\lambda$ is a hydrodynamic coefficient
and $(T^5\bar\lambda)$ is renormalization-group invariant.

The PCAC relation can then be interpreted as an equation of motion for the axial charge density,
\be
\partial_0 A^{0,a} = - \nabla\cdot \vec A^a + 2m_q \phi^a
=  D_A\triangle A^{0,a} -  {\bar\lambda} m_q^2 G_s^2 A^{0,a} + \dots\,.
\ee
Linear response theory then leads to a prediction for the form of the spectral function
for the axial charge density (see e.g. Ref.\,\cite{Meyer:2011gj}),
\be\la{eq:rhoAhydro}
\frac{\rho_A(\omega,\vec k)}{\omega} = \frac{\chi_A(\vec k)}{\pi}\;
\frac{\bar\lambda m_q^2 G_s^2 + D_A\vec k^2}{\omega^2+(\bar\lambda m_q^2 G_s^2  +D_A\vec k^2)^2}.
\ee
Close to the chiral limit, $\chi_A(\vec k)\simeq \chi_s$ is close to the isospin susceptibility corresponding to the vector current,
and similarly, $D_A=D_I$ corresponds to the diffusion coefficient of isospin.
Thus, in the chiral regime the only new and independent transport coefficient is $\bar\lambda$.
The qualitative difference in the spectral functions of the broken phase and the restored phase at $\vec k=0$ is illustrated
in Fig.\ \ref{eq:rhoA0A0quali}.
We also note that for non-interacting quarks, $\bar\lambda=0$; see Appendix \ref{sec:freerhoA0}.

\begin{figure}
  \centerline{\includegraphics[width=0.6\textwidth]{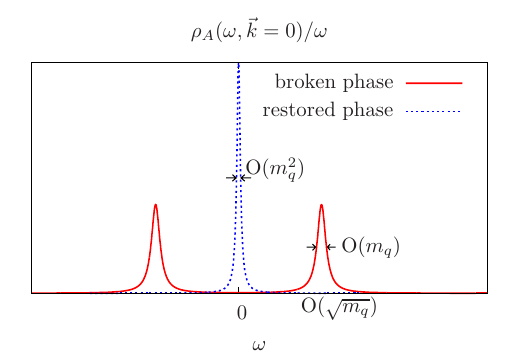}}
  \vspace{-0.4cm}
  \caption{\la{eq:rhoA0A0quali} Qualitative features of the low-energy poles in the spectral function of the axial charge,
    in the broken and in the restored phase.}
\end{figure}

From the result (\ref{eq:rhoAhydro}), one deduces the spectral funtions
for the $\<A^0 P\>$ and $\<P P\>$ correlators at vanishing spatial momentum,
\ba
\rho_{AP}(\omega,\vec 0) &=&  \frac{\omega}{2m_q} \rho_A(\omega,\vec 0)
=\frac{\chi_A(0)}{2\pi} \;\frac{\bar\lambda m_q G_s^2 \omega^2}{\omega^2 + (\bar\lambda m_q^2 G_s^2)^2},
\\
\frac{\rho_{P}(\omega,\vec 0)}{\omega} &=&
-\frac{1}{2m_q} \,\rho_{AP}(\omega,\vec 0)
= -\frac{\chi_A(0)}{4\pi} \;\frac{\bar\lambda G_s^2 \omega^2}{\omega^2 + (\bar\lambda m_q^2 G_s^2)^2}.
\ea
We note that in the chiral limit,
\be
\lim_{\omega\to 0} \frac{\rho_{P}(\omega,\vec 0)}{\omega} 
= -\frac{\chi_s}{4\pi} \;\bar\lambda \,G_s^2 .
\la{eq:Kubolda}
\ee
By contrast, at large frequencies we have $\frac{\rho_{P}(\omega,\vec 0)}{\omega}\stackrel{|\omega|\to\infty}{\sim} - \frac{N_c}{16\pi^2}|\omega|$.
Eq.\,(\ref{eq:Kubolda}) represents a Kubo formula for the transport coefficient $\bar\lambda$.

Finally, it is useful to record the parametric size of the Euclidean correlators $G_{A_0}$ and $G_P$
at $x_0=\beta/2$; see Table \ref{tab:CorrBeta2}. This information determines whether the real part of the pole can be determined
without having to address an inverse problem. Clearly, this is possible in the broken phase near the chiral limit, because
both $G_{A_0}$ and $G_P$ are dominated by the pole contribution.
In the restored phase, however, the pseudoscalar correlator is not dominated by this contribution.

\begin{table}
  \centerline{\begin{tabular}{c|cc||cc}
      \hline      \hline
      & pion pole & rest & diffusion pole & rest \\
            \hline
  $G_{A_0}$ & O(1) &  O($m_q^2$) & O(1) & O($m_q^2$) \\ 
      $G_P$ &  O($m_q^{-1}$) & O(1) & O(1) &O(1)  \\
            \hline\hline
  \end{tabular}}
  \caption{\la{tab:CorrBeta2} Parametric size, in terms of the average up-down quark mass $m_q$,
    of the low-energy contribution vs.\ the rest in the axial charge density and
    in the pseudoscalar density Euclidean-time dependent correlators at zero spatial momentum and $x_0=\beta/2$.
  The left part of the table concerns the chirally broken phase, and the right-hand part the restored phase.}
  \end{table}

\subsection{Lattice implementation of the correlators}
\label{sec:lat_impl}
In this work we use exclusively the local discretizations of the operators introduced in the previous subsection.
Therefore, the expression of the bare operators in the lattice theory is the same as in Eq.\,(\ref{eq:ps. density, vector curr, ax. curr}).
These bare operators are first O($a$)-improved and then renormalized.
While the bare pseudoscalar density is by itself O($a$)-improved, the details of the improvement of the vector and axial-vector currents are given in Appendix~\ref{app:impr_process}.

The finite renormalization of the vector and the axial-vector currents
is performed with the non-perturbatively determined renormalization
factors $Z_V(g_0^2)$ and $Z_A(g_0^2)$, supplemented by a quark-mass
dependent factor in order to fully realize O($a$) improvement; details
are provided in Appendix~\ref{app:ren}.

The pseudoscalar density $P^a(x)$ acquires a scale (and scheme) dependence via the process of renormalization.
The renormalization factor is notated $Z_P(g_0^2,a\mu)$.
Here, we renormalize $P^a(x)$ in the (non-perturbative)
gradient-flow (GF) scheme at the renormalization scale $\mu$ where the corresponding coupling $\bar g^2_{\rm GF}=9.25$;
this corresponds to a low scale of  $\mu\approx230\;$MeV \cite{Campos:2018ahf}.
While none of our physics applications relies on the choice of a specific scheme, we note that in the latter publication,
the scale dependence of the renormalization factor has been computed up to perturbative scales $\mu$;
thereby the connection to the renormalization-group invariant (RGI) operator is known.

The chosen renormalization of the PCAC mass preserves the axial Ward identity Eq.\,(\ref{eq:PCAC relation}). Thus, all renormalization-scale dependent quantities in this paper are quoted in the aforementioned gradient-flow scheme.
In particular the PCAC mass is renormalized by multiplying it with $Z_A/Z_P$, and the combination $m_\pi^2 f_\pi^2 / m_{\rm PCAC}$ considered
in Sec.\,\ref{sec:GOR} via the factor $ Z_A Z_P$. The numerical values of the renormalization factors are collected in Appendix~\ref{app:ren}.

\subsection{Numerical setup}

\sloppy Our calculations are performed on three $N_{\mathrm{f}}=2+1$ ensembles with tree-level $\mathcal{O}(a^2)$-improved Lüscher-Weisz gauge action and non-perturbatively $\mathcal{O}(a)$-improved Wilson fermions\,\cite{Bulava:2013cta}. The action corresponds to the choice of the Coordinated Lattice Simulations (CLS) initiative\,\cite{Bruno:2014jqa} and the bare parameters match those of the CLS zero-temperature ensemble E250\,\cite{Mohler:2017wnb}.
The latter are listed in Table\,\ref{tab:E250params}, together with the lattice spacing as determined in Ref.\,\cite{Bruno:2016plf}. 
The spatial extent is $N_s=96$ on all three ensembles and the time direction admits thermal boundary conditions with $N_{\tau}\in\{24,20,16\}$, which is the only difference relative to the zero-temperature ensemble, resulting in the temperatures 
\begin{align}
T_{24} &= \frac{1}{24a} = 127.9(1.5)\,{\rm MeV}\,,\\
T_{20} &= \frac{1}{20a} = 153.5(1.8)\,{\rm MeV}\,,\\
T_{16} &= \frac{1}{16a} = 191.9(2.3)\,{\rm MeV}\,.
\end{align}
With a pseudocritical temperature $T_{pc}=156.5(1.5)$\,MeV obtained in $(2+1)$-flavor QCD \cite{HotQCD:2018}, our temperature scan covers the range $\{T_{24},T_{20},T_{16}\}/T_{pc} \approx \{0.82,0.98,1.23\}$. Notably, these temperature choices place one ensemble in the hadronic phase, another in the high-temperature or quark-gluon plasma phase, and the third at the chiral crossover.

For reference, we also quote the zero-temperature pseudoscalar masses and pion decay constant, determined in Ref.\,\cite{Ce:2022kxy},
\begin{align}
\label{eq:mpimKT0}
T=0:\qquad  m_\pi^0 &= 130.0(0.7)(1.5) \,{\rm MeV}, \qquad m_K^0 = 489.36(0.25)(5.8)\, {\rm MeV},\\
\label{eq:f_pi_0}
f_{\pi}^0 &= 87.1(0.3)(1.0)\,{\rm MeV}
\end{align}
where the first error is from the corresponding quantity in lattice units, and the second is from
the lattice spacing determination of Ref.\,\cite{Bruno:2016plf}.

\begin{table}[tb]
\caption{\label{tab:E250params} Parameters and lattice spacing of the ensembles analyzed in this work.
    The lattice spacing determination is from Ref.\,\cite{Bruno:2016plf}.
  }
  \begin{tabular}{c@{~~~}c@{~~~}c@{~~~}c@{~~~}c@{~~~}c@{~~~}c@{~~~}c@{~~~}c}
    \hline
    \hline
    $N_{\tau}/a$ & $L/a$  & $6/g_0^2$ & $\kappa_l$ & $\kappa_s$ & $a\,[{\rm fm}]$ &$N_{\text{conf}}$ & $\frac{\text{MDUs}}{\text{conf}}$ & Label\\
    \hline
    24  & 96  & 3.55  & 0.137232867 & 0.136536633  &  0.06426(76) &  1200 & 4 & E250Nt24\\
    \hline
    20  & 96  & 3.55  & 0.137232867 & 0.136536633  &  0.06426(76) & 1000 & 4 & E250Nt20\\
    \hline
    16  & 96  & 3.55  & 0.137232867 & 0.136536633  &  0.06426(76) & 1500 & 4 & E250Nt16\\
    \hline
    \hline
  \end{tabular}
  \end{table}

The ensembles have been generated using version 2.0 of the openQCD package\,\cite{Luscher:2012av}, applying a small twisted mass to the light quark doublet for algorithmic stability. 
The correct QCD expectation values are obtained including the reweighting factors
for the twisted mass and for the rational approximation of the propagator used to simulate the strange quark\,\cite{Clark:2006fx,Luscher:2012av,Mohler:2020txx,Kuberski:2023zky}.
Errors are estimated using the $\Gamma$ method\,\cite{Madras:1988ei,Wolff:2003sm,Ramos:2018vgu,Joswig:2022qfe}.

Measurements are performed using stochastic wall sources in the implementation of Renwick~J.~\!Hudspith's Witnesser code.
One improvement of this calculation over our previous one~\cite{Ce:2022dax} is the use of stochastic wall sources for the two-point functions. Here we use $Z_2\otimes Z_2$ noise \cite{Foster:1998vw}, inverting on the spin indices only. These sources, also known as ``Z2SEMWall'' \cite{Boyle:2008rh} or ``linked'' sources \cite{ETM:2008zte} only require 4 inversions of the Dirac matrix as compared to the usual 12 for a point source. As the wall-source has support on a large volume, it is less sensitive to local fluctuations of the gauge field, and typically gives a smaller variance to two-point functions at fixed cost in comparison to point sources. On top of this we make heavy use of the Truncated Solver Method \cite{Bali:2009hu} with a large number of low-precision solves, making the calculations presented here fairly cheap to perform.

The results for the E250Nt24 ensemble presented in Ref.\,\cite{Ce:2022dax} were obtained using point sources. Comparing the result for the pion screening decay constant, $f_{\pi}/T=0.559(11)$, therein to our result quoted here, $f_{\pi}/T=0.569(4)$, we have gained a factor of $\approx 8$ in statistics [see Sec.\,\ref{sec:extraction_screening_quant}].


\section{Results on the pseudoscalar sector}
\label{sec:results}

In this section, we present our lattice results on observables in the pseudoscalar sector, i.e.\ those related to pion properties.
As an important benchmark, we begin with the determination of the light quark doublet PCAC masses.
\subsection{The PCAC mass: a control quantity}
\label{sec:PCAC}

\begin{figure}[tp]
	\includegraphics[scale=0.65]{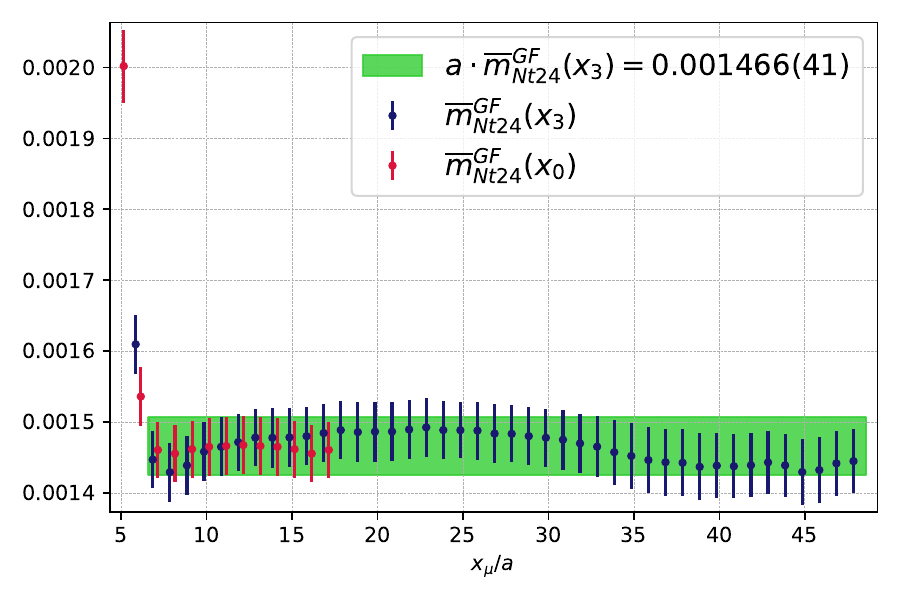}
	\caption{Renormalized PCAC mass in the E250Nt24 ensemble along the $x_0$ and $x_3$ direction. The final result --- obtained from a fit along the $x_3$ direction --- is also shown with a $1-\sigma$ band. We have used the improved axial current together with the symmetrized derivative (see Eqs.\,(\ref{eq: imp_axial_corr}-\ref{eq:symm_deriv_forw_backw})).
	\label{fig:PCACNt24}}
\end{figure}

\begin{figure}[h!]
\center
	\includegraphics[scale=0.54]{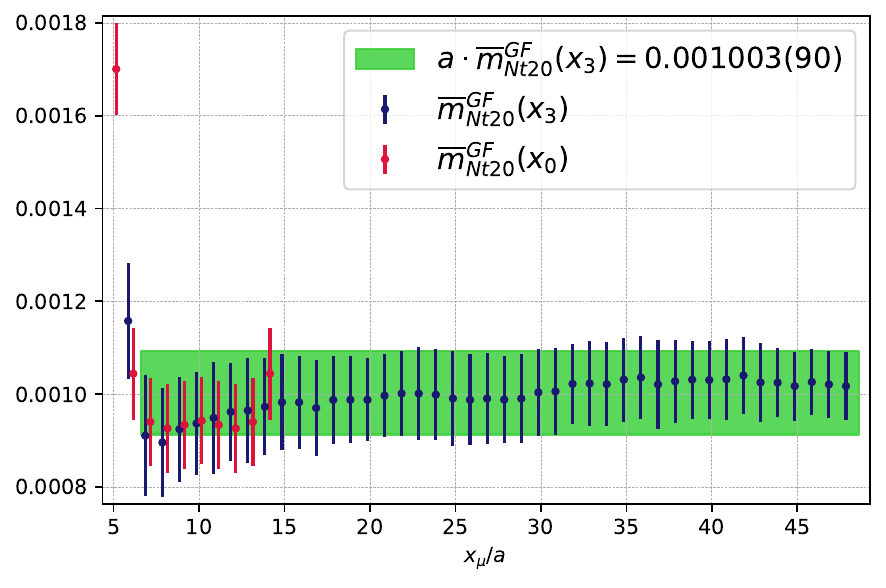}
	\includegraphics[scale=0.54]{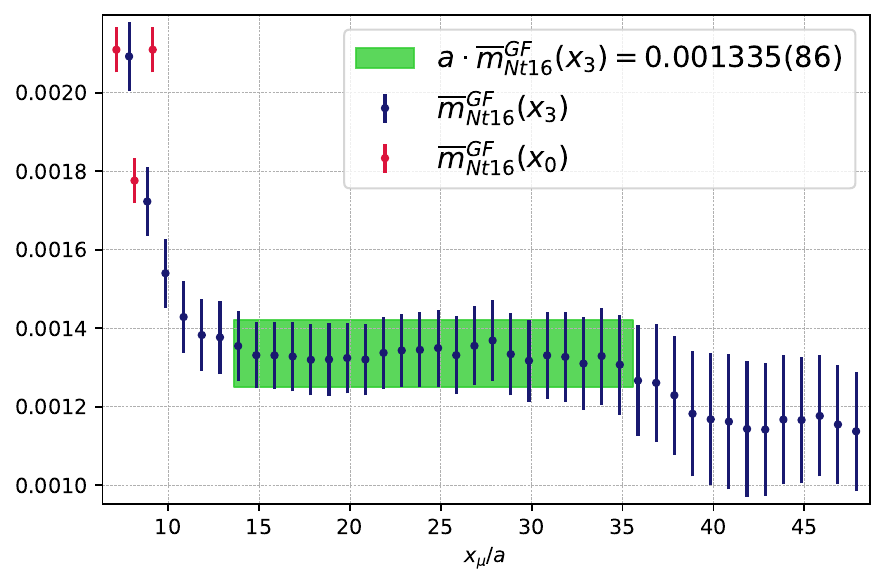}
        \caption{ 
        Renormalized PCAC masses in the E250Nt20 ensemble (\textbf{left}) and E250Nt16 ensemble (\textbf{right}) along the $x_0$ and $x_3$ direction. For the E250Nt20 ensemble we have used $\mathcal{O}(a)$-improvement of the axial current and the symmetrized derivative (see Eqs.\,(\ref{eq: imp_axial_corr}-\ref{eq:symm_deriv_forw_backw})). In contrast, the PCAC mass on E250Nt16 (\textbf{right}) was determined using the unimproved version of the axial current.}
        \label{fig:PCACNt20and16}
\end{figure}

The extraction of the PCAC masses as defined in Eq.\,(\ref{eq:mPCAClat})
is carried out by performing a fit to a constant in the range where a plateau is observed; see Figs.\,\ref{fig:PCACNt24}-\ref{fig:PCACNt20and16}.
Here we benefit from the long plateaus available in the $x_3$ direction. We obtain
\begin{align}
    \label{eq:PCAC_masses}
    \begin{split}
a\cdot m_{\text{PCAC}}^{vac} &= 0.001428(16)\,,\\ 
a\cdot m_{\text{PCAC}}^{N_t24} &= 0.001466(41)\,,\\
a\cdot m_{\text{PCAC}}^{N_t16} &= 0.001335(86)\,.
\end{split}
\end{align}
For the boxes in the hadronic phase and at the chiral phase transition the extracted PCAC masses obtained from the $x_0$ direction are compatible with the ones obtained from the $x_3$ direction, pointing to small cutoff effects at this value of the lattice spacing. However, the time extent\footnote{Due to periodic boundary conditions, we are dealing with effectively eight points in time direction.} of the high temperature E250Nt16 box is too small to match the value of the PCAC mass obtained from the spatial direction.
Furthermore, for the E250Nt16 ensemble we have not used the improved axial current. The $AP$-correlator is an order parameter for chiral symmetry restoration and consequently approaches zero in the chirally symmetric phase. Thus, the correction term proportional to $c_A$ in Eq.\,(\ref{eq: imp_axial_corr}) becomes of the same order of magnitude as the correlator itself in the chirally restored phase.

\begin{figure}[h!]
\center
	\includegraphics[scale=0.53]{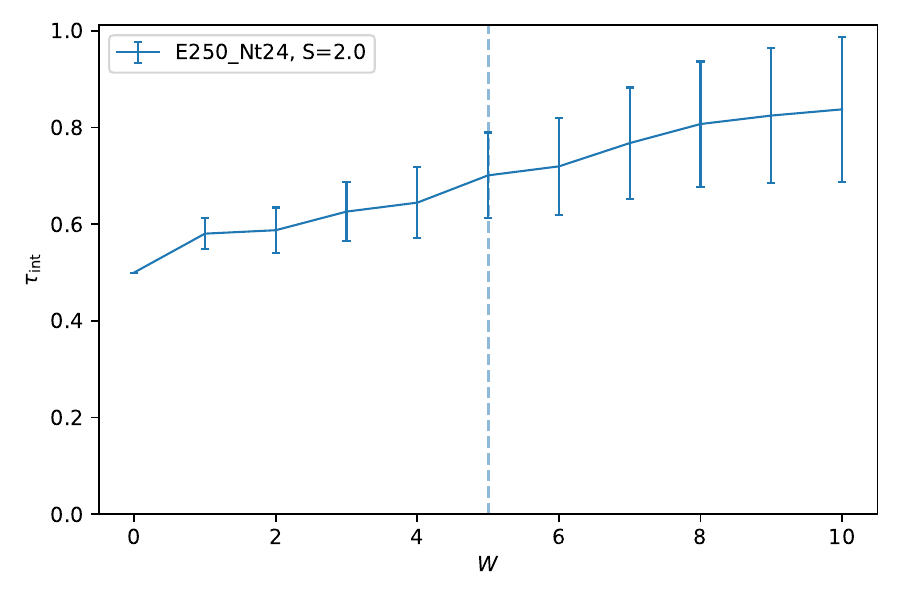}
	\includegraphics[scale=0.53]{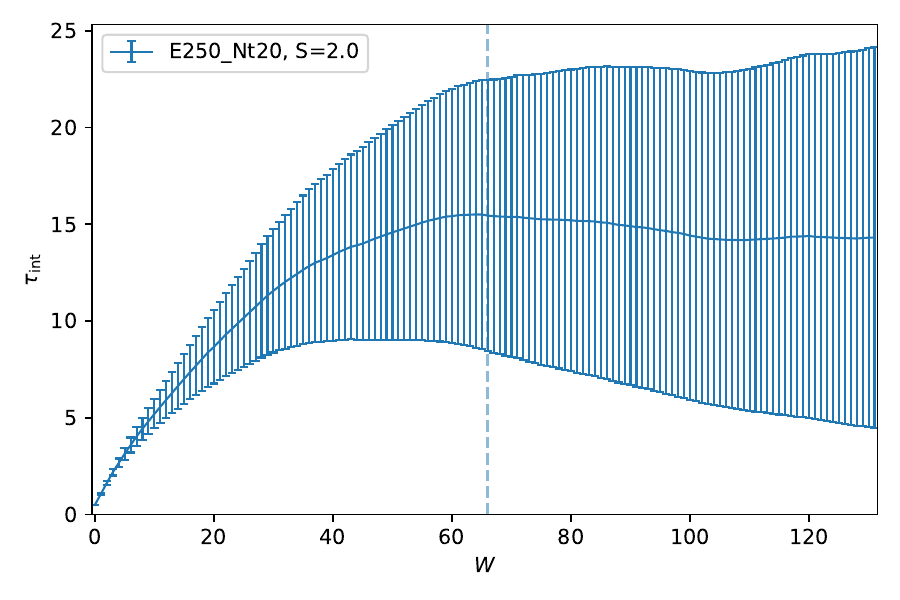}
        \caption{ 
        Comparison of the integrated autocorrelation times $\tau_{int}$ for the $AP$-correlator at fixed source sink seperation $x_3/a=10$. The l.h.s. shows the correlator on the hadronic E250Nt24 ensemble and the r.h.s. shows the correlator on the chiral crossover ensemble E250Nt20.}
        \label{fig:tau_int_Nt20and24}
\end{figure}

Since the PCAC mass is based on an operator identity, it should be independent of the temperature. However, in lattice units the value of the PCAC mass, $a\cdot m_{\text{PCAC}}^{N_t20} = 0.001003(90)$, is off by a factor of $\approx 1.5$ relative to the corresponding vacuum box. As the PCAC mass is proportional to the derivative of the $AP$-correlator, we have compared the integrated autocorrelation time $\tau_{int}$ at fixed source-sink seperation $x_3/a=10$ on the E250Nt20 box with the E250Nt24 box. As can be seen in Fig.\,\ref{fig:tau_int_Nt20and24}, we are dealing with severe autocorrelation in this observable at the chiral crossover (More details in Appendix~\ref{app:error_analysis}).
As a consequence, we are going to use the value of the PCAC mass extracted from the E250Nt24 ensemble in our further analysis, since it is compatible within errors with the PCAC mass on the vacuum ensemble.

\subsection{Static correlators: the pion screening mass and decay constant}
\label{sec:extraction_screening_quant}

In order to extract the pion screening mass and decay constant,
we fit simultaneously the static screening pseudoscalar and the axial-pseudoscalar correlator
In Appendix~\ref{sec:PSscreenApdx}, we provide the details of the calculation, including effective mass plots.

The results are listed in Table\,\ref{tab:results} in units of the respective temperature.
In physical units, they correspond to
\ba
\nonumber
m_{\pi}^{Nt24} &=& 145(2)\,{\rm MeV}\,,
\\ \la{eq:mpi}
m_{\pi}^{Nt20} &=& 196(4)\,{\rm MeV}\,,
\\ m_{\pi}^{Nt16} &=& 555(8)\,{\rm MeV}\,.
\nonumber
\ea
The pion screening decay constant is significantly reduced on all three ensembles compared to the decay constant $f_{\pi}^0=87.4(1.0)$\,MeV on the corresponding zero-temperature ensemble. We obtain
\ba
f_{\pi}^{Nt24} &=& 73(1)\,{\rm MeV}\,,
\\ f_{\pi}^{Nt20} &=& 26(2)\,{\rm MeV}\,,
\\ f_{\pi}^{Nt16} &=& 5.9(4)\,{\rm MeV}\,.
\ea
The observables characterizing the properties of the pion quasiparticle across the chiral phase transition are summarized in Table\,\ref{tab:results}.


\subsection{Properties of the pion quasiparticle}
\label{sec:velocity}

The results for the estimators $u_m$ and $u_f$ defined in Eqs.\,(\ref{eq:u_m},\ref{eq:u_f})
together with those of the screening quantities are presented in Table\,\ref{tab:results}.
Good agreement is found for the two independent estimators $u_f$ and $u_m$ of the pion velocity, in the hadronic as well as high-temperature phase. Both of them differ significantly from unity,
which clearly represents a breaking of Lorentz invariance due to thermal effects. However, well in the chirally restored phase, the interpretation of $u$ as the pion velocity is no longer valid: still,
the ratio $u_f/u_m$ is expected to remain finite in the chiral limit, since both $u_f$ and $u_m$ are of order the
quark mass\footnote{Indeed, $m_\pi$ is of order unity  but $f_\pi$ is of order $m_q$.
In this point, we rectify an uncorrect remark written in Ref.\,\cite{Brandt:2014qqa}.}.
At the chiral crossover, $u_f$ and $u_m$ differ significantly, $u_f/u_m \approx 0.64$; in the latter estimate, we use the PCAC mass obtained from
E250Nt24, since it is more reliable.
Thus, the interpretation as a pion velocity is already questionable at the chiral crossover. 

Additionally, we found that the zero-temperature pion mass given in Eq.\,(\ref{eq:mpimKT0}) `splits' into a lower pion quasiparticle mass and a higher pion screening mass. For the former, we obtain from the estimator $\omega_{\vec 0}=u_m\,m_\pi$
\ba
\omega_{\bold{0}}^{Nt24} &=& 117(2)\,{\rm MeV}\,,
\\
\omega_{\bold{0}}^{Nt20} &=& 81(2)\,{\rm MeV}\,,
\\
\omega_{\bold{0}}^{Nt16} &=& 26(1)\,{\rm MeV}\,;
\ea
these values can be compared to the pion screening masses given in Eq.\,(\ref{eq:mpi}).
Clearly, the drop in $\omega_{\bold{0}}$ is particularly rapid between the crossover point and our ensemble in the high-temperature phase.

The quasiparticle decay constant in the hadronic phase, $f^{t}_{\pi} = f_{\pi}/u_m = 90(1)$\,MeV, is much closer to the vacuum decay constant $f_{\pi}^0=87(1)$\,MeV than $f_{\pi}^{Nt24}$. At the chiral crossover,  we obtain $f^t_{\pi}=62(3)$\,MeV ($f^t_{\pi}=96(3)$\,MeV using $u_m$ determined from the PCAC mass on the E250Nt20 ensemble) and in the high-temperature phase $f^t_{\pi}=127(10)$\,MeV.


\begin{table}[tb]
\caption{Summary of the results of the E250 thermal ensembles with $N_{\tau} \in \{24,20,16\}$. The pion quasiparticle mass $\omega_{\bold{0}}$ is calculated using $\omega_{\bold{0}} = u_m m_{\pi}$.} 
\begin{tabular}{c@{~~~}|c@{~~~}c@{~~~}c}
\hline
\hline
Ensemble                                    & E250Nt24    &E250Nt20     &E250Nt16 \\
\hline
$m_{\pi}/T$                         & 1.136(16)   &1.250(24)&  2.891(43)  \\ 
$f_{\pi}/T$                         & 0.569(4)    &0.172(12)&  0.0305(21)  \\ \hline
$u_{f}$                             & 0.831(3)    &0.274(11)&  0.0462(31)  \\
$u_{m}$                             & 0.807(9)    &0.425(13)&  0.0462(16)  \\
$u_{f}/u_m$                         & 1.028(10)   &0.644(28)&  1.001(76) \\ \hline
$\omega_{\bold{0}}/T$               & 0.917(15)   &0.531(16)&  0.133(4)  \\
$f_{\pi}^t/T$                       & 0.705(7)    &0.405(18)&  0.661(50)  \\
$\text{Res}(\omega_{\bold{0}})/T^4$ & 0.418(14)   &0.0463(60)& 0.0078(11)  \\
\hline
\hline
\end{tabular}
\label{tab:results}
\end{table}

\subsection{Dependence of the pion velocity \texorpdfstring{$u_f$}{uf} on a finite pion thermal width \texorpdfstring{$\Gamma(T)$}{Γ(T)}}
\label{sec:thermal_width}

The analysis of Son and Stephanov~\cite{Son:2001ff, Son:2002ci} concluded that at temperatures below the chiral phase transition, the  imaginary part of the pion pole is parametrically small compared to its real part.
In this subsection, we investigate the sensitivity of our results for the pion quasiparticle mass and velocity parameter $u$
to the assumption of a negligible thermal width of this quasiparticle.
In order to examine the consequences of a finite thermal pion width on the pion velocity $u$ we replace the $\delta$-distribution in Eq.\,(\ref{app:eq:spec}) by a Breit-Wigner peak of width $\Gamma(T)$ resulting in
\begin{equation}
    \label{app:eq:mod_spec}
    \rho_{A_0}(\omega, T) = f_{\pi}^2m_{\pi}^2\frac{\Gamma(T)}{\pi}\frac{1}{2\omega_{\bold{0}}}\left(\frac{1}{(\omega-\omega_{\bold{0}})^2+\Gamma(T)^2}-\frac{1}{(\omega+\omega_{\bold{0}})^2+\Gamma(T)^2}\right)+\dots\,,
\end{equation}
where the second term is needed to ensure the antisymmetry of the spectral function in $\omega$\,\cite{Meyer:2011gj}. Expressing the correlator midpoint of the time-dependent Euclidean correlator $G_{A_0}(\beta/2,T)$ in terms of the spectral function $\rho_{A_0}$ with the help of Eq.\,(\ref{eq:axial_spec_func}) and using $\omega_{\bold{0}}=u_f m_{\pi}$ one can extract the pion velocity $u_f$ for different thermal pion widths. The results are shown in Table\,\ref{tab:thermal_width}.

\begin{table}[th]
\caption{Dependence of the pion velocity $u_f$ on a finite pion thermal width $\Gamma(T)$ on the E250 thermal ensembles with $N_{\tau} \in \{24,20,16\}$.} 
\begin{tabular}{c@{~~~}|c@{~~~}c@{~~~}c}

\hline
\hline
{$\Gamma(T)\,[\text{MeV}]$}  & E250Nt24    &E250Nt20     &E250Nt16 \\
\hline
5  & & &0.0454(31)\\ 
10    & & &0.0426(33)  \\
15    &0.825(3) & 0.262(20) &0.0375(38) \\
20    & 0.819(3)  &0.251(21) &0.0289(50)\\
30    & 0.806(3) & 0.224(23) &\\ 
40    & 0.785(3) & 0.177(28)&   \\
80    & 0.626(6) \\
100   & 0.472(10)& &  \\
\hline
\hline
\end{tabular}
\label{tab:thermal_width}
\end{table}
We find that the extracted estimator of the pion velocity in the hadronic phase, $u_f = 0.831(3)$ (assuming the presence of a discrete delta term in the spectral function), is consistent with a Breit-Wigner approach up to pion thermal widths $\Gamma(T) \approx 15\,\text{MeV}$. At the chiral crossover the estimate assuming a vanishing width, $u_f = 0.274(11)$, is consistent with the estimation of $u_f$ assuming a pion thermal width of up to $\approx 20\,\text{MeV}$.

\subsection{Spectral function reconstruction with the Backus-Gilbert method}
\label{sec:Backus}

In order to extract the spectral function $\rho_{A_0}(\omega)$ at zero momentum from the corresponding temporal Euclidean correlator, $G_{A_0}(\tau_i,T),\ \tau\equiv x_0$, one has to invert the analogue for $G_{A_0}$ of Eq.\,(\ref{eq:axial_spec_func}) with a kernel $K(\tau_i,\omega) = \text{cosh}(\omega(\beta/2-\tau_i))/\text{sinh}(\omega\beta/2)$, encountering a numerically ill-posed problem. 
A possible approach to surmount this is the Backus-Gilbert method\,\cite{Backus:1968}. We adopt the notation of Ref.\,\cite{Brandt:2015sxa}, where the method was applied to lattice QCD for the first time. It should be emphasized that, with this approach, no particular ansatz is needed for the spectral function. The Backus-Gilbert method provides an estimator for the smeared axial spectral function, $\hat{\rho}$.
It is useful to introduce a rescaling function $f(\omega)$, with $\lim_{\omega\to0}f(\omega)/\omega$ finite and $f(\omega)>0$ for $\omega>0$,
and to write $K^f(\tau,\omega) = f(\omega)K(\tau,\omega)$. 
The values of $\hat{\rho}$ are then constructed linearly from the lattice correlator data $G(\tau_i)$,
\begin{equation}
    \label{eq:smeared_sf}
    \frac{\hat{\rho}(\bar{\omega})}{f(\bar\omega)} = \sum_{i=1}^{N_p} q_i(\bar{\omega})G(\tau_i)
 = \int_0^{\infty}\mathrm{d}\omega\,\hat{\delta}(\bar{\omega},\omega)\frac{\rho(\omega)}{f(\omega)}\,.
\end{equation}
Note that the coefficients $q_i$ depend on some reference value $\bar{\omega}$ around which the resolution function,
\begin{equation}
    \label{eq:resolution_function}
    \hat{\delta}(\bar{\omega},\omega) = \sum_{i=1}^{N_{\tau}} q_i(\bar{\omega})K^f(\tau_i,\omega)\,,
\end{equation}
is concentrated. The coefficients $q$ will be chosen such that the resolution function is normalized according to
\begin{equation}
    \label{eq:norm_cond}
    \int_0^{\infty}\mathrm{d}\omega\,\hat{\delta}(\bar{\omega},\omega)=1\,.
\end{equation}
Although the final coefficients $q$ we use depend on the choice of $f$, and 
the functions $\hat{\rho}$ and $\hat{\delta}$ depend on this choice as well, we do not indicate this explicitly,
in order to avoid overcluttering the notation.
The desirable resolution function would be a Dirac delta distribution centered at $\bar{\omega}$.
In order to make the resolution function as sharply centered around $\bar{\omega}$ as possible,
we follow the strategy of minimizing the second moment of its square, subject to the constraint in Eq.\,(\ref{eq:norm_cond}).
However, in order to avoid widely oscillating coefficients $q_i(\bar\omega)$, the quadratic form to be minimized is regulated by the
covariance matrix $\text{Cov}[G]$ of the Euclidean correlator:
defining 
\ba
\label{eq:functional_to_minimize}
R_i &=& \int_0^\infty \mathrm{d}\omega\, K^f(\tau_i,\omega),
\\
W_{ij}(\bar{\omega}) &=& \int_0^\infty \mathrm{d}\omega\, K^f(\tau_i,\omega)(\omega-\bar{\omega})^2K^f(\tau_j,\omega),
\\
W^{\text{reg}}(\bar{\omega}) &=& \lambda W(\bar{\omega}) + (1-\lambda)\text{Cov}[G] \phantom{\Big|}
\ea
with $0<\lambda<1$, the function to be minimized is ${\cal F}_\alpha(q) = (q,W^{\rm reg}q) - \alpha ((q,R)-1)$, with the result
\ba
    \label{eq:coeff_q}
 q(\bar{\omega}) &=&  \frac{1}{(R,W^{\text{reg}}(\bar{\omega})^{-1}R)}\; W^{\text{reg}}(\bar{\omega})^{-1}R.
\ea

\begin{figure}[t]
	\includegraphics[scale=0.43]{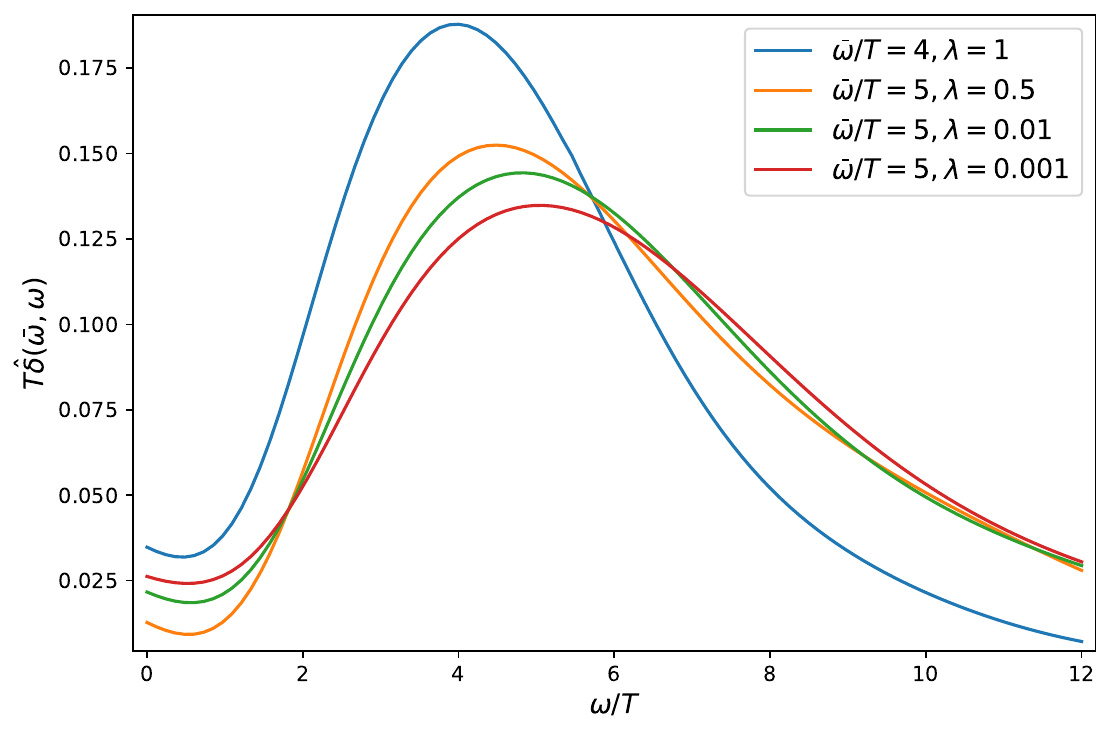}
	\includegraphics[scale=0.43]{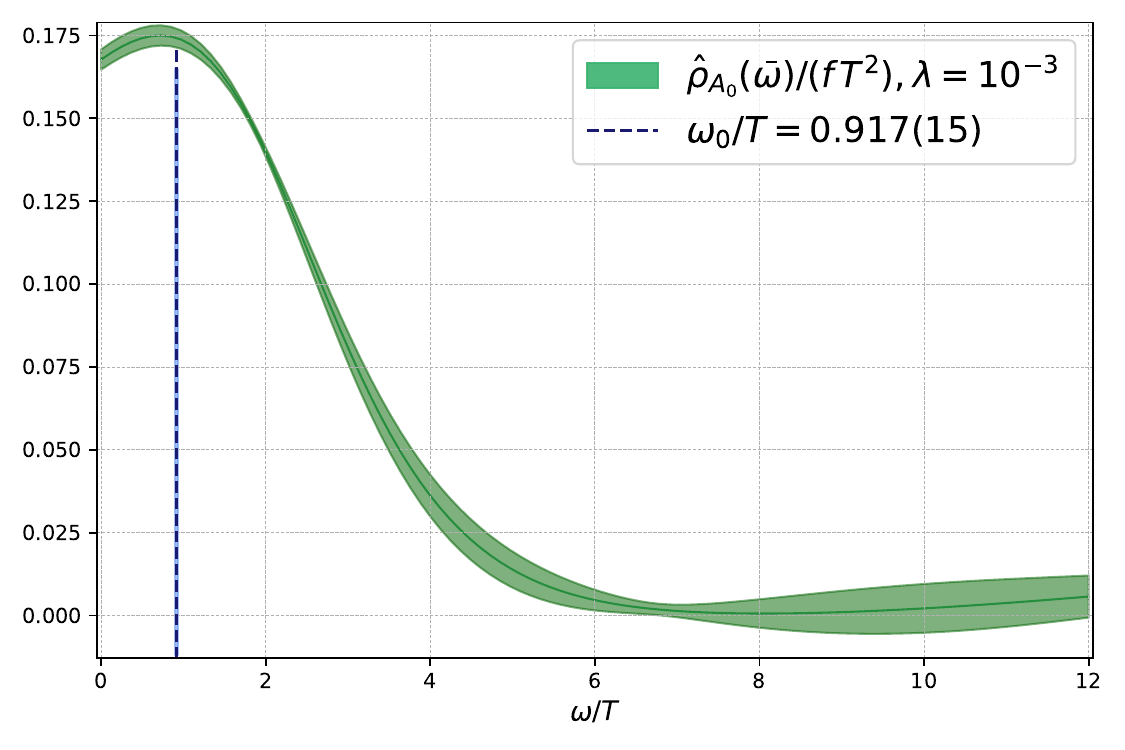}
\caption{
  {\bfseries Left panel}: Some examples of resolution functions for different values of $\lambda$, centered around $\bar{\omega}/T$. {\bfseries Right panel}: Estimator of the spectral function $\hat{\rho}_{A_0}(\omega,T=128\,\text{MeV})/(f\,T^2)$. The blue dashed line corresponds to the location of the expected position of the pole $\omega_{\bold{0}}$ according to Eq.\,(\ref{eq:dispersion}). The temporal $G_{A_0}(x_0,T)$-correlator is $\mathcal{O}(a)$-improved.
}
\label{fig:res_and_spec_Nt24}
\end{figure}

\begin{figure}[t]
	\includegraphics[scale=0.42]{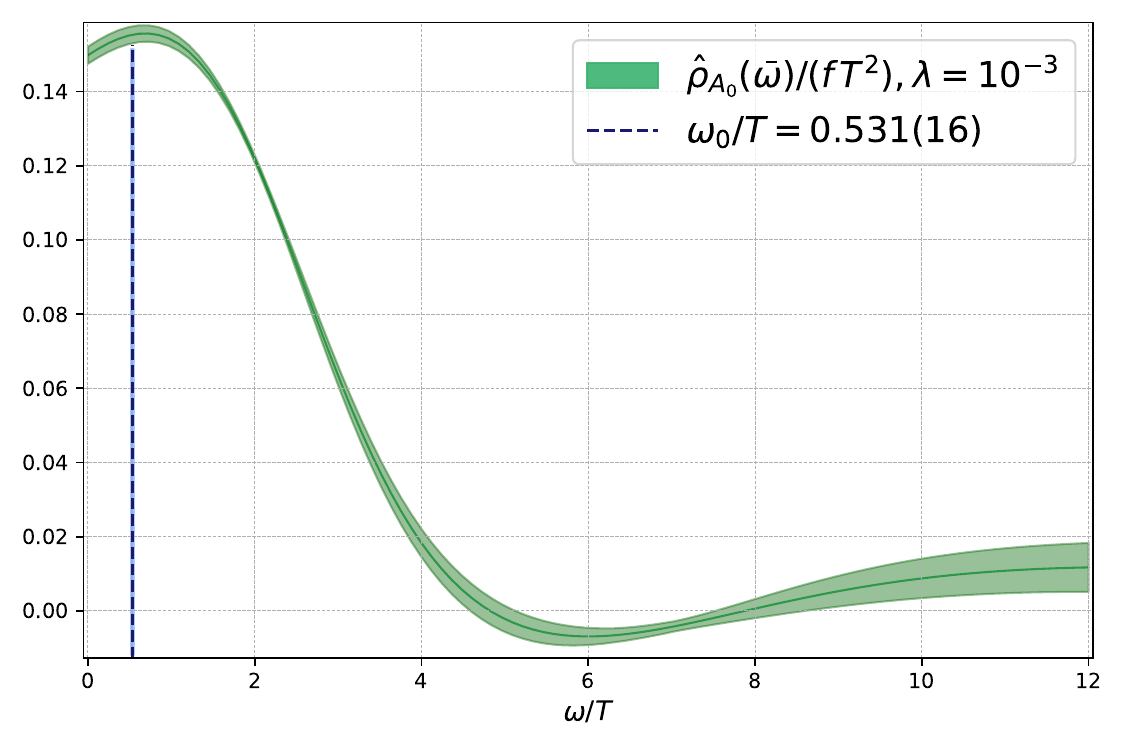}
	\includegraphics[scale=0.42]{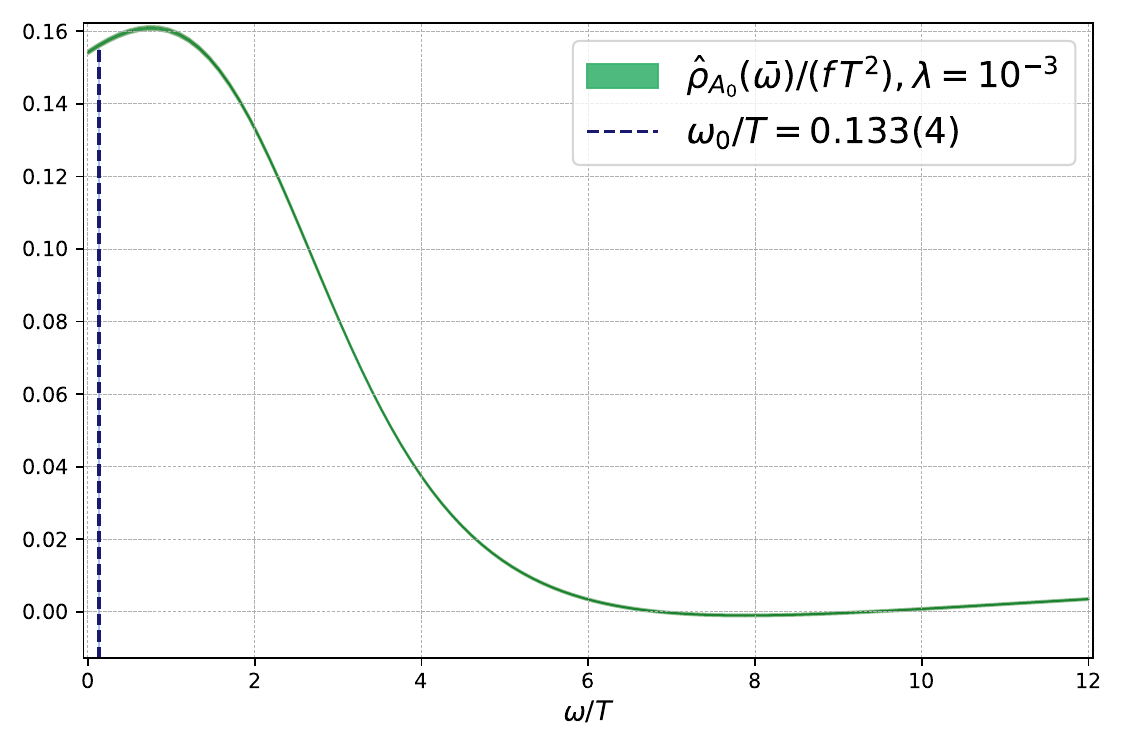}
\caption{
  {\bfseries Left panel}: Estimator of the spectral function $\hat{\rho}_{A_0}(\omega,T=154\,\text{MeV})/(f\,T^2)$. The blue dashed line corresponds to the location of the expected position of the pole $\omega_{\bold{0}}$ according to Eq.\,(\ref{eq:dispersion}). The temporal $G_{A_0}(x_0,T)$-correlator is $\mathcal{O}(a)$-improved. {\bfseries Right panel}: Estimator of the spectral function $\hat{\rho}_{A_0}(\omega,T=192\,\text{MeV})/(f\,T^2)$. The blue dashed line corresponds to the location of the expected position of the pole $\omega_{\bold{0}}$ according to Eq.\,(\ref{eq:dispersion}). 
}
\label{fig:res_and_spec}
\end{figure}

For the analysis of $\rho_{A_0}$, we make the choice
\be
f(\omega) := \tanh(\beta\omega/2).
\ee
The values of $\lambda$ quoted below refer to units in which all dimensionful quantities are turned into dimensionless ones by appropriate powers of the temperature.
Some examples for the resolution function $T \hat{\delta}(\bar{\omega},\omega)$ for different values of $\lambda$ are shown in the left panel of Fig.\,\ref{fig:res_and_spec_Nt24}. The right panel of Fig.\,\ref{fig:res_and_spec_Nt24} shows the smeared and rescaled  axial spectral function $\hat{\rho}_{A_0}(\bar{\omega},T)/(f\,T^2)$ with $\lambda=10^{-3}$ for the hadronic E250Nt24 ensemble, while Fig.\,\ref{fig:res_and_spec} shows the results on the E250Nt20 (left panel) and E250Nt16 ensemble (right panel). It demonstrates model-independently that the axial-charge correlator is dominated by low frequencies. On the other hand, for $\omega/T>8$ we observe the expected perturbative slight increase proportional to the quark mass squared. Furthermore, the predicted position of the quasiparticle mass $\omega_{\bold{0}}$ is close to the peak of the smeared spectral function in the hadronic regime.

Up until now, we have determined the pion quasiparticle mass $\omega_{\bold{0}}$ indirectly with the aid of screening quantities and compared it to the location of the pion pole position in the smeared axial spectral funtion $\hat{\rho}_{A_0}$. As a next step, we want to predict the temporal axial correlator $G_{A_0}(x_0,T=128\,\text{MeV})$ in terms of the screening quantities $f_{\pi}$ and $m_{\pi}$. Using Eqs.\,(\ref{eq:axial_spec_func}) and (\ref{app:eq:spec}-\ref{eq:def_res}) we obtain:
\begin{equation}
    \label{eq:pred_A0}
    G_{A_0}(x_0,T) = \frac{\text{Res}(\omega_{\bold{0}})}{2\omega_{\bold{0}}}\frac{\text{cosh}(\omega_{\bold{0}}(\beta/2-x_0))}{\text{sinh}(\omega_{\bold{0}}\beta/2)}+\cdots .
\end{equation}
The result of this prediction together with the lattice data of the temporal $G_{A_0}(x_0,T)$-correlator and its fit result are displayed in the right panel of Fig.\,\ref{fig:A0A0_E250Nt24_pred_fit} in Appendix~\ref{app:numerical_values}. As can be seen the prediction shows excellent agreement with the lattice data for $x_0/a \geq 7$. The direct fit result reads $\omega_{\bold{0}}/T = 0.91(9)$. As a further crosscheck we also have determined the effective cosh mass directly from the temporal correlator, obtaining $m_{\text{cosh}}(x_0)/T = 0.83(6)$ [see left panel of Fig.\,\ref{fig:A0A0_E250Nt24_pred_fit}].

We want to emphasize that all three estimators for the pion pole mass in the hadronic phase, namely
\begin{itemize}
    \item $\omega_{\bold{0}} = u_m m_{\pi} = 117(2)$\,MeV,\quad\,obtained from screening quantities,
    \item $\omega_{\bold{0}} = 116(12)$\,MeV,\quad\quad\quad\quad \ \ obtained from a direct fit of the temporal correlator,
    \item $m_{\text{cosh}}(x_0)/T = 106(8)$\,MeV,\quad obtained as the effective mass of the temporal correlator,
\end{itemize}
are all consistent within errors of each other. They all give significantly reduced values in comparison to the \textit{in vacuo} pion mass value of $m_{\pi}^0=130(2)$\,MeV.

\subsection{Comparison with results from the literature}
\label{sec:comparison_lit}
Comparing our pion quasiparticle mass $\omega_{\bold{0}}(T)$ and quasiparticle decay constant $f_{\pi}^t(T)$ at $T=128$\,MeV to the matching quantities at the corresponding zero temperature ensemble we get $\omega_{\bold{0}}(T)/ m_{\pi}^0 = 0.892(14)$ and $f_{\pi}^t(T)/ f_{\pi}^0 = 1.023(10)$. Thus, the quasiparticle mass decreases at finite temperature while the quasiparticle decay constant experiences a slight increase. This behavior is similar to what is found in a ChPT calculation (see Ref.\,\cite{Toublan:1997rr}, Fig.\,3 and Fig.\,4) at two loops, where the reduction of the quasiparticle mass therein is $\approx 0.9$. On the other hand the quasiparticle decay constant increases by a factor of approximately 1.03. That the pion pole mass decreases with temperature is also supported by a recent lattice-improved soft-wall AdS/QCD model study\,\cite{Wen:2024hgu}.
On the chiral crossover ensemble, we obtain $\omega_{\bold{0}}(T)/ m_{\pi}^0 = 0.63(2)$ and $f_{\pi}^t(T)/ f_{\pi}^0 = 0.69(3)$. In the high temperature phase, we obtain $\omega_{\bold{0}}(T)/ m_{\pi}^0 = 0.17(1)$ and $f_{\pi}^t(T)/ f_{\pi}^0 = 1.57(5)$.

Regarding the screening pion mass $m_{\pi}$, we found that it increases with temperature compared to $m_{\pi}^0$. The ratio is $m_{\pi}/m_{\pi}^0 = 1.105(14)$ at $T=128\,$MeV. This increase is also supported qualitatively by the study of Son and Stephanov near the chiral phase transition\,\cite{Son:2001ff}. 
On the chiral crossover ensemble we obtain $m_{\pi}/ m_{\pi}^0 = 1.460(27)$, while the high temperature phase ensemble gives $m_{\pi}/ m_{\pi}^0 = 4.218(63)$.

In Ref.\,\cite{Goderidze:2022vlm}, the authors work out the pion damping width and the pion spectral function in the framework of a $SU(2)$ Nambu-Jona-Lasinio (NJL) model for a few temperatures below the critical temperature $T_c^{\text{NJL}}$ = 190\,MeV. They observe that the position of the peak of the pion spectral function at vanishing momentum $\bold{p}$ moves to the right for increasing temperatures $T/T_c^{\text{NJL}} \in \{0,0.79,0.89,0.97\}$ (see Fig.\,(3) in Ref.\,\cite{Goderidze:2022vlm}). This contradicts our observation of the pion pole mass being reduced at finite temperature.

\section{Static mesonic screening masses}
\label{sec:comp_HotQCD}


\begin{figure}[tp]
	\includegraphics[scale=0.52]{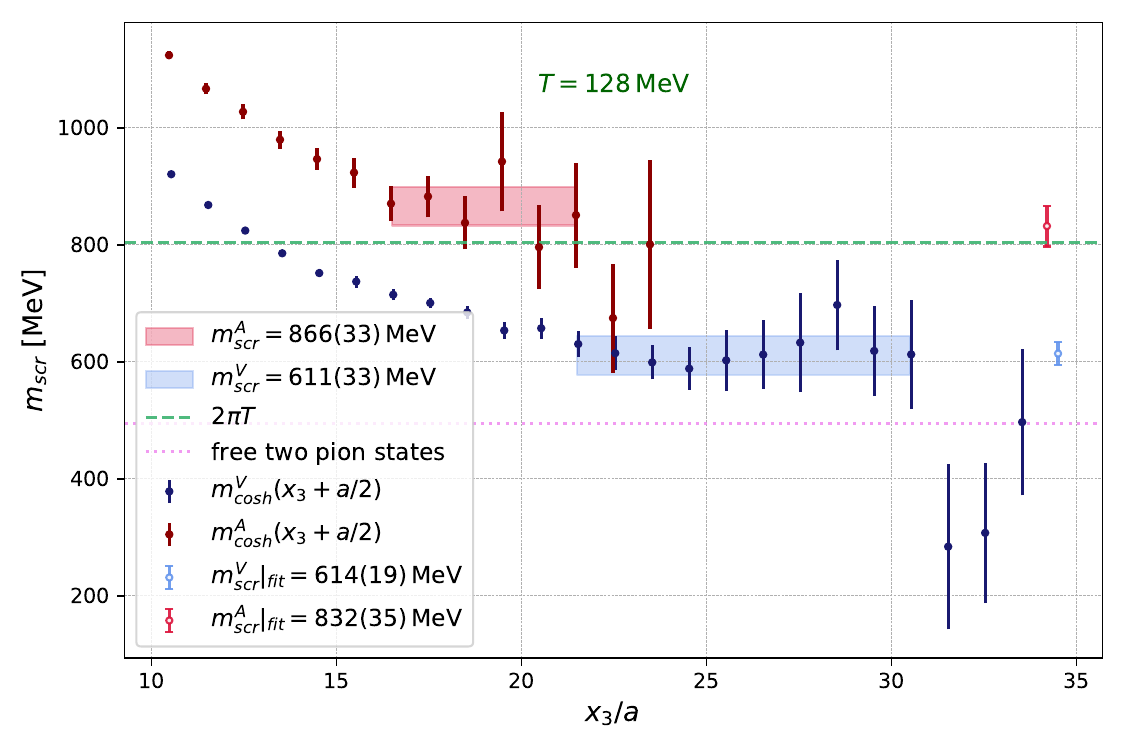}
	\includegraphics[scale=0.52]{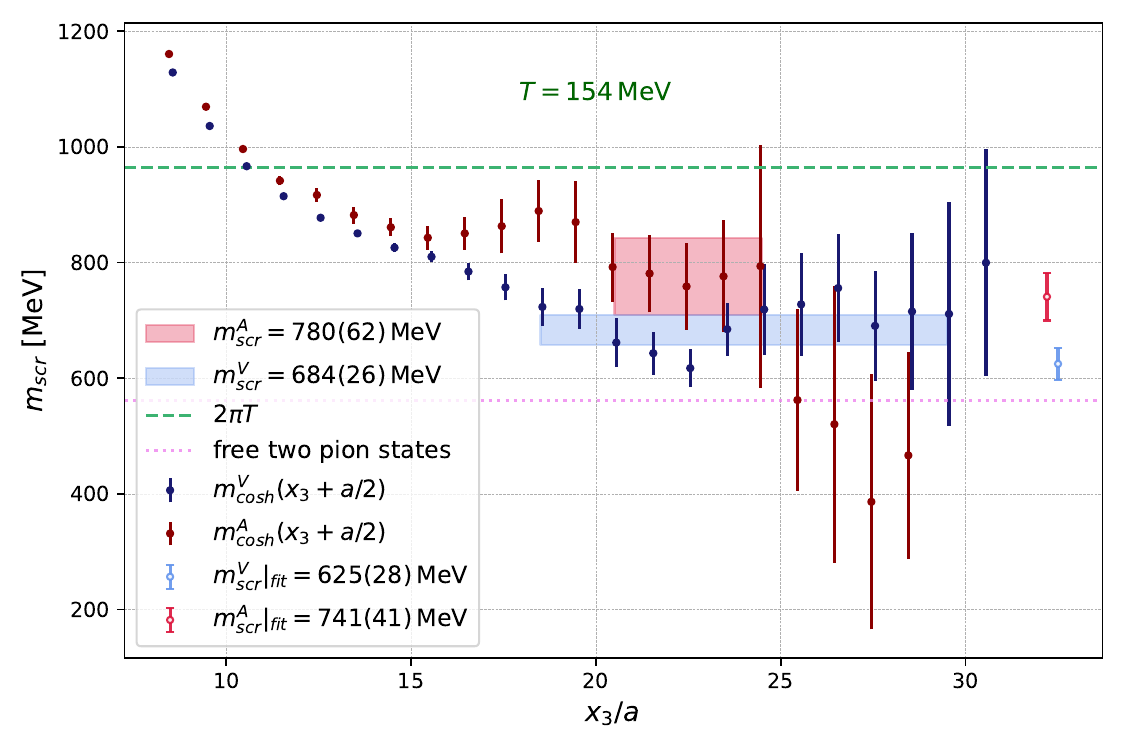}
	\includegraphics[scale=0.52]{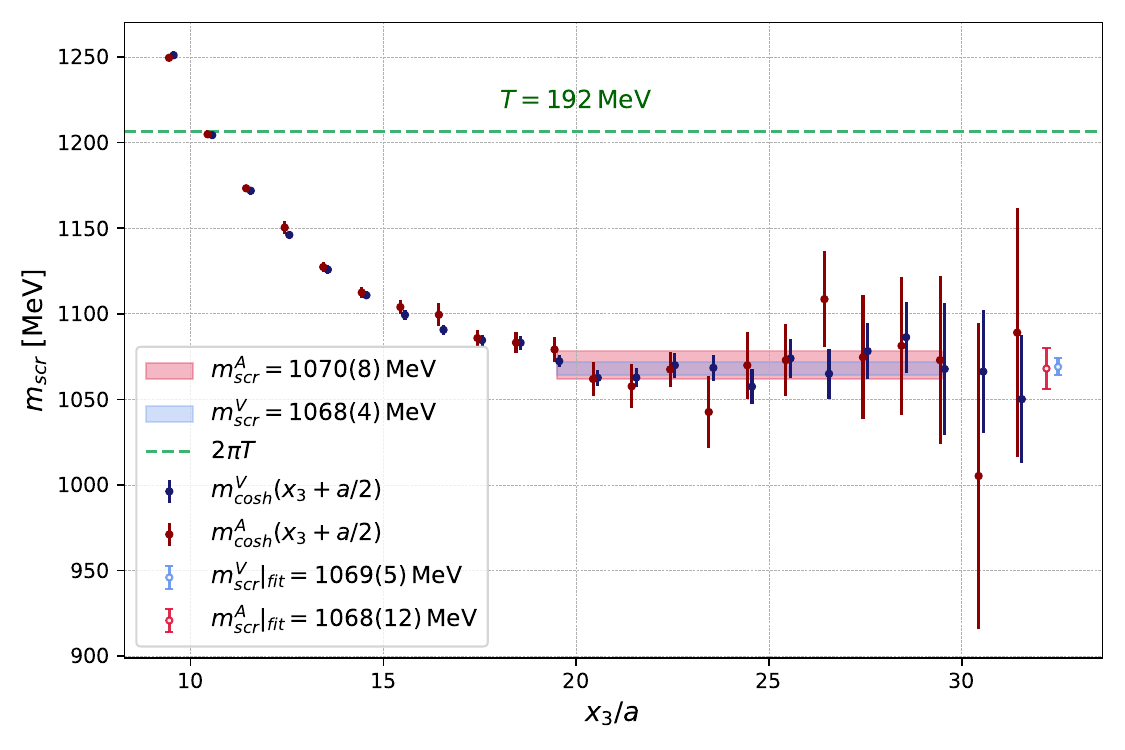}
\caption{
{\bfseries Top panel}: Screening vector and axial-vector effective masses  at $T=128\,$MeV. {\bfseries Middle panel}: Screening vector and axial-vector effective masses  at $T=154\,$MeV. {\bfseries Bottom panel}:  Screening vector and axial-vector effective masses  at $T=192\,$MeV. The bands represent correlated fits to the plateau, while the green dashed line represents the $2\pi T$ limit. Furthermore, the value for the screening masses obtained from a direct fit of the axial-vector respectively vector correlator are also included and represented by a single light bar. The violet dotted line represents the free two pion states (for $T=128\,$MeV and $T=154\,$MeV).
}
\label{fig:eff_masses_V_and_A_screening}
\end{figure}

In this section we present our results for the screening masses in the pseudoscalar, vector and axial-vector channels and compare them to those of the HotQCD collaboration\,\cite{Bazavov:2019www}. The latter have been obtained using the highly improved staggered quark (HISQ) action and a continuum extrapolation has been performed from a range of lattice spacings corresponding to $N_{\tau} = 6-16$. In contrast to Ref.\,\cite{Bazavov:2019www} we only have results for mesons of the type $u \Bar{d}$. A direct comparison of the numerical values in the three channels is summarized in Table\,\ref{tab:comparison_hotqcd}.

At $T=128\,$MeV, we obtain a pseudoscalar screening mass of $145(2)\,$MeV, which is increased by $\approx 12\,\%$ compared to the pion mass on the corresponding vaccuum ensemble [see Eq.\,(\ref{eq:mpimKT0})] and also compared to the HotQCD result of $129(5)\,$MeV. The latter has been obtained at a slightly higher temperature $T=132\,$MeV. The results of the HotQCD collaboration for the vector and axial vector screening masses at the aforementioned temperature read $0.7(2)\,$GeV and $1.0(2)\,$GeV, respectively and are compatible within errors with our results of $614(19)\,$MeV and $832(35)\,$MeV, respectively. The lightest vector and axial vector mesons with $u\Bar{d}$ valence quarks are the $\rho$ meson with a physical mass of $775\,$MeV, respectively the $a_1$ meson with a physical mass of $1.23\,$GeV. It is notable that both masses are significantly larger $[m_{\rho}/m_{scr}^{V}\approx 1.26, m_{a_1}/m_{scr}^{A}\approx 1.48]$ than the screening masses extracted on our $N_{\tau}=24$ ensemble at $T=128\,$MeV in the hadronic phase. However, since the vector and axial-vector screening spectra start with a continuum, not with an isolated pole, we also show the free two pion states with a violet dotted line in Fig.\,\ref{fig:eff_masses_V_and_A_screening}.

\begin{table}[bt]
\caption{Comparison of the screening masses in the $u\Bar{d}$ channel extracted at temperatures $T\in \{128,154,192\}\,$MeV to the continuum-extrapolated results obtained by the HotQCD collaboration\,[see Ref.\,\cite{Bazavov:2019www}, Table\, X].} 
\begin{tabular}{c@{~~~}|c@{~~~}c@{~~~}|c@{~~~}c@{~~~}|c@{~~~}c@{~~~}c}
\hline
\hline
& \multicolumn{2}{c}{$m_{scr}^{PS}\,$[MeV] } & \multicolumn{2}{c}{$m_{scr}^{V}\,$[MeV] } & \multicolumn{2}{c}{$m_{scr}^{A}\,$[MeV] } \\
Temperature\,[MeV]    & HotQCD        & this work       & HotQCD        & this work        &HotQCD & this work   \\
\hline
128      &            & 145(2)        &                 & 614(19)       &                  & 832(35)     \\ 
132      & 129(5)     &               & 700(200)        &               & 1000(200)        &             \\ 
\hline
152      & 187(2)     &               & 730(50)         &               & 850(40)          &             \\ 
154      &            & 196(4)        &                 & 625(28)       &                  & 741(41)     \\ 
156      & 202(3)     &               & 750(60)         &               & 830(60)          &             \\ 
\hline
192      & 540(10)    & 555(8)        & 1020(40)        & 1069(5)       & 1040(30)         & 1068(12)    \\

\hline
\hline
\end{tabular}
\label{tab:comparison_hotqcd}
\end{table}

Next, we  compare our results of the screening masses obtained on our $N_\tau=20$ ensemble near the pseudocritical temperature to the continuum extrapolated HotQCD results. As far as the pseudoscalar screening mass is concerned our result $m_{scr}^{P}(T=154\,\text{MeV})=196(4)\,$MeV is in very good agreement with the HotQCD results $187(2)\,$MeV and $202(3)\,$MeV at the temperatures $T=152\,$MeV and $T=156\,$MeV, respectively. However, our vector screening mass $m_{scr}^{V}(T=154\,\text{MeV})=625(28)\,$MeV differs by $\approx 2\sigma$ compared to the HotQCD results 730(50)\,MeV and 750(60)\,MeV at the temperatures $T=152\,$MeV and $T=156\,$MeV, respectively. Furthermore, it is compatible within errors with our result obtained in the hadronic phase. Our result for the axial-vector screening mass $m_{scr}^{A}(T=154\,\text{MeV})=741(41)\,$MeV differs again by $\approx 2\sigma$ compared to the HotQCD result and is reduced by a factor of $\approx 1.12$ compared to our result at $T=128\,$MeV. Thus, the ratio of axial vector screening mass and vector screening mass decreases from $\approx 1.36$ at $T=128\,$MeV to $\approx 1.19$ at $T=154\,$MeV.

Finally, we confront our screening masses extracted from a $N_{\tau=16}$ ensemble in the chirally-symmetric phase with the results of the HotQCD collaboration. The pseudoscalar screening mass  $m_{scr}^{P}(T=192\,\text{MeV})=555(8)\,$MeV is compatible within errors with the HotQCD result $540(10)\,$MeV obtained at the same temperature. For the vector and axial-vector screening masses we obtain $1069(5)\,$MeV and $1068(12)\,$MeV, respectively. Both screening masses are compatible within errors with the HotQCD collaboration results $1020(40)\,$MeV and $1040(30)\,$MeV and their ratio is precisely one. Inspecting effective mass plots displayed in the lower panel of Fig.\,\ref{fig:eff_masses_V_and_A_screening}, we notice that the degeneracy of the vector and axial-vector screening mass at $T=192\,$MeV is not only evident in the plateau region, but can also be seen (within errors) individually over a long range of source-sink seperations. However, at this temperature, the vector and axial vector screening masses are still $\approx 11\%$ below the $2\pi T$ limit of screening masses. This high-temperature limit would correspond to the propagation of two quark quasiparticles in the thermal medium, each carrying a `mass' of $\pi T$ corresponding to the first Matsubara mode.

\section{Isovector quark number susceptibility}
\label{sec:suscep}

The quark number susceptibility (QNS) measures the response of the net quark number density
to an infinitesimal change in the quark chemical potential.
On the lattice, we determine the QNS via
\begin{align}
\label{QNS}
 \delta^{ab}\chi_q(x_0, T) = 2 Z_V^2(g_0^2)\, \beta\, \int \mathrm{d}^3x\, \langle V_{0}^a(x_0,\bold{x})\,V_{0}^b(0,\bold{0}) \rangle\,, \quad \quad x_0 \neq 0\,.
\end{align}
For the QNS, no additive improvement of the vector current is needed.

\begin{figure}[tp]
\includegraphics[scale=0.65]{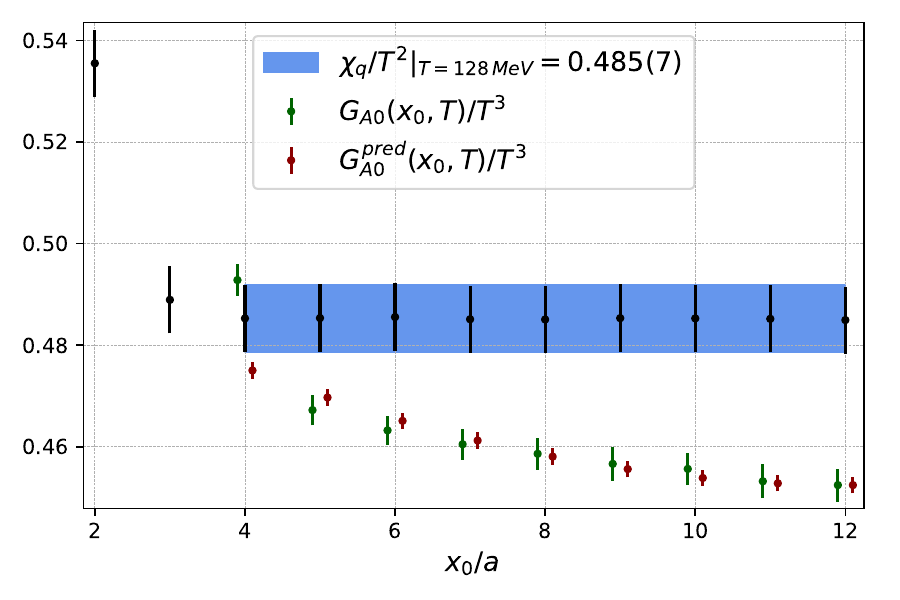}
 \includegraphics[scale=0.65]{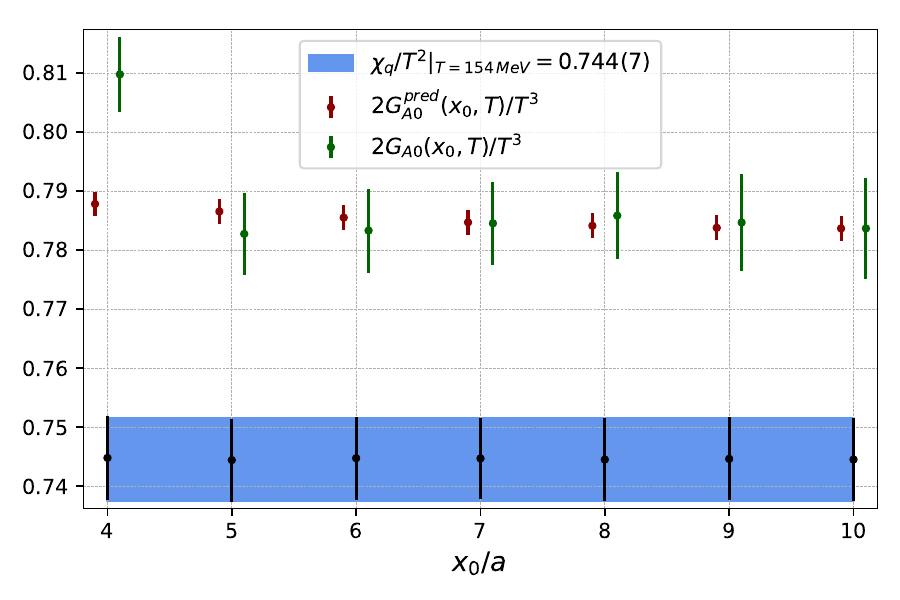}
 \includegraphics[scale=0.65]{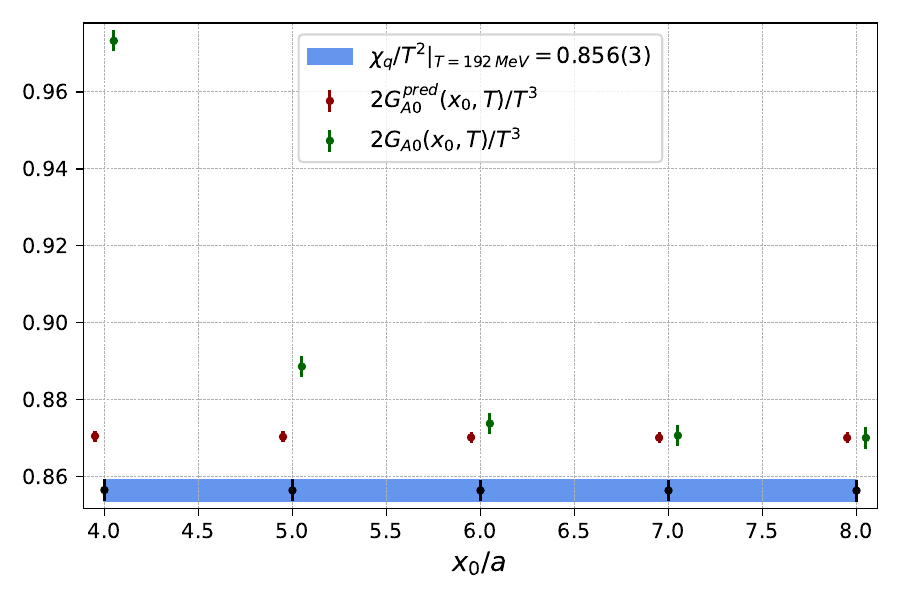}
 \caption{Quark number susceptibility (QNS) extracted from the local vector current correlator, Eq.\,(\ref{QNS}). For comparison also $G_{A_0}(x_0,T)$ and its prediction in terms of screening quantities is shown. The top panel shows the QNS for a temperature $T=128\,$MeV, the middle panel for a temperature $T=154\,$MeV and the bottom panel for a temperature $T=192\,$MeV. The correlators in these plots follow the normalization
   given in Eq.\ (\ref{QNS}), with the exception of $G_{A_0}(x_0,T=128\,\text{MeV})$, as indicated by the legend, for better visibility.
	\label{fig:QNS}}
\end{figure}
The results are shown in Fig.\,\ref{fig:QNS} in temperature units. Note that computing the isospin susceptibility does not involve any contributions from disconnected diagrams. In Ref.\,\cite{Borsanyi:2011sw} (see Table\,I) the QNS was determined as a function of the temperature using 2+1 dynamical staggered
quark flavors and, additionally, a continuum extrapolation was done. Their result is $\chi_q(T)/T^2 = 0.432(92)$ and 
$\chi_q(T)/T^2 = 0.481(87)$ for the temperatures $T = 125$\,MeV and $T = 130$\,MeV, respectively (see Table\,I of Ref.\,\cite{Borsanyi:2011sw}).
Our result,
\be\la{eq:chiNt24}
\chi_q(T)/T^2 = 0.485(7)\,, \qquad (T=128\,{\rm MeV})
\ee
 is compatible with both of these results. At the crossover, our estimate of the QNS
\be
\chi_q(T)/T^2 = 0.744(7)\,, \qquad (T=154\,{\rm MeV})
\ee
overshoots theirs, $\chi_q(T)/T^2 = 0.669(47)$ at a temperature $T=155\,$MeV by $\approx 1.6\sigma$. Also in the high temperature phase  our estimate
\be 
\chi_q(T)/T^2 = 0.856(3)\,, \qquad	(T=192\,{\rm MeV})
\ee
overshoots their result of $0.810(17)$ by $\approx 2.7\sigma$. Note however that our results are not continuum-extrapolated. For comparison also the axial charge correlator $G_{A_0}(x_0,T)$ is shown in the plots. Already at $T=128\,$MeV the correlator is quite flat, but a factor of $\approx 2$ bigger in magnitude compared to the isospin charge correlator. Around the crossover at $T=154\,$MeV the axial charge correlator is only $6\%$ bigger than the isospin charge correlator. Finally, at $T=192\,$MeV the two correlators are near-degenerate. 
Next, we are going to compare our lattice estimate for the QNS with the hadron resonance gas (HRG) model and also test an alternative
`quasiparticle gas' model employing our modified dispersion relation for the pion quasiparticle.

\subsection{Comparison with the hadron resonance gas model}
\label{subsec:HRG}

The HRG model\,\cite{Hagedorn:1984hz, Karsch:2003vd} describes the thermodynamic properties and the quark number
susceptibilities of the low-temperature phase rather well. It assumes that the thermodynamic properties of the system are given by the sum of the partial contributions of non-interacting hadron species, i.e.
\begin{align}
    \label{eq:HRG partition function}
    \text{ln}[Z(T,V)] = -\frac{V}{2\pi^2}\sum_i\int_0^{\infty}\,\mathrm{d}p\, p^2\,\text{ln}[1-\eta_i\,e^{-\sqrt{m_i^2+\bold{p}^2}/T}]\,,
\end{align}
where $\eta_i = \pm 1$ takes into account bosons and fermions respectively. The sum extends over all resonances up to a mass of 2.0\,GeV, since for most of them the width is not large compared to the temperature.

The QNS can be obtained as the sum\,\cite{Brandt:2015aqk}
\begin{equation}
    \chi_q(T) = (\chi_q)_{\text{mesons}} + (\chi_q)_{\text{baryons}}\,,
\end{equation}
where
\begin{align}
    \label{eq:QNS HRG mesons}
    \frac{(\chi_q)_{\text{mesons}}}{T^2} &= \frac{2\beta^3}{3}\sum_{\text{multiplets}}(2J+1)I(I+1)(2I+1)\int\frac{\mathrm{d^3}\bold{p}}{(2\pi)^3}\,f_{\bold{p}}^B(1+f_{\bold{p}}^B)\,,\\
    \label{eq:QNS HRG baryons}
    \frac{(\chi_q)_{\text{baryons}}}{T^2} &= \frac{2\beta^3}{3}\sum_{\text{multiplets}}(2J+1)I(I+1)(2I+1)\int\frac{\mathrm{d^3}\bold{p}}{(2\pi)^3}\,f_{\bold{p}}^F(1-f_{\bold{p}}^F)\,,
\end{align}
and $f_{\bold{p}}^{B/F} = 1/[e^{\beta\omega_{\bold{p}}}\mp 1]$ are the Bose-Einstein and Fermi-Dirac distributions, respectively.
The sums are carried out over all multiplets of spin $J$ and isospin $I$ that are not identical. Especially particles and antiparticles have to be considered separately. This results in an additional factor of two in the baryon case and for mesons with strange quark constituents.

\begin{figure}[tp]
\begin{tikzpicture}
 \pie[color = {
        cyan!=40!, 
        green!45!, 
        lime!40, 
        yellow!50,
        lightgray!50},
        text = legend 
        ]
        {77.9/pion,
    8.1/vector- and pseudoscalar meson octet without $\rho$,
    7.3/ $\rho$ vector meson,
    3.8/baryon octet and decuplet,
    2.9/heavier meson and baryon resonances up to $2.0\,\text{GeV}$
    }
 \end{tikzpicture}
 \caption{Relative composition of the total quark number susceptibility predicted by the hadron resonance gas model in the hadronic phase.
	\label{fig:pie diagram}}
\end{figure}
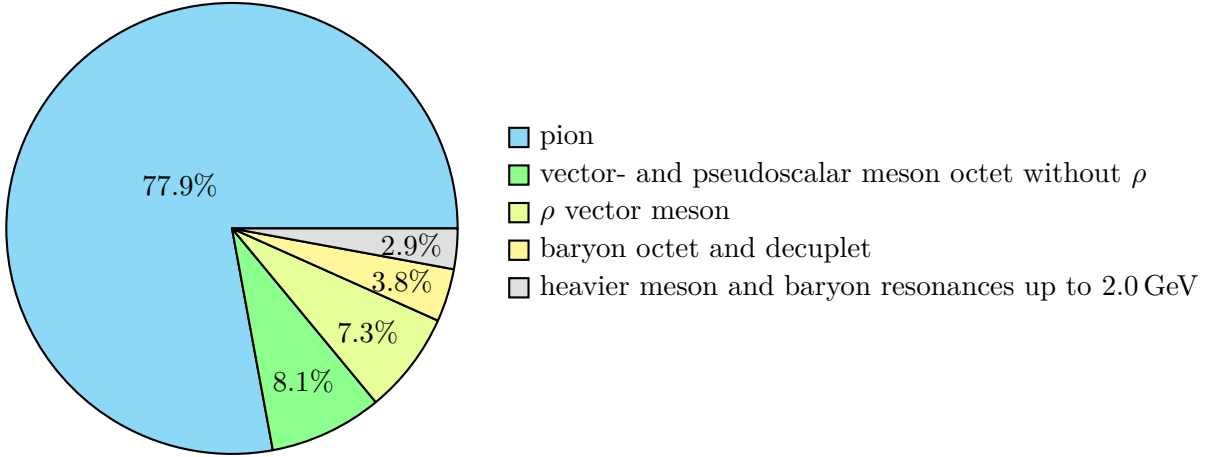

Employing Eqs.\,(\ref{eq:QNS HRG mesons}-\ref{eq:QNS HRG baryons}) within the HRG model and summing all resonances up to a mass of 2\,GeV, we obtain $\chi_q(T)/T^2 = 0.486$ which agrees perfectly with our lattice estimate, Eq.\,(\ref{eq:chiNt24}). The relative composition of the total QNS is summarized in Fig.\,\ref{fig:pie diagram} 
It is questionable whether resonances like the $K_0^*(700)$, whose full Breit-Wigner width $(478\pm50\,\text{MeV})$ is higher than the temperature considered should be taken into account.

An alternative to the HRG model is to only include the
pion contribution, however taking into account the modified dispersion relation (\ref{eq:dispersion}) at low momenta,
\begin{align}
    \label{eq:QNS with modified dispersion relation}
    \frac{\chi_q}{T^2} = 4\beta^3\int_{\abs{\bold{p}}<\Lambda_p}\frac{\mathrm{d^3}\bold{p}}{(2\pi)^3}\,f_{\bold{p}}^B(\omega_{\bold{p}})(1+f_{\bold{p}}^B(\omega_{\bold{p}}))\,, 
\end{align}
At this point we only integrate up to the momentum cut off $\Lambda_p = 400\,\text{MeV}$ since it is not clear if the thermal width of the pion is still negligible for $\abs{\bold{p}}>\Lambda_p$ and, as a consequence, including contributions from higher momenta may not be justified. In Ref.\,\cite{Brandt:2015sxa} the authors have performed fits to the axial-charge density correlator at nonvanishing momenta. They then have compared the residue obtained from the fit parameters to the chiral prediction
\begin{equation}
    \label{residuum_with_add_param}
    \text{Res}(\omega_{\bold{p}}) = f_{\pi}^2(m_{\pi}^2+\bold{p}^2)(1+b(\bold{p}))\,.
\end{equation}
For $\abs{\bold{p}}\approx 400\,$MeV they obtained a small $b=-0.08(3)$ indicating that the chiral effective theory description still works. To estimate our error, we thus quote the results for 
$\Lambda_p \in \{400,500\}\,\text{MeV}$. Making use of Eq.\,(\ref{eq:QNS with modified dispersion relation}) with $u_m=0.807$, a screening pion mass $m_{\pi} = 145\, \text{MeV}$ and $\Lambda_p=400\,$MeV we obtain $\chi_q(T)/T^2 = 0.402$ which is $\approx 17\%$ below the lattice estimate. On the other hand, for $\Lambda_p=500\,$MeV we obtain $\chi_q(T)/T^2 = 0.512$ which is $\approx 5\%$ above the lattice estimate.
Note that in this model the sum over the resonances is absent. The contributions of the other hadrons are taken into account indirectly via the modified dispersion relation, since the collisions of the pions among themselves and with other hadrons give rise to the modified pion dispersion relation.

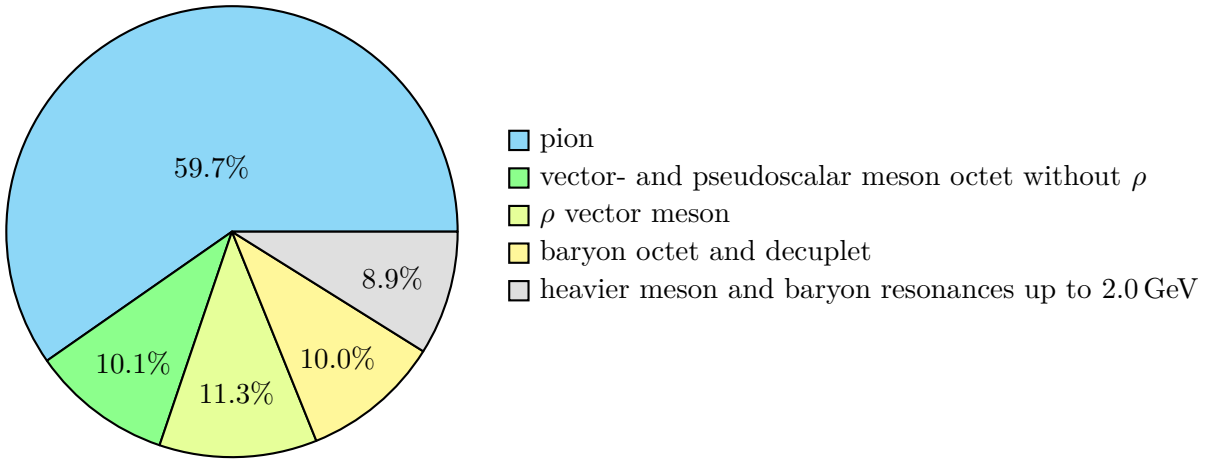
\begin{figure}[tp]
\begin{tikzpicture}
 \pie[color = {
        cyan!=40!, 
        green!45!, 
        lime!40, 
        yellow!50,
        lightgray!50},
        text = legend 
        ]
        {59.7/pion,
    10.1/vector- and pseudoscalar meson octet without $\rho$,
    11.3/ $\rho$ vector meson,
    10.0/baryon octet and decuplet,
    8.9/heavier meson and baryon resonances up to $2.0\,\text{GeV}$
    }
 \end{tikzpicture}
 \caption{Relative composition of the total quark number susceptibility predicted by the hadron resonance gas model at the chiral crossover.
	\label{fig:pie diagram_Nt20}}
\end{figure}

Employing the HRG for the ensemble at the chiral crossover, we get $\chi_q(T)/T^2 = 0.701$, a result that undershoots the lattice estimate by $5.8\%$. The relative composition of the total QNS predicted by the HRG model in this case is given in Fig.\,\ref{fig:pie diagram_Nt20}. At both temperatures the pion clearly dominates the contribution to the total QNS. 

\section{Order parameters for chiral symmetry restoration}
\label{sec:order_param}

In this section, several order parameters for chiral symmetry restoration are investigated. Based on the pion screening quantities $m_{\pi}^2$ and $f_{\pi}^2$ presented in Sec.\,\ref{sec:results}, we first evaluate an ‘effective chiral condensate’ based on the Gell-Mann--Oakes--Renner relation.
Additionally, we explore two Euclidean-time dependent thermal correlation functions that are order parameters for chiral symmetry
and compare them to their zero-temperature counterparts.
We begin with the $(PA_0)$-correlator, which contains the pion pole that we have studied in Sec.\,\ref{sec:results}.
We then consider the difference of the (isovector) vector and axial-vector correlators.
In the QCD vacuum, the corresponding spectral functions are measured experimentally in $\tau$ decays~\cite{Davier:2005xq}.
They become degenerate in the chirally restored phase of QCD. Their temperature dependence in the chirally broken phase
has been studied extensively in the framework of hadronic models supplemented by sum rules~\cite{Rapp:1999ej,Kapusta:1993hq,Hohler:2013eba}.

\subsection{The Gell-Mann--Oakes--Renner relation}
\label{sec:GOR}

Following Ref.\,\cite{Brandt:2014qqa}, we introduce an ‘effective chiral condensate’ based on the Gell-Mann--Oakes--Renner (GOR) relation,
\begin{equation}
    C_{\rm GOR}^{\rm GF}
     \equiv -\frac{f_{\pi}^2m_{\pi}^2}{m_{\text{q}}}\,. 
\end{equation}
In the chiral limit, $C_{\rm GOR}^{\rm GF}\to \<\bar\psi\psi\>$. Additionally, since above $T_c$, $m_{\pi} \sim T$ and $f_{\pi} \sim m_{\text{q}}$,
$C_{\rm GOR}^{\rm GF}$ is of $\mathcal{O}(m_{\text{q}}\,T^2)$. Thus, it serves as an order parameter for chiral symmetry. Using $m_{\text{q}} = m_{\text{PCAC}}$ and the screening quantities of Table\,\ref{tab:results} we obtain
\begin{align}
    \left|C_{\rm GOR}^{\rm GF}\right|^{1/3}_{T=128\,\text{MeV}} &= 291(3)\, \text{MeV}\,,\nonumber \\ 
    \left|C_{\rm GOR}^{\rm GF}\right|^{1/3}_{T=154\,\text{MeV}} &= 180(8)\, \text{MeV}\,,\\ \nonumber
    \left|C_{\rm GOR}^{\rm GF}\right|^{1/3}_{T=192\,\text{MeV}} &= 133(6)\, \text{MeV}\,. 
\end{align}
The value of the chiral condensate has been extracted in the gradient flow scheme just like the PCAC mass (see Sec.\,\ref{sec:lat_impl}).
Comparing with the chiral condensate on the corresponding zero-temperature ensemble \cite{Ce:2022kxy}, we get
\begin{align}
    \left[\frac{C_{T\approx 128\,\text{MeV}}}{C_{T\approx 0\,\text{MeV}}}\right]_{\text{GOR}} &\equiv \frac{(f_{\pi}^2m_{\pi}^2)_{T\approx 128\,\text{MeV}}}{(f_{\pi}^2m_{\pi}^2)_{T\approx 0\,\text{MeV}}} = 0.84(3)\,, \nonumber\\
    \label{eq:reduction_quark_cond}
    \left[\frac{C_{T\approx 154\,\text{MeV}}}{C_{T\approx 0\,\text{MeV}}}\right]_{\text{GOR}} &\equiv \frac{(f_{\pi}^2m_{\pi}^2)_{T\approx 154\,\text{MeV}}}{(f_{\pi}^2m_{\pi}^2)_{T\approx 0\,\text{MeV}}} = 0.21(3)\,, \\ \nonumber
    \left[\frac{C_{T\approx 192\,\text{MeV}}}{C_{T\approx 0\,\text{MeV}}}\right]_{\text{GOR}} &\equiv \frac{(f_{\pi}^2m_{\pi}^2)_{T\approx 192\,\text{MeV}}}{(f_{\pi}^2m_{\pi}^2)_{T\approx 0\,\text{MeV}}} = 0.08(1)\,. 
\end{align}
which corresponds to a reduction by 16\% in the hadronic phase. Notably, during the temperature increase from $T=128\,$MeV to $T=154\,$MeV the effective chiral condensate (normalized to the zero-temperature equivalent) decreases by a factor of four down to $0.21(3)$. Finally,  at a temperature of $T=192\,$MeV the effective chiral condensate reaches a value of $\approx 8 \%$ of its vacuum value. The reduction in the low-temperature phase is compatible within the scope of the error with a three-loop result of Gerber and Leutwyler (see Ref.\,\cite{Gerber:1988tt}, Fig.\,5). 

\subsection{The \texorpdfstring{$(PA_0)$}{(PA0)}-correlator}

In Ref.\,\cite{Brandt:2014qqa} it was shown that the  temporal $(PA_0$)-correlator can be predicted exactly in the chiral limit,
\begin{equation}
  \label{eq:PA0_corr}
        G_{PA_0}(x_0,T) = \frac{\langle\bar{\psi}\psi\rangle}{2\beta}    \left(x_0-\frac{\beta}{2}\right)\,.
\end{equation}
As can be seen from Eq.\,(\ref{eq:PA0_corr}) the $(PA_0)$-correlator is antisymmetric around $\beta/2$. Consequently, we set the point $x_0=\beta/2$ to zero. In our analysis we have averaged over the $(PA_0)$- and $(A_0P)$-correlator. The latter can be obtained interchanging source and sink.\footnote{This is only true for our vacuum correlators, which are obtained from point-sources. For our finite temperature correlators obtained from stochastic wall-sources these are equivalent, as we use a symmetric source-sink setup.} We find that the estimator based on the last expression of Eq.\,(\ref{eq:A0P}) has the better signal-to-noise ratio.
Since this correlator is proportional to the chiral condensate $\langle\bar{\psi}\psi\rangle$, it can serve as an order parameter for chiral symmetry restoration as well. Looking at the ratio of the thermal over the reconstructed correlator obtained via Eq.\,(\ref{eq:reconstr}),
we find the reductions of the $(PA_0)$ correlator to be in satisfactory agreement with the reductions we have obtained for the chiral condensate using the Gell-Mann–Oakes–Renner relation (see Eq.\,(\ref{eq:reduction_quark_cond})) for all three temperatures, namely
\be
\lim_{x_0\to\beta/2}\frac{G_{PA_0}(x_0,T)}{G_{PA_0}^{\rm rec}(x_0,T)} = \left\{ \begin{array}{ll}
 0.817(21)\,,   & N_{\tau}=24
\\ 0.276(23)\,, & N_{\tau}=20
\\ 0.088(5)\,,  & N_{\tau}=16.
  \end{array}\right.
\ee
The corresponding plots can be found in Fig.\,\ref{fig:ratio_A0P} in Appendix~\ref{app:numerical_values}.


\begin{figure}[tp]
	\includegraphics[scale=0.5]{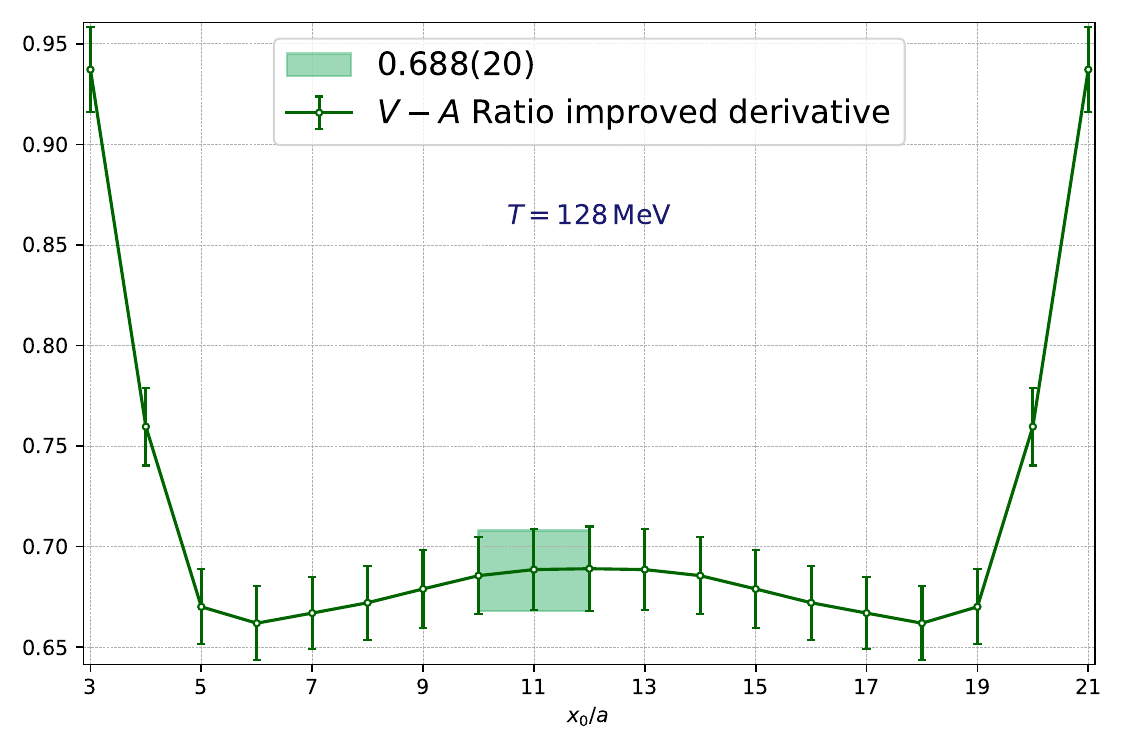}
	\includegraphics[scale=0.5]{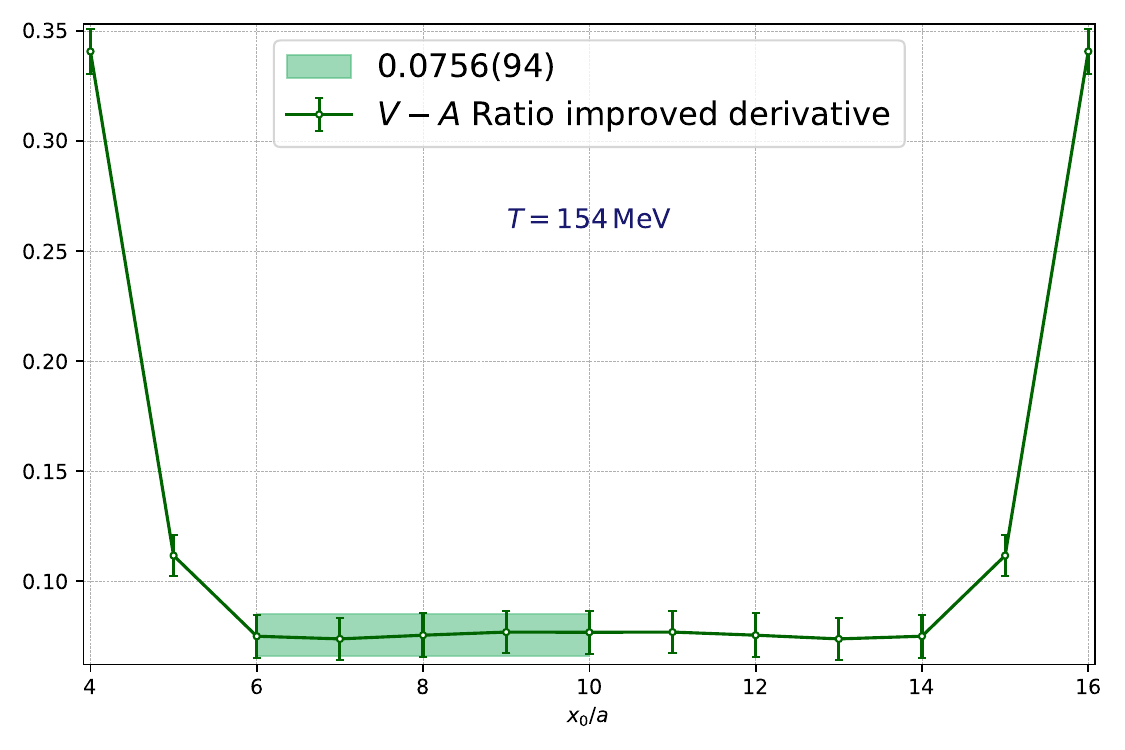}
	\includegraphics[scale=0.5]{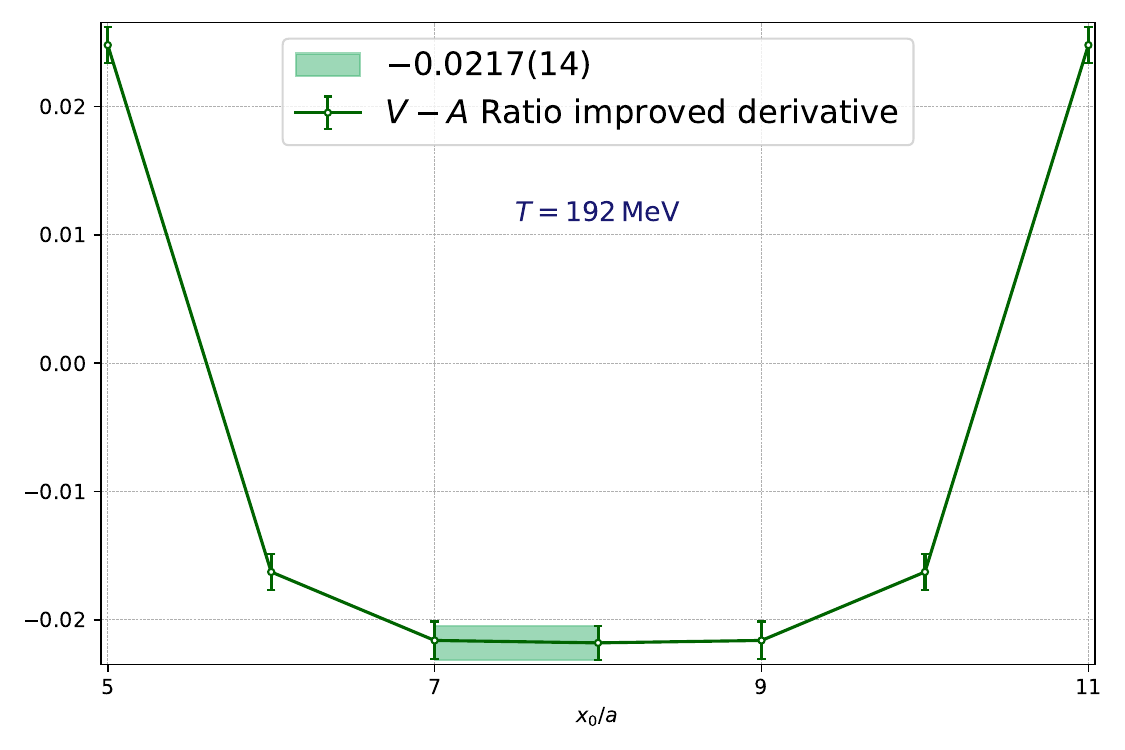}
\caption{
{\bfseries Top panel}: Ratio of the difference ‘$V-A$’ and the difference of the reconstructed correlator ‘$(V-A)_{\text{rec}}$’ at $T=128\,$MeV. {\bfseries Middle panel}: Same quantity at $T=154\,$MeV. {\bfseries Bottom panel}:  Same quantity at $T=192\,$MeV. The green band shows the result of a correlated fit to the plateau.
}
\label{fig:V-A}
\end{figure}

\begin{figure}[tp]
  \includegraphics[width=0.495\textwidth]{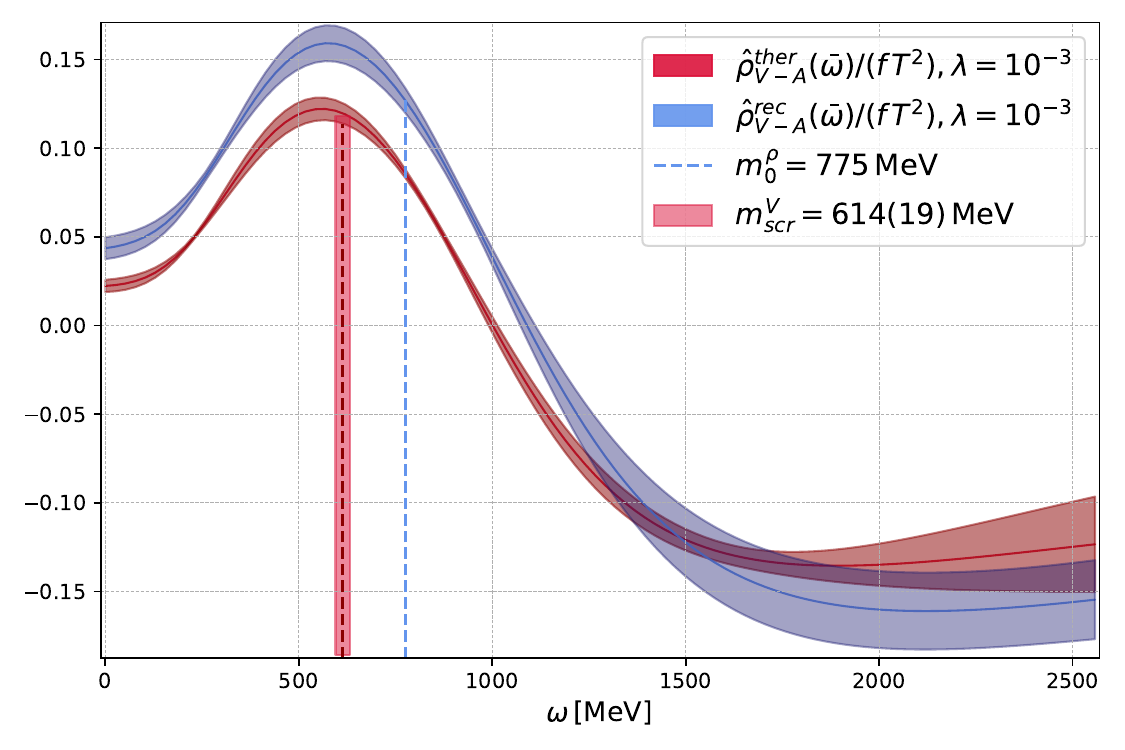}
  	\includegraphics[width=0.495\textwidth]{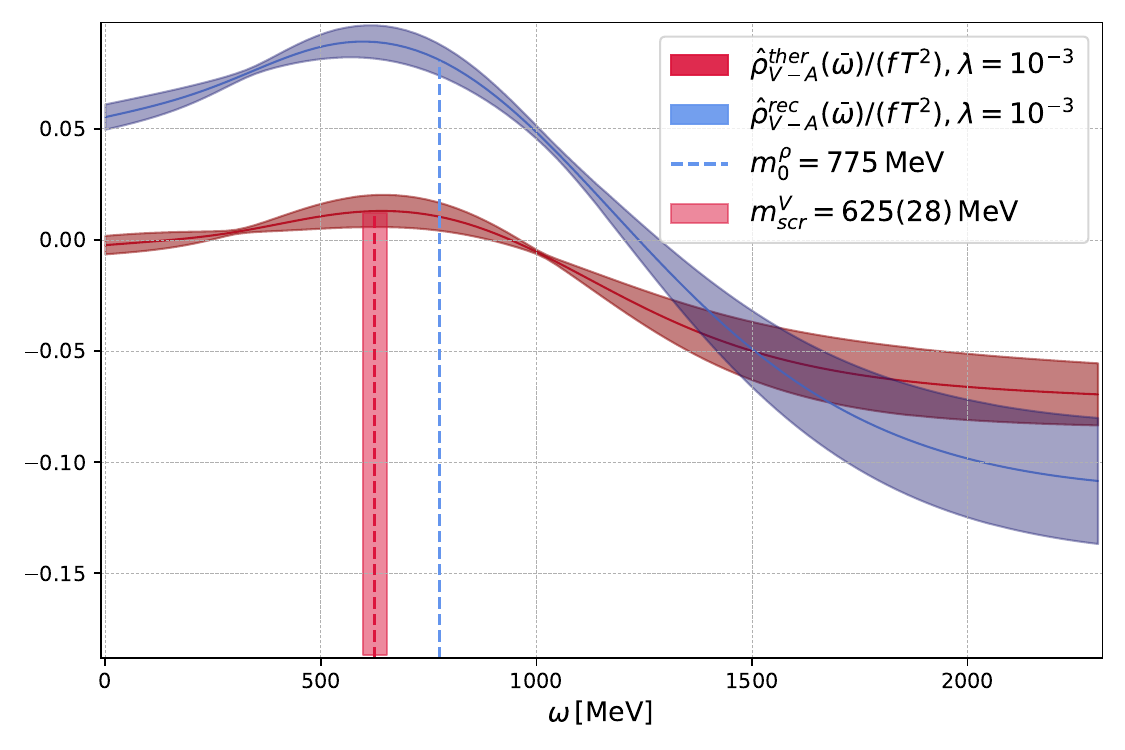}
\caption{
The Backus-Gilbert spectral function $\hat{\rho}_{V-A}^{\text{rec}}(\omega)/(f\,T^2)$ obtained from the reconstructed $V-A$ correlator is displayed as a blue band, at a temperature of $T=128\,$MeV and $154\,$MeV respectively for the left and right panel. The corresponding estimator of the thermal spectral function $\hat{\rho}_{V-A}^{\text{ther}}(\omega)/(f\,T^2)$  is shown as a red band. For orientation, the blue vertical line corresponds to the physical mass of the $\rho$ meson, while the vertical red line corresponds to the vector screening mass at the respective temperature [see Table\,\ref{tab:comparison_hotqcd}]. The underlying temporal vector correlators are $\mathcal{O}(a)$ improved.
}
\label{fig:E250Nt24_V-A_spectral_function}
\end{figure}

\subsection{Dey-Eletsky-Ioffe mixing theorem at finite quark mass}
\label{sec:Ioffe}


In Refs.\,\cite{Dey:1990ba,Eletsky:1992ay,Eletsky:1994rp} it was demonstrated, using PCAC current algebra, that near the chiral limit the finite-temperature vector and axial-vector correlators can be described with the help of their vacuum counterparts. In terms of the corresponding spectral functions, this statement reads
\begin{align}
    \label{eq:diff_spectral}
    \rho_V(\omega,\bold{p},T)-\rho_A(\omega,\bold{p},T)=(1-2\epsilon(T))\left[\rho_V(\omega,\bold{p},T=0)-\rho_A(\omega,\bold{p},T=0)\right]\,,
\end{align}
while the sum $(\rho_V+\rho_A)$ receives no thermal correction at this order. The coefficient
\begin{equation}
    \label{eq:mixing_param_epsilon}
    \epsilon(T) \equiv \frac{T^2}{6(f_{\pi}^0)^2}
\end{equation}
is a temperature dependent expansion parameter in powers of the pion density. 
Thus, the difference in Eq.\,(\ref{eq:diff_spectral})  serves as an order parameter for chiral symmetry restoration.

Inspired by the Dey-Eletsky-Ioffe mixing theorem, in the following we consider the difference
$(G_V(x_0,\bold{p}=0,T)-G_A(x_0,\bold{p}=0,T))$ at vanishing spatial momentum.
 Using the corresponding quasi zero-temperature E250 ensemble of size $192\times96^3$,
we  calculate the ‘reconstructed’ correlator $G_V^{\text{rec}}-G_A^{\text{rec}}$ for the difference. This correlator can be interpreted as the thermal Euclidean correlator that would be realized if the spectral function was unaffected by thermal effects, during the transition from $T\approx 0$ to $T\in\{128,154,192\}\,$MeV. Following a method first proposed in Ref.\,\cite{Meyer:2010}, we obtain the reconstructed correlator as
\begin{align}
    \label{eq:reconstr}
    G_J^{\text{rec}}(x_0,T,\bold{p}) = \sum_{m\in \mathbb{Z}}\,G_J(\abs{x_0+m\beta},0,\bold{p}) \ \ \ \ \ \ \ (J = \{V,A,\dots\}). 
\end{align}
It is based on the identity of the kernel function 
\begin{align}
    \frac{\text{cosh}(\omega(\beta/2-x_0))}{\text{sinh}(\omega \beta/2)} = \sum_{m \in \mathbb{Z}}\, e^{-\omega\abs{x_0+m\beta}}\, .
\end{align}
The difference ‘$V-A$’ for the thermal ensembles are shown in the l.h.s. of Fig.\,\ref{fig:thermal_vs_rec} in Appendix \ref{app:numerical_values}, while the r.h.s. displays the corresponding reconstructed correlators obtained from the $T\approx 0$ ensemble using Eq.\,(\ref{eq:reconstr}). In both cases renormalization factors are included following Eqs.\,{(\ref{eq:renorm_V})-(\ref{eq:renorm_A})}. Furthermore, the vector-current has been $\mathcal{O}(a)$-improved\footnote{The axial-vector current is automatically $\mathcal{O}(a)-$improved since the spatially integrated improvement term vanishes in this case. On the other hand, for the vector current the term $\propto T_{i0}$ contributes in the improvement process [see Eqs.\,(\ref{eq: imp_axial_corr})-(\ref{eq:imp_vec})].} using the improved derivative defined in Eq.\,(\ref{eq:improved_derivative}). The top two panels show the relevant correlator difference ‘$V-A$’ at $T=128\,$MeV, the middle panels ‘$V-A$’ at $T=154\,$MeV and the bottom panels ‘$V-A$’ at $T=192\,$MeV. It is notable that we are completely dominated by the error on the corresponding vacuum ensemble. 
The ratios of the thermal and reconstructed correlators are shown in Fig.\,\ref{fig:V-A}. It is immediately apparent that the ratio differs significantly from one at all three temperatures, even though one expects the ratio to tend to unity for $x_0\to0$ for $m_q\neq 0$. Remembering the physical interpretation of the reconstructed correlator given above Eq.\,(\ref{eq:reconstr}), we can conclude that also the corresponding spectral function must have changed during the transition from $T \approx 0$ to $T\in\{128,154,192\}\,$MeV.   

As Eq.\,(\ref{eq:diff_spectral}) is obtained in the chiral limit, one would not necessarily expect to see a drop to a constant value in the ratio even for finite quark masses. However, as can be seen in Fig.\,\ref{fig:V-A}, we observe a flat behavior around the midpoint at $T=128\,$MeV and $T=154\,$MeV. Especially in the $N_{\tau}=20$ ensemble the ratio is almost constant in the range $6\leq x_0/a \leq 10$. In the hadronic phase, we read off a reduction to a factor $0.69(2)$ pointing to chiral symmetry restoration being already at an advanced stage in the hadronic phase [see top panel of Fig.\,(\ref{fig:V-A})]. During the transition from $T=128\,$MeV to $T=154\,$MeV the ratio in the ‘$V-A$’-channel drops drastically by a factor of $\approx 9$ down to a reduction factor of $\approx 0.08(1)$.
This is confirmed by inspecting the smoothed $(V-A)$ spectral function obtained
from the Backus-Gilbert method in Fig.\,\ref{fig:E250Nt24_V-A_spectral_function}.
Increasing the temperature further to $T=192\,$MeV well in the chirally symmetric phase, the ratio even becomes slightly negative ($\approx -0.02$).
Due to the short extent in the time direction, cut-off effects could be responsible for this behavior.

Summarizing our results, we find that at temperatures $T\in\{128,154,192\}\,$MeV the 'effective chiral condensate', which is extracted from static screening quantities, is reduced to \{$0.84(3), 0.21(3),0.08(1)\}$ times the value at the corresponding vacuum ensemble. These values are compatible within errors with the reduction to $\{0.817(21),0.276(23),0.088(5)\}$ times the value at the corresponding vacuum ensemble observed in the temporal $(PA_0)$-correlator, which, in the chiral limit, is proportional to the chiral condensate. Especially in the low- and high-temperature phase these two predictions are in good agreement. This reduction factors can be compared to the reduction in the temporal $(V-A)$ sector. While the reduction to $0.69(2)$ times the vacuum value at $T=128\,$MeV is in the same ballpark, we find that during the transition to the crossover region chiral symmetry restoration appears to happen more rapidly. At $T=154\,$MeV the reduction factor in the $V-A$ sector is smaller by a factor of $\approx 4$. Increasing the temperature further to $T=192\,$MeV the ratio in the $PA_0$-channel drops down to $\approx 0.09$, while the ratio in the temporal $(V-A)$-channel is compatible with zero.  

The different rate at which different order parameters of chiral symmetry breaking approach zero already shows up in predictions from chiral perturbation theory at vanishing quark masses. Specifically the suppression of the chiral condensate to leading order in $T^2$ goes like $1-\frac{T^2}{8(f_{\pi}^0)^2}$\,(see Ref.\,\cite{Gasser:1986vb, Gerber:1988tt}) and is therefore less pronounced relative to Eq.\,(\ref{eq:diff_spectral}) with $1-2\epsilon = 1- \frac{T^2}{3(f_{\pi}^0)^2}$~\cite{Dey:1990ba}. \\

\section{Vacuum subtracted vector and axial-vector spectral functions}
\label{sec: vector_and_axial_vector_spectral_functions}
By inspection of Eq.\,(\ref{eq:axial_spec_func}) it becomes evident that the Euclidean correlators depend on the temperature in a dual way: implicitly through the occurrence of the inverse temperature $\beta$ in the kernel function $\text{cosh}(\omega(\beta/2-x_0))/\text{sinh}(\omega\beta/2)$ as well as through the explicit temperature dependence of the spectral function. The first effect leads to a temperature dependence of the Euclidean correlator even if the spectral function does not change. It is important to isolate this effect from a genuine temperature dependence resulting from a change in the thermal medium (e.g. the quark-gluon plasma). Consequently, differences in the thermal and reconstructed correlator can be traced back to a change in the spectral function, and thus the physical situation. The inverse is not necessarily true\,\cite{Aarts:2020dda}.

\begin{figure}[p]
	\includegraphics[scale=0.42]{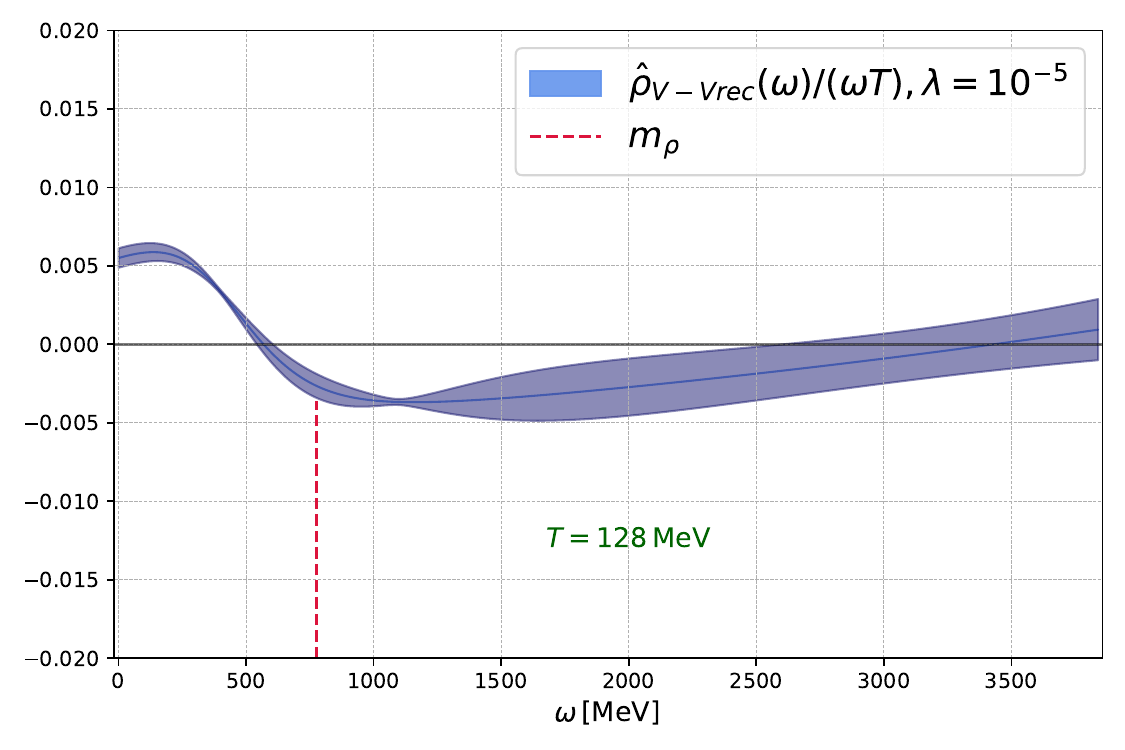}
	\includegraphics[scale=0.42]{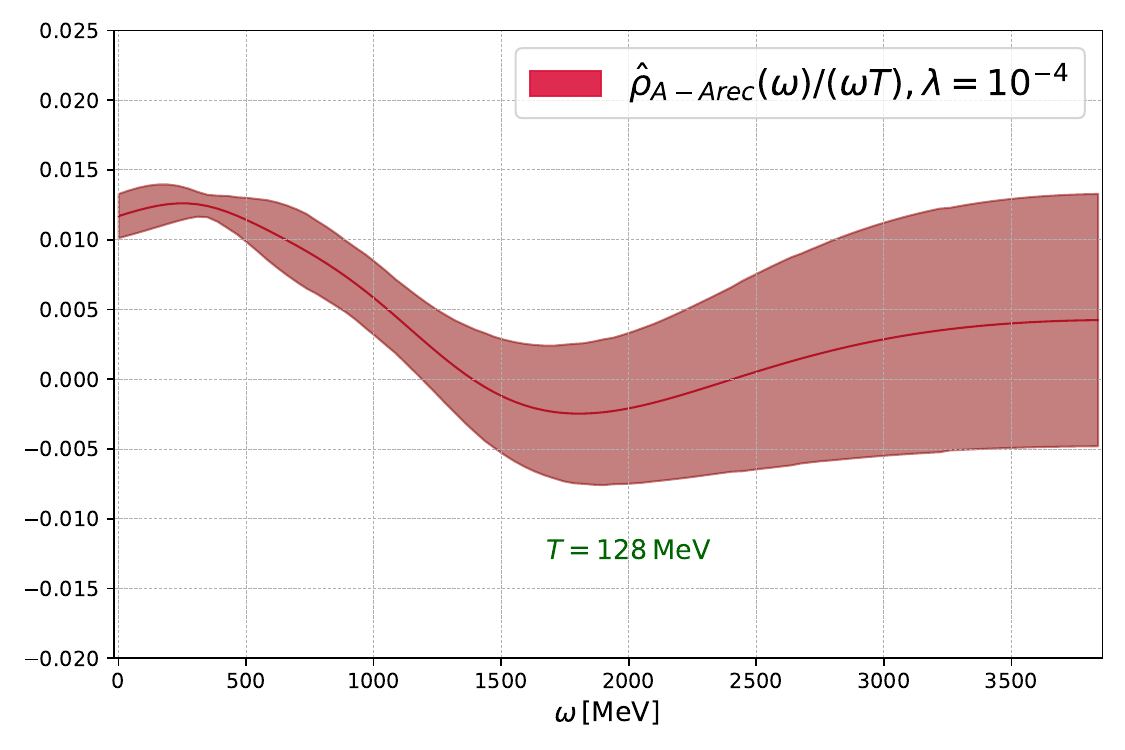}
	\includegraphics[scale=0.42]{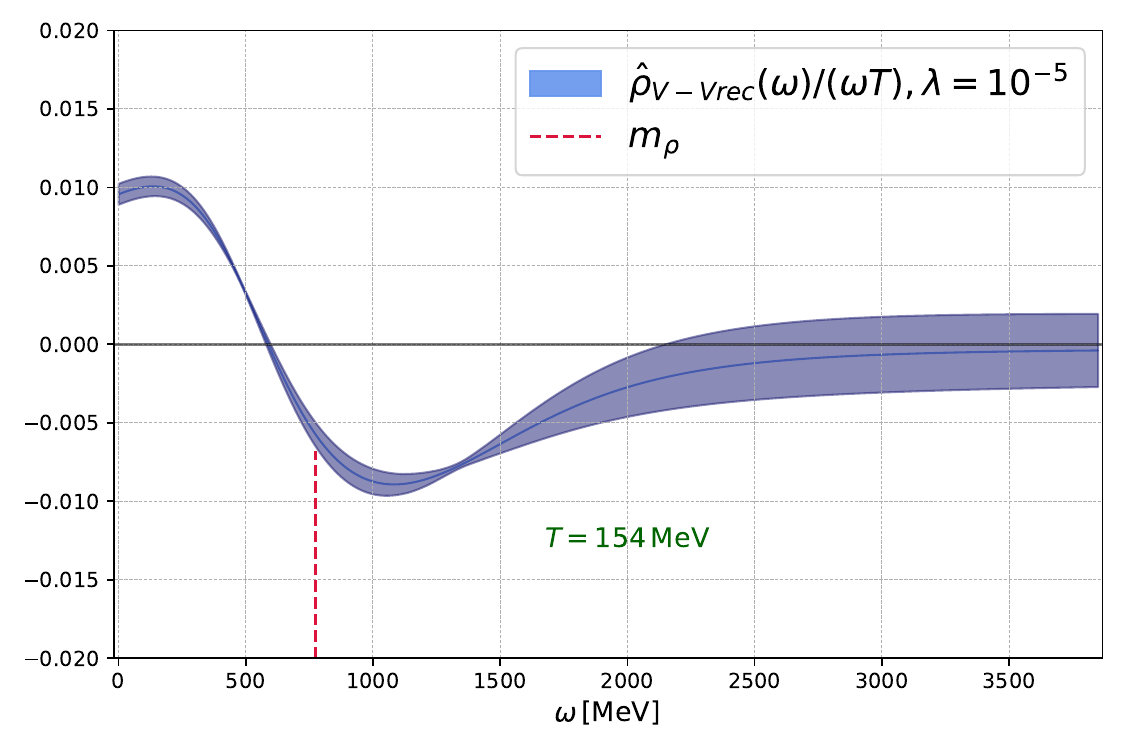}
        \includegraphics[scale=0.42]{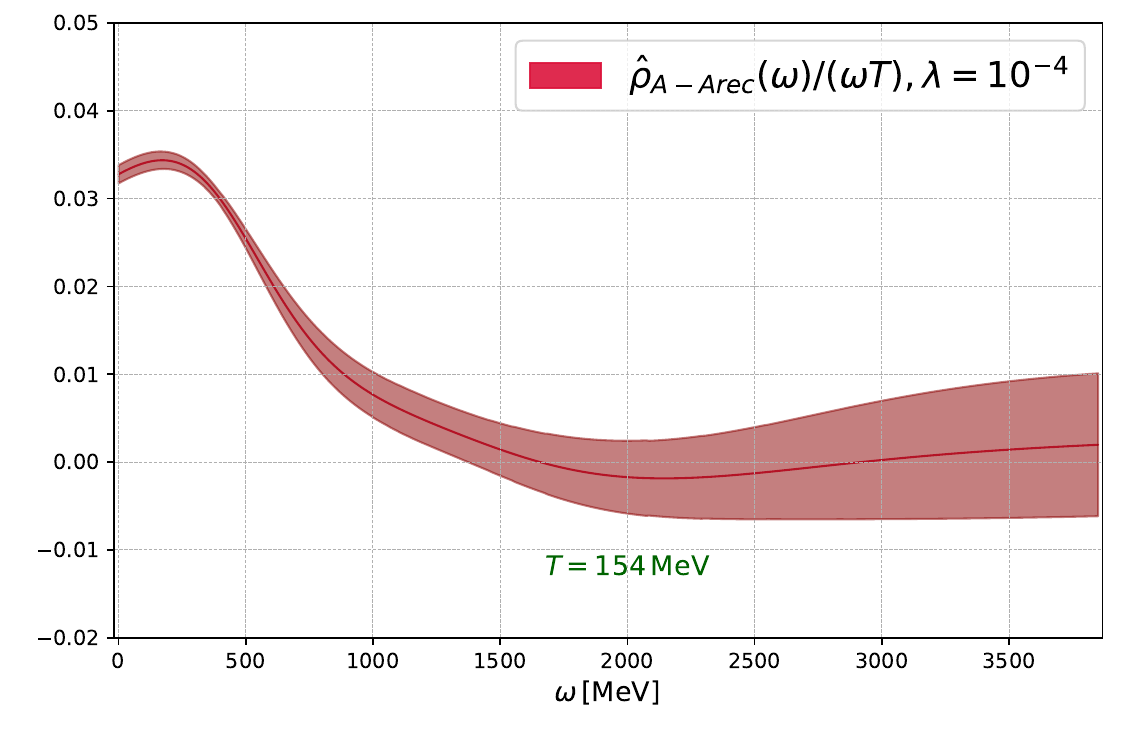}
	\includegraphics[scale=0.42]{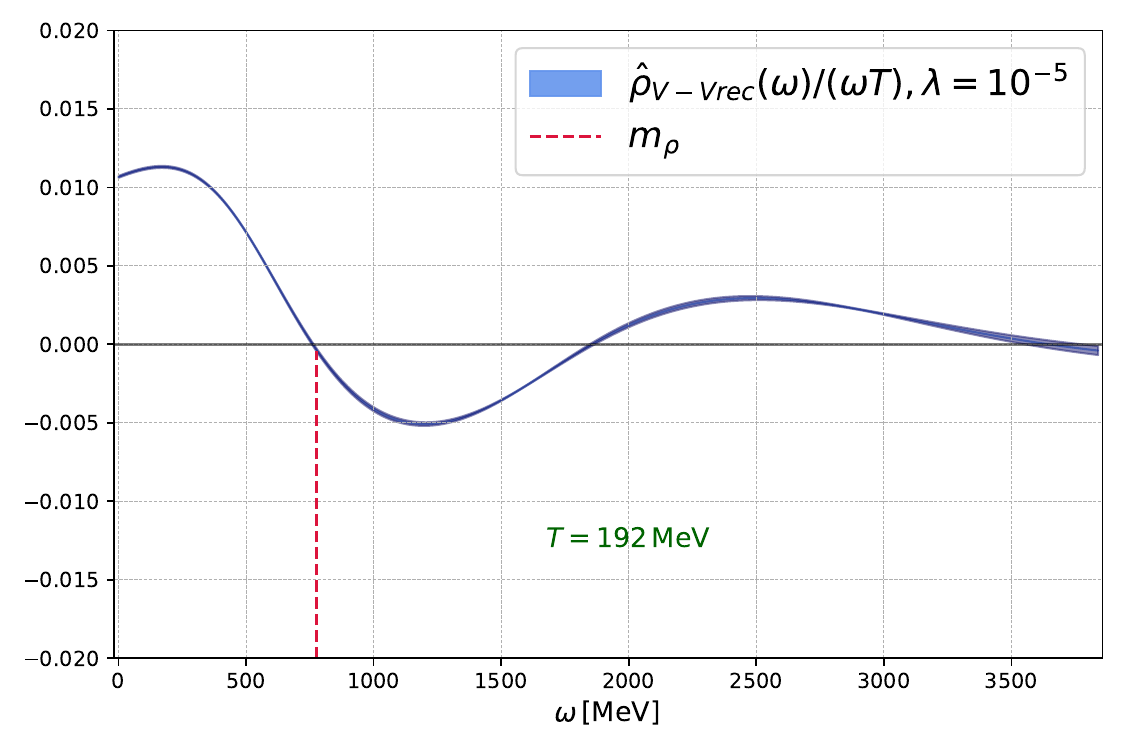}
        \includegraphics[scale=0.42]{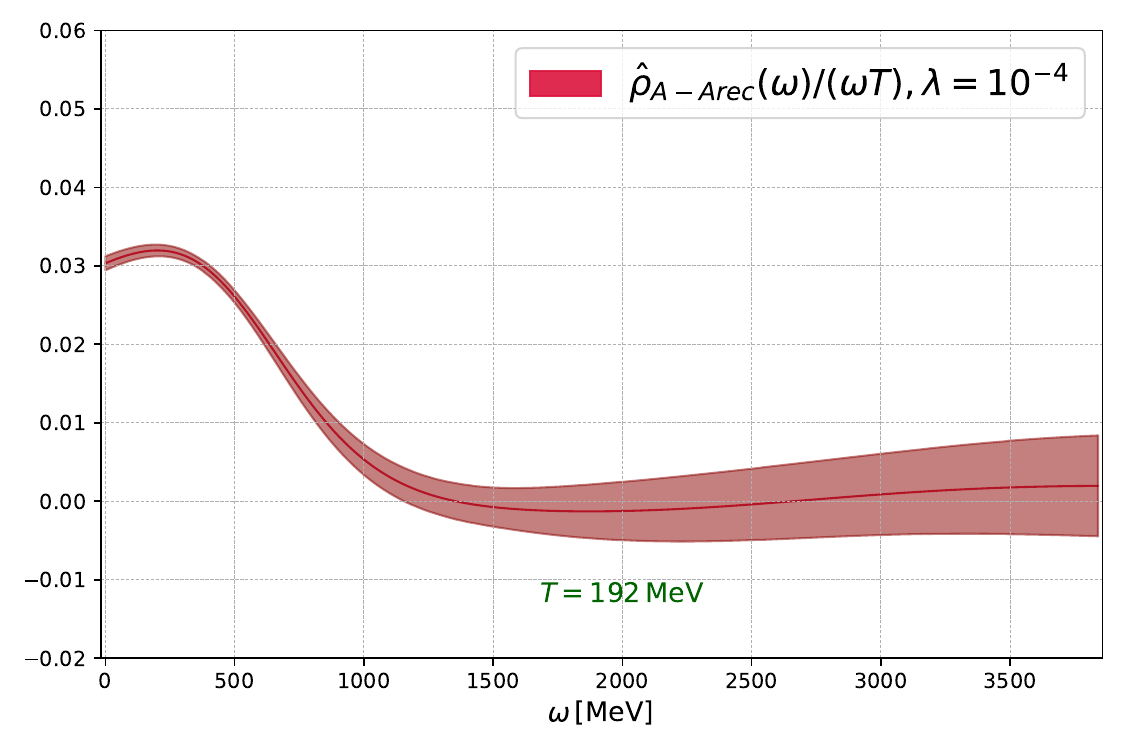}
\caption{
{\bfseries Top panels}: The l.h.s. shows the smeared spectral function of the difference between the finite temperature and reconstructed zero-temperature temporal vector correlators projected to zero momentum at a temperature $T=128\,$MeV. The r.h.s shows the same quantity for the temporal axial-vector correlators projected to zero momentum. {\bfseries Middle panels}: The same quantities at a temperature $T=154$\,MeV. {\bfseries Bottom panels}: Again the same quantities at a temperature $T=192$\,MeV. 
}
\label{fig:spe_func_vector_and_axial_vector}
\end{figure}

Using the Backus-Gilbert method described in Sec.\,\ref{sec:Backus}, we provide an estimator for the smeared spectral function of the difference between the finite temperature and reconstructed zero-temperature temporal vector correlators, projected to zero momentum. For the analysis of vacuum subtracted spectral functions we use 
\begin{equation}
    \label{eq:scaling_func_vac_sub}
    f_2(\omega) := \omega
\end{equation}
as a rescaling function. 
The results of the smeared and rescaled spectral functions are shown in Fig.\,\ref{fig:spe_func_vector_and_axial_vector}.

The difference
\begin{equation}
\label{eq:diff_spec_vector}
    \Delta \rho_V(\omega,\vec p, T) \equiv \rho_{V}(\omega, \vec p, T) - \rho_{V}(\omega, \vec p, T=0)\,,
\end{equation}
between the spatial components of the finite and zero-temperature spectral function satisfies the sum rule~\cite{Bernecker:2011gh}
\begin{equation}
    \label{eq:sum_rule}
    \int_0^\infty \frac{\mathrm{d}\omega}{\omega}\,\Delta \rho_V(\omega,\vec 0,T) = 0
\end{equation}
in the thermodynamic limit. From the operator-product expansion (OPE) it is known that the difference in Eq.\,(\ref{eq:diff_spec_vector}) falls of like $1/\omega^2$ for large frequencies $\omega$\,\cite{Caron-Huot:2009ypo}. Applying the sum rule to our smeared spectral function obtained from the Backus-Gilbert method, Eq.\,\ref{eq:sum_rule} evaluates to $-0.30(29)$ at $T=128\,$MeV, $-0.46(37)$ at $T=154\,$MeV and $0.44(1)$ at $T=192\,$MeV. Thus, the sum rule is satisfied within the large errors at the lower two temperatures. In the high temperature phase the sum rule is not satisfied. However, it should be stressed that due to the short time extent of the E250Nt16 ensemble, we could only use $4\leq x_0/a \leq 8$ for the spectral function reconstruction. Therefore, this result should be interpreted with caution.

For the thermal modification of the axial-vector channel, we have the spectral representation
\be
2 \int_0^\infty \frac{\mathrm{d}\omega}{\omega} \, \Delta\rho_A(\omega,\vec p,T)
= \int_0^\beta \mathrm{d} x_0 \,G_A(x_0,\vec p,T) - \int_{-\infty}^\infty \mathrm{d} x_0 \,G_A(x_0,\vec p,0).
\ee
Close to the chiral limit, we can saturate the correlation functions on the r.h.s. at the zero four-momentum by the screening pion, leading to the sum rule
\be\la{eq:axialsr}
2 \int_0^\infty \frac{\mathrm{d} \omega}{\omega} \, \Delta\rho_A(\omega,\vec 0,T) =
f_\pi^2(T=0) - f_\pi^2(T)  + {\mathcal{O}}(m_q^2).
\ee
Obviously, this is similar to the first Weinberg sum rule~\cite{Weinberg:1967kj} and its finite-temperature generalisation~\cite{Kapusta:1993hq},
however, the integral on the l.h.s. of Eq.\ (\ref{eq:axialsr}) is convergent even at $m_q\neq 0$.
We have seen that $f_\pi^2(T)$ is already significantly reduced at $T=128$\,MeV, and it is of ${\mathcal{O}}(m_q^2)$ in the chirally restored phase.
Thus the r.h.s of Eq.\,(\ref{eq:axialsr}) soon becomes only weakly temperature-dependent.
This qualitative behaviour is confirmed by the numerical data shown in the right panels of Fig.\,\ref{fig:spe_func_vector_and_axial_vector}.
Since $\Delta\rho(\omega,\vec p,T)\sim 1/\omega^2$ at large $\omega$,
the sum rule suggests that thermal effects lead to the build-up of spectral weight below the mass of the vacuum $a_1(1260)$ meson.

\section{Conclusion}
\label{sec:conclusion}

Using lattice simulations at physical $(u,d,s)$ quark masses,
we have shown that the real part of the low-energy pole in the two-point function of the axial charge
is reduced as the temperature increases throughout the hadronic phase of QCD.
We interpret this as a reduction of the pion quasiparticle mass in the thermal medium.
At our lowest temperature of 128\,MeV, the effect amounts to a ten percent reduction.
This finding is possible because the pole describing the pion quasiparticle
parametrically dominates both the Euclidean two-point functions of the axial charge and of the
pseudoscalar density at $x_0=\beta/2$.

Above the crossover temperature, the pole in the axial-charge
correlator is expected to be of diffusive nature, i.e.\ its imaginary
part dominates over the real part.  Since the curvature in the
Euclidean correlator of the axial charge is not parametrically
dominated by the pole in the chirally restored phase, one faces an
inverse problem if one wants to determine the precise position of the
pole. What is certain is that the pole tends to the origin,
$\omega=0$, in the chiral limit. Indeed our calculation shows that the axial-charge
correlator is near-degenerate with the (temporally constant) isospin charge correlator.

We have found that the static pseudoscalar correlation length, $m_\pi^{-1}$,
shrinks steadily with increasing temperature, consistently with previous lattice calculations.
The associated `decay constant' $f_\pi$, which we defined (in Eq.\,(\ref{eq:asymp}))
as the coupling of the spatial, longitudinal axial current to
this correlation length, decreases steadily with temperature. Parametrically,
it is of order the quark mass in the chirally restored phase.
Similarly, we have computed the spatially transverse vector and axial-vector
static screening lengths, finding mostly consistency with results by the HotQCD
collaboration.

The combination $f_\pi^2 m_\pi^2/m_q$ is an estimator for the
condensate $\<-\bar\psi\psi\>$ in the chiral limit, and we have
computed the ratio of $f_\pi^2 m_\pi^2$ to its value in the vacuum as
an estimator for the `melting' of the chiral condensate.  That ratio
was compared to a different estimator based on the temporal
correlation function of the axial charge with the pseudoscalar
density, yielding consistent results.

We have constrained the thermal modification of the vector and the
axial-vector spectral functions by constructing the difference of the
thermal and the `reconstructed' correlator, i.e.\ the correlator that
would be realised at finite temperature if no change took place in the
spectral function. This strategy has been used before in a two-flavour
calculation with $m_\pi\simeq 270\,$MeV~\cite{Brandt:2015aqk}, but the
present calculation is the first one to operate at physical quark
masses around the crossover temperature.  Using the Backus-Gilbert
method to directly analyze the thermal change in the spectral
function, we conclude that in the vector channel, excess spectral
weight appears at low frequencies, while a depletion occurs around
energies of one GeV. Such a change is consistent with the sum rule
(\ref{eq:sum_rule}). In the axial-vector case, we observe a
significant excess thermal spectral weight up to one GeV -- see
Fig.\,\ref{fig:spe_func_vector_and_axial_vector}.

Finally, we have investigated the difference of the thermal correlators
for spatial components of the vector and axial-vector currents 
(at zero spatial momentum).
Phenomenologically, it is often studied in the context of the dilepton
production by the thermal medium~\cite{Hohler:2013eba}.
This difference, too, is an order parameter
for chiral symmetry, and it provides additional information on its restoration.
This correlator is already suppressed relative to its vacuum analogue 
by about at factor of two thirds at our lowest temperature, $T\simeq 128$\,MeV, and by more than
an order of magnitude at the crossover, $T\simeq 154$\,MeV.
The flat behaviour of this ratio is consistent with an overall suppression
factor of the $V-A$ spectral function up to fairly high energies.
This qualitative behaviour was the one originally predicted at low temperatures
in the chiral limit~\cite{Dey:1990ba}, and we have found it to be a good approximation
at the correlator level up to the crossover.

Overall, we have found simulations at physical quark masses with
$\mathcal{O}(a)$-improved Wilson fermions to be entirely feasible with the
current algorithmic techniques. We have observed long auto-correlation
times for the two-point correlation function of the axial current with the pseudoscalar
density at the chiral crossover, which is not surprising for an order
parameter in the vicinity of a phase transition.
Future simulations could be performed at finer lattice spacing
in order to perform continuum extrapolations of the most important
observables such as the pion quasiparticle mass.

\clearpage

\acknowledgments{We thank Tim Harris for many helpful discussions related to this project.
This work was supported by the European Research Council (ERC) under the European
Union’s Horizon 2020 research and innovation program through Grant Agreement
No.\ 771971-SIMDAMA, as well as by the Deutsche Forschungsgemeinschaft 
(DFG, German Research Foundation) through the Cluster of Excellence “Precision Physics,
Fundamental Interactions and Structure of Matter” (PRISMA+ EXC 2118/1) funded by
the DFG within the German Excellence strategy (Project ID 39083149).
The research of M.C. is funded through the MUR program for young researchers ``Rita Levi Montalcini''.
The authors gratefully acknowledge the Gauss Centre for Supercomputing e.V. (www.gauss-centre.eu) for funding this project by providing computing time on the GCS Supercomputer HAWK at Höchstleistungsrechenzentrum Stuttgart (www.hlrs.de), as well as 
through the John von Neumann Institute for Computing (NIC) on the GCS Supercomputer JUWELS at Jülich Supercomputing Centre (JSC) under project SIMCHIC.
The measurements of the two-point functions have been partly performed on our local machine MogonII and partly on  Noctua2 at PC$^2$ in Paderborn as part of an NHR proposal.
A.K. wishes to thank Simon Kuberski for providing the pion decay constant on the E250 vacuum ensemble, as well as for sharing high-precision vector correlator data on the aforementioned box\,\cite{Kuberski:2023qgx}. He also acknowledges Konstantin Ottnad for sharing his data in the axial-pseudoscalar channel on the E250 vacuum ensemble.
We made use of the following libraries: GNU Scientific Library\,\cite{Galassi:2019},
GNU MPFR\,\cite{10.1145/1236463.1236468} and GNU MP\,\cite{Granlund:2020}.
}

\clearpage
\appendix

\section{Improvement process}
\label{app:impr_process}
Lattice derivatives lead to cutoff effects. Therefore,
in this section we specify the lattice discretized (improved) derivatives used in this work, as well as the $\mathcal{O}(a)-$improvement process of the axial-vector and vector current. Defining a forward and backward derivative as
\begin{align}
   \label{eq:forw_backw_deriv}
    \partial_{\mu} f(x) &= \frac{1}{a}\left(f(x+a\hat{\mu}) - f(x)\right) + \mathcal{O}(a)\,, \\
    \partial^{\,*}_{\mu} f(x) &= \frac{1}{a}\left(f(x) - f(x-a\hat{\mu})\right) + \mathcal{O}(a)\, ,
\end{align}
a symmetrized version with reduced discretization errors can be obtained by combining forward and backward derivative
\begin{align}
   \label{eq:symm_deriv_forw_backw}
    \widetilde\partial_{\mu} f(x) = \frac{1}{2}\left(\partial_{\mu}f(x)+\partial^{\,*}_{\mu}f(x)\right)  = \frac{1}{2a}\left(f(x+a\hat{\mu}) - f(x-a\hat{\mu})\right)  + \mathcal{O}(a^2)\, .
\end{align}
Note the absence of $\mathcal{O}(a)$ dicretization errors. 
The above symmetrized derivative can be further improved by making the replacement\,\cite{Guagnelli:2000jw}
\begin{align}
    \label{eq:improved_derivative}
    \notag \widetilde\partial_{\mu} f(x) &\rightarrow  \widebar\partial_{\mu} f(x) := \widetilde\partial_{\mu} f(x)\left(1-\frac{1}{6}a^2\partial^{\,*}_{\mu}\partial_{\mu} f(x)\right)\,,\\ 
    &=\frac{1}{12a^2}\left[f(x-2a\hat{\mu}) - 8f(x-a\hat{\mu}) + 8f(x+a\hat{\mu}) -f(x+2a\hat{\mu})\right]\,.
\end{align}
When acting on smooth functions this improved lattice derivative has $\mathcal{O}(a^4)$ discretization errors.

Next, we define the $\mathcal{O}(a)-$improved axial-vector and vector current as
\begin{align}
\label{eq: imp_axial_corr}
A_{\mu}^{\text{imp},b}(x) &= A_{\mu}^b(x) + ac_A(g_0^2)\widetilde\partial_{\mu} P^b(x)\, ,
\\
    \label{eq:imp_vec}
    V^{\text{imp},b}_{\mu}(x) &= V_{\mu}^b(x) + ac_V(g_0^2)\widebar\partial_{\nu}\,T_{\mu\nu}(x)\, ,
\end{align}
where 
\begin{equation}
    \label{eq:tensor_current}
    T_{\mu\nu}^a(x)\equiv - \frac{1}{2} \bar\psi [\gamma_\mu,\gamma_\nu]\frac{\tau^a}{2}\psi
\end{equation}
is the tensor current.

The non-perturbatively calculated coefficient $c_A$ was taken from Ref.\,\cite{Bulava:2015bxa},
and the coefficient $c_V$ from Ref.\,\cite{Heitger:2020mkp}. 
For the analysis of the $V-A$ correlators in Sec.\,{\ref{sec:Ioffe}}, we use an improved version of the symmetric derivative defined in Eq.\,{\ref{eq:improved_derivative}}. 

\section{Renormalization process}
\label{app:ren}

Following \cite{Korcyl:2016ugy}, we renormalize the correlators in the following way:
\begin{align}
    \label{eq:renorm_V}
    G_V^{\text{ren.}} &= Z_V^2(g_0^2)\left[1+2am_lb_V(g_0^2)+6am_{\text{av}}\widetilde b_{V}(g_0^2)+\mathcal{O}(a^2)\right]G_V\, ,\\
    \label{eq:renorm_A}
    G_A^{\text{ren.}} &= Z_A^2(g_0^2)\left[1+2am_lb_A(g_0^2)+6am_{\text{av}}\widetilde b_{A}(g_0^2)+\mathcal{O}(a^2)\right]G_A\, ,\\
    G_{A_0}^{\text{ren.}} &= Z_A^2(g_0^2)\left[1+2am_lb_A(g_0^2)+6am_{\text{av}}\widetilde b_{A}(g_0^2)+\mathcal{O}(a^2)\right]G_{A_0}\, ,\\
        G_{PA_0}^{\text{ren.}} &= Z_P(g_0^2) Z_A(g_0^2)\left[1+am_lb_A(g_0^2)+3am_{\text{av}}\widetilde b_{A}(g_0^2)+\mathcal{O}(a^2)\right]G_{PA_0}\, ,\\
    \label{eq:renorm_P}
    G_P^{\text{ren.}} &= Z_P^2(g_0^2)G_P\, ,
    \end{align}
    
with $g_0^2 = 6/\beta$ being the bare gauge coupling and 
\begin{align}
    \label{eq:m_q_and_m_av}
    am_q &= \frac{1}{2}\left(\frac{1}{\kappa_{\text{q}}}-\frac{1}{\kappa_{\text{cr.}}}\right)\,,\quad q\in \{l,s\}\, , \\
    m_{\text{av}} &= \frac{1}{3}\left(2m_l+m_s\right)\,,
\end{align}
denoting the bare subtracted quark mass and the `average quark mass', respectively.
The values for the renormalization constants $Z_J$ and the finite quark mass parameters $b_J$ are given in Table\,\ref{tab:ren_par}.

\begin{table}[tb]
\caption{Summary of the renormalization parameters.} 
\vspace{-0.4cm}
\begin{tabular}{l @{\hspace{17mm}} c @{\hspace{17mm}} S[table-format=1.9(2)]}
\\
\hline
\hline
$Z_V(g_0^2)$                & \cite{Heitger:2020mkp}    & 0.73571(10) \\ 
$Z_A(g_0^2)$                & \cite{DallaBrida:2018tpn} & 0.76900(42)\\ 
$Z_P(g_0^2)$                & \cite{Campos:2018ahf}     & 0.34768 \\ \hline
$b_V(g_0^2)$                & \cite{Bali:2023sdi}       & 1.388(17) \\
$b_A(g_0^2)$                & \cite{Bali:2023sdi}       & 1.232(15) \\ 
$\widetilde b_V(g_0^2)$     & \cite{Bali:2023sdi}       & 0.069(42) \\
$\widetilde b_A(g_0^2)$     & \cite{Bali:2023sdi}       & \ -0.03(13) \\ 
\hline
$\kappa_{\text{cr.}}$       & \cite{Gerardin:2018kpy}   & 0.1371726(13) \\
$\kappa_{\text{l}}$         & \cite{Ce:2022eix}         & 0.137232867\\
$\kappa_{\text{s}}$         & \cite{Ce:2022eix}         & 0.136536633\\
\hline
\hline
\end{tabular}
\label{tab:ren_par}
\end{table}



\section{Error analysis}
\label{app:error_analysis}

\begin{figure}[h!]
\center
	\includegraphics[scale=0.6]{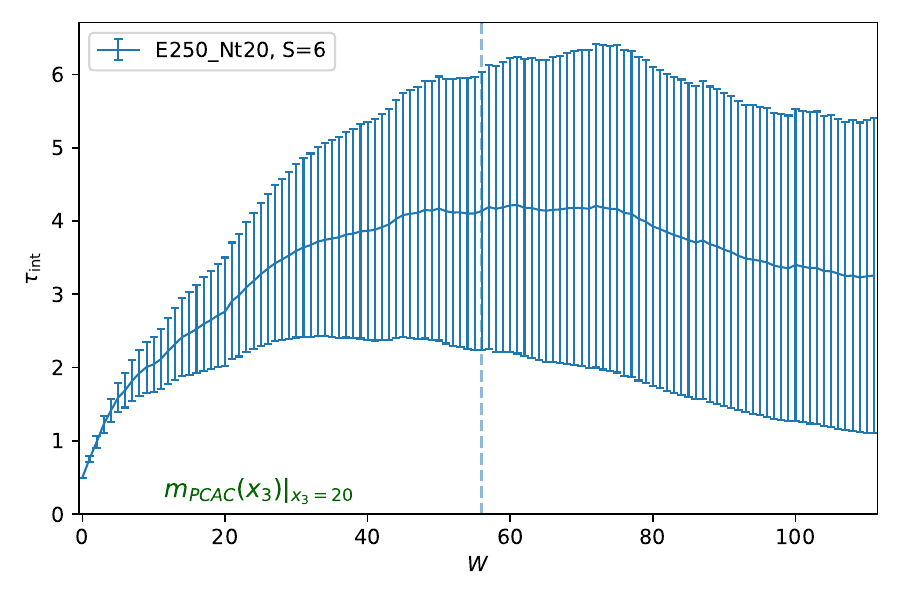}
        \caption{ 
         Integrated autocorrelation time $(S=6)$ of the PCAC mass at fixed source-sink seperation $x_3/a=20$ on the E250Nt20 at the chiral crossover. }
        \label{fig:autocorr_PCAC}
\end{figure}

For the estimation of the statistical errors of our (derived) observables we use the $\Gamma$ method \cite{Madras:1988ei,Wolff:2003sm,Ramos:2018vgu} in the implementation of the pyerrors package introduced in Ref.\,\cite{Joswig:2022qfe}.
If $\tau_{\text{exp}}$ (\textit{exponential autocorrelation time}) denotes the largest autocorrelation time in the system, the autocorrelation function typically exhibits an asymptotic exponential behavior,
\begin{equation}
    \label{eq:tau_exp}
    \frac{\Gamma_{F}(t)}{\Gamma_{F}(0)} \sim \text{exp}\left(-\frac{t}{\tau_{\text{exp}}}\right)\,.
\end{equation}
For any (derived) observable $F$, one defines
\begin{equation}
    \label{eq:int_autocorr}
    \tau_{\text{int},F}(W)=\frac{1}{2} + \sum_{t=1}^W  \frac{\Gamma_{F}(t)}{\Gamma_{F}(0)}\,,
\end{equation}
as the \textit{integrated autocorrelation time}. In practice one needs to choose an appropriate window $W$. In doing so, one is confronted with an trade-off between capturing all the relevant information in the chosen window and simultaneously not being dominated by noise contributions to $\tau_{\text{int}}$ for large values of $t$. In Ref.\,\cite{Wolff:2003sm} an automatic windowing procedure was introduced to determine $W$ in actual data analysis. Based on the hypothesis $\tau_{\text{exp}} =S\,\tau_{\text{int},F}(W)$ one chooses the factor $S$ such that the sum of systematic and statistical error is minimized. The default value is $S=2$. However, for the correct error estimation of the PCAC mass in the E250Nt20 ensemble, we choose $S=6$ [see Fig.\,\ref{fig:autocorr_PCAC}]. The final error based on the integrated autocorrelation time reads
\begin{equation}
    \label{eq:error_in_terms_of_tau_int}
    \sigma_F=\sqrt{2\tau_{\text{int},F}\frac{\Gamma_F(0)}{N}}\,.
\end{equation}

We utilize the robust median and median absolute deviation (MAD) described to identify two (out of 1000)
outliers on the E250Nt20 ensemble~\cite{Agadjanov:2023efe,Krasniqi:2024inr}.

\section{Chiral effective theory Lagrangian of Son and Stephanov}
\label{app:Son}
In the chiral effective theory approach of Son and Stephanov \cite{Son:2001ff}\cite{Son:2002ci} the dynamics of the pions at finite temperature is described by the Lagrangian,
\begin{align}
    \label{eq:effective Lagrangian}
    \mathcal{L}_{\text{eff}} = \frac{f_t^2}{4}\langle\nabla_0\Sigma\nabla_0\Sigma^{\dagger}\rangle - \frac{f_{\pi}^2}{4}\langle\partial_i\Sigma\partial_i\Sigma^{\dagger}\rangle + \frac{m_{\pi}^2f_{\pi}^2}{2}\text{Re}\langle\Sigma\rangle\,,
\end{align}
where $\Sigma$ denotes an $SU(2)$ matrix whose phase describes the pions, $\nabla_0\Sigma = \partial_0\Sigma -\frac{i}{2}\mu_{I5}(\tau_3\Sigma+\Sigma\tau_3)$ is the covariant derivative, $\mu_{I5}$ denotes the axial isospin chemical potential and the trace is taken in flavor space. Note that in the presence of a thermal medium Lorentz invariance is broken resulting in two independent decay constants which are related through the pion velocity \cite{Son:2001ff},
\begin{align}
    \label{eq:relation f and f_t}
    u = \frac{f_{\pi}}{f_t}\,.
\end{align}

\section{Determination of the static screening properties in the pseudoscalar channel}
\la{sec:PSscreenApdx}

\begin{figure}[h!]
\center
	\includegraphics[scale=0.6]{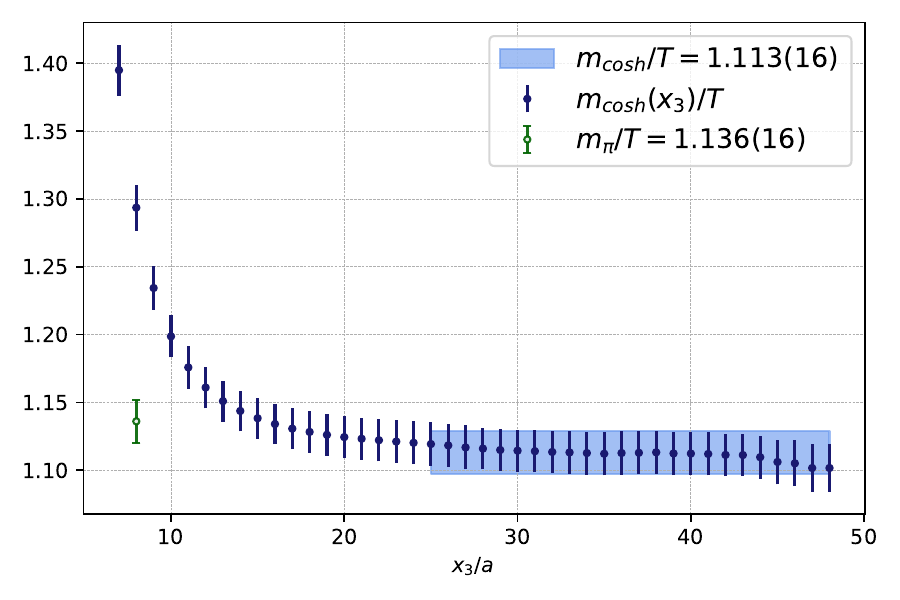}
	\includegraphics[scale=0.6]{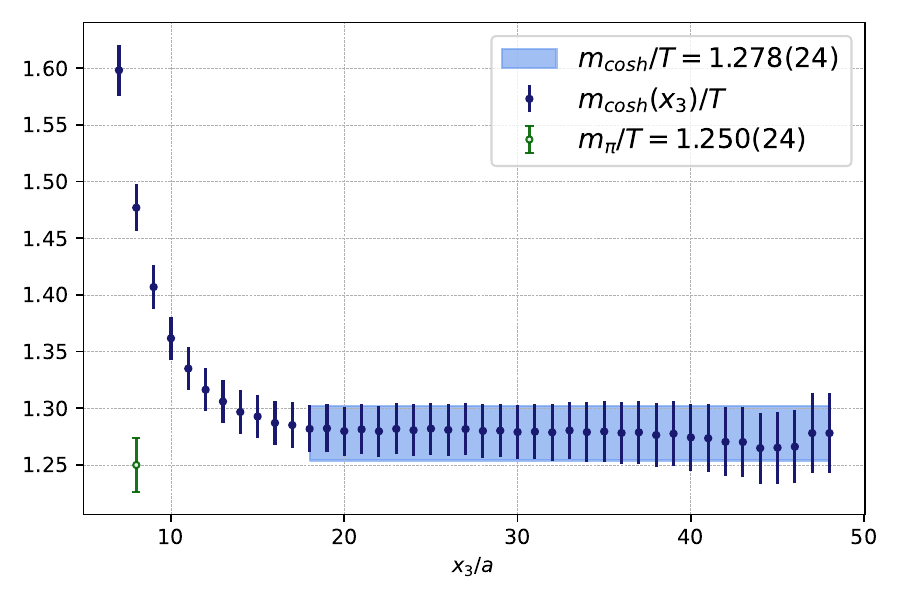}
        \includegraphics[scale=0.6]{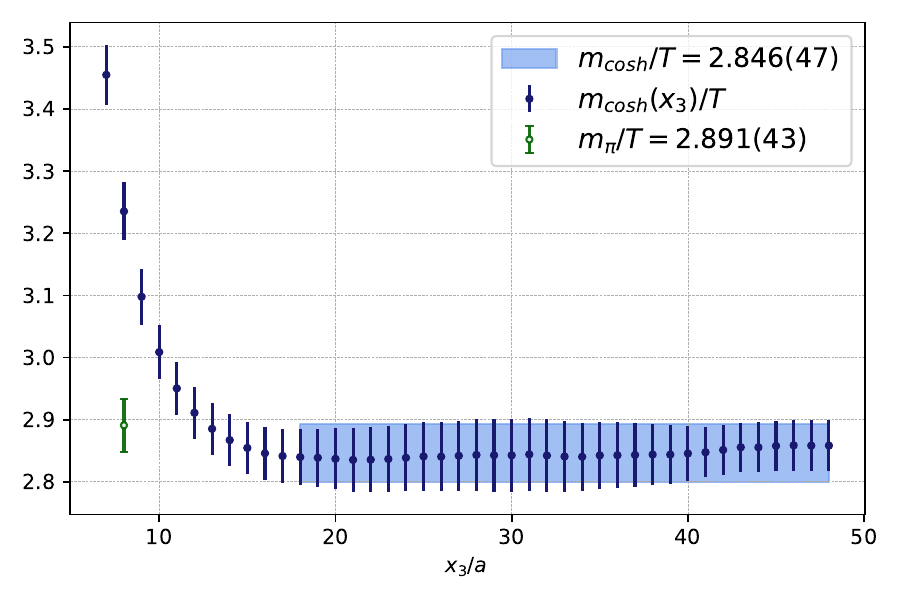}
        \caption{ 
        Effective mass plots for the cosh mass $m_{\text{cosh}}(x_3)/T$ as a function of the $x_3$-coordinate in temperature units, obtained from the pseudoscalar screening correlation function at zero spatial momentum $G_P^s(x_3,T)$. For comparison the value for the screening pion mass $m_{\pi}/T$, obtained from a simultaneous fit of the pseudoscalar and axial-pseudoscalar correlator is also included (green bars). 
        {\bfseries Top panel}: Hadronic E250Nt24 ensemble.
        It is assumed that the effective mass plateau starts at $x_3/a = 25$. 
        {\bfseries Middle panel}: E250Nt20 ensemble at the chiral crossover. It is assumed that the effective mass plateau starts at $x_3/a = 18$.
        {\bfseries Bottom panel}: E250Nt16 ensemble in the high-temperature phase. It is assumed that the effective mass plateau starts at $x_3/a = 18$.
        For all three ensembles he result of the fit to the effective mass values is represented by a $1-\sigma$ band.}
        \label{fig:cosh_masses}
\end{figure}

In this appendix, we describe how the screening pion mass $m_{\pi}$ and the screening decay constant $f_{\pi}$ can be calculated. In order to accomplish this, we perform a simultaneous fit of the static screening pseudoscalar and the axial-pseudoscalar correlator [see Eqs.\,(\ref{eq:def_static_PP_corr})-(\ref{eq:def_static_AP_corr})].

The spectral decomposition of the aforementioned correlators reads
\begin{align}
    \label{eq:spectral_dec_PP}
    G_{P}^s(x_3,T) &= \frac{\bra{0}\Bar{u}\gamma_5 d \ket{\pi}\bra{\pi}\Bar{d}\gamma_5 u\ket{0}}{2\,m_{\pi}} \left(e^{-m_{\pi}x_3}+e^{-m_{\pi}\cdot(L-x_3)}\right)\,,\\
    \label{eq:spectral_dec_AP}
    G_{AP}^s(x_3,T) &= \frac{\bra{0}\Bar{u}\gamma_3\gamma_5 d \ket{\pi}\bra{\pi}\Bar{d}\gamma_5 u\ket{0}}{2\,m_{\pi}} \left(e^{-m_{\pi}x_3}-e^{-m_{\pi}\cdot(L-x_3)}\right)\,,
\end{align}
which suggests the following one-state fit ansatz:
\begin{align}
    \label{eq:fit_PP}
    G_{P}^{\text{fit}}(x_3,T) &= \abs{A}^2 \left(e^{-C x_3}+e^{-C\cdot(L-x_3)}\right)\,,\\
    \label{eq:fit_AP}
    G_{AP}^{\text{fit}}(x_3,T) &= \abs{A}\cdot\abs{B} \left(e^{-C x_3}-e^{-C\cdot(L-x_3)}\right)\,,
\end{align}
$A$, $B$ and $C$ being free fit parameters. The pion screening mass $m_\pi$ and the screening pion decay constant $f_\pi$ are obtained from the fit parameters via
\beq
m_\pi \overset{\circ}{=} C, \qquad f_\pi \overset{\circ}{=}\sqrt{\frac{2\abs{B}^2}{C}}\;.
\eeq

\begin{figure}[p]
        \includegraphics[scale=0.53]{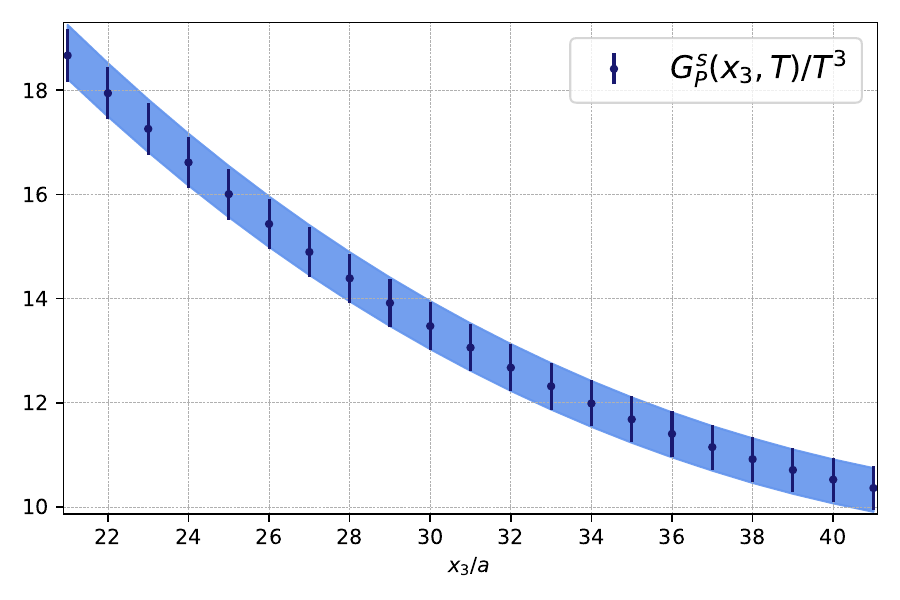}  
	\includegraphics[scale=0.53]{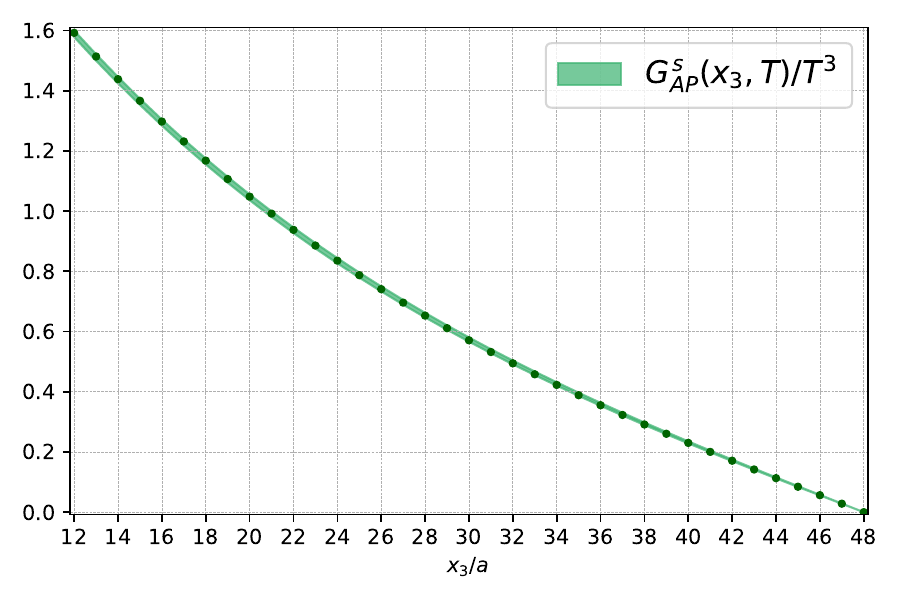}
        \includegraphics[scale=.53]{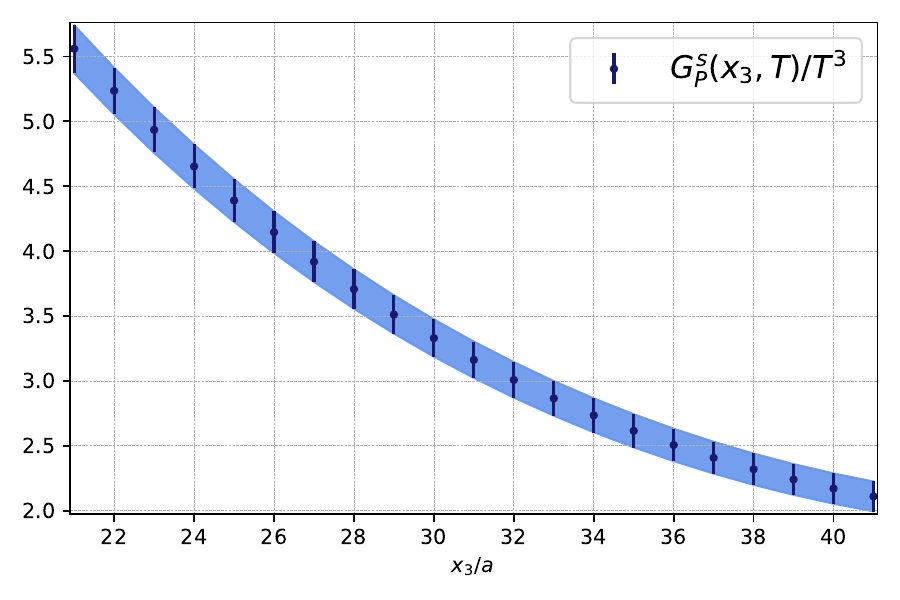}
        \includegraphics[scale=0.53]{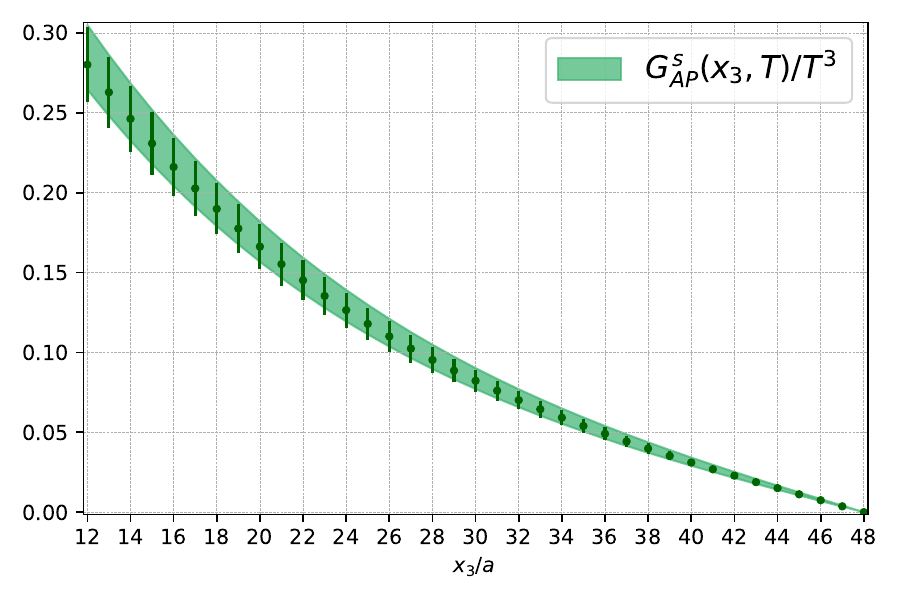}
        \includegraphics[scale=.53]{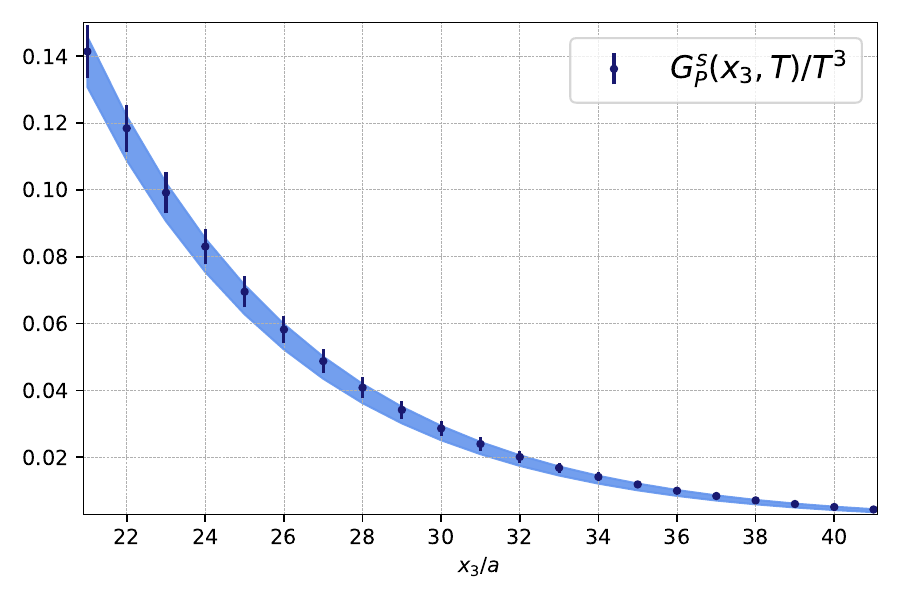}
        \includegraphics[scale=.53]{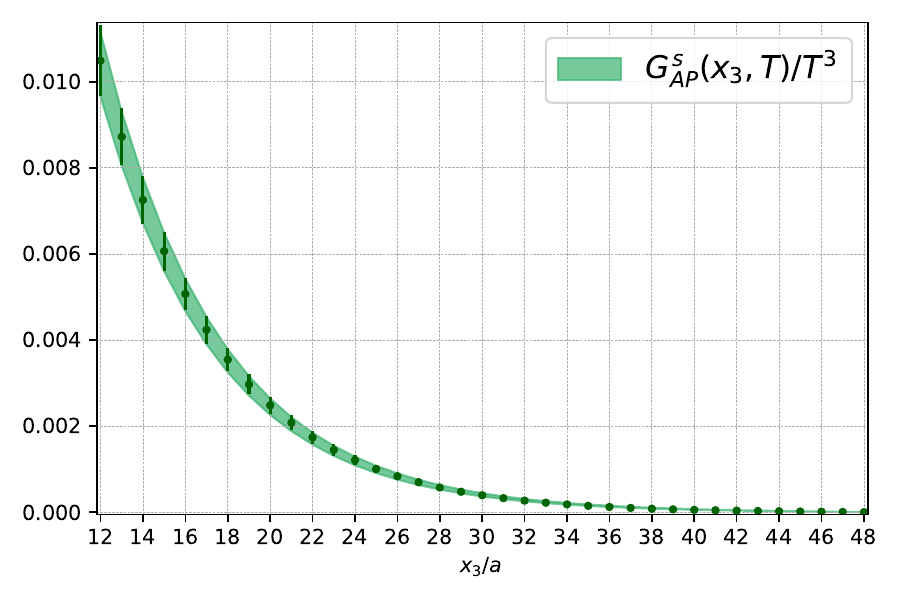}
\caption{Renormalized screening correlation functions in temperature units $G_P^s(x_3,T)/T^3$ (l.h.s) and $G_{AP}^s(x_3,T)/T^3$ (r.h.s.) and the result of the simultaneous fit. {\bfseries Top panels}: Correlators measured on the hadronic E250Nt24 ensemble.  {\bfseries Middle panels}: Correlators measured on the chiral crossover E250Nt20 ensemble.  {\bfseries Bottom panels}: Correlators measured on the high temperature E250Nt16 ensemble.
}
\label{fig:corr}
\end{figure}

The `cosh mass’ is defined as the solution of the algebraic equation,
\begin{align}
    \frac{G_P^s(x_3+a,T)}{G_P^s(x_3,T)} = \frac{\text{cosh}[m_{\text{cosh}}(x_3+a-L/2)]}{\text{cosh}[m_{\text{cosh}}(x_3-L/2)]}\,.
\end{align}
It is visualized in Fig.\,\ref{fig:cosh_masses}. Note that there is a different equation and a different solution for $m_{\text{cosh}}$ for each value of $x_3$.
The fits are performed using the Levenberg-Marquardt’s method \cite{press2007numerical}
and the results are shown in Fig.\,\ref{fig:corr}.

We fit the screening correlators $G_{P}^s(x_3,T)$ and $G_{AP}^s(x_3,T)$ simultaneously using Eqs.\,(\ref{eq:fit_PP})-(\ref{eq:fit_AP}). To assign fit qualities in view of different fit windows, we apply the Akaike information criterion\,\cite{Akaike:1998zah} and the model averaging method from Ref.\,\cite{Jay:2020jkz}. We assign a weight
\begin{equation}
    \label{eq:Akaike}
    w_i = N\cdot\text{exp}\left[-\frac{1}{2}\left(\chi_i^2+ 2p_i - 2n_i\right)\right]\,,
\end{equation}
where $p_i$ is the number of fit parameters and $n_i$ the number of data points in the fit\footnote{Since we are performing simultaneous one-state fits according to Eqs.\,(\ref{eq:fit_PP})-(\ref{eq:fit_AP}) the number of fit parameters is three and we vary only the starting points and lengths of the fit intervals.} with minimized $\chi_i^2$. $N$ is chosen such that the weights $w_i$ are normalized according to $\sum_i w_i=1$. The results of the fits are shown in Fig.\,\ref{fig:corr}.

\subsection{Alternative determination of the screening pion decay constant}

\begin{figure}[tp]
        \includegraphics[scale=0.42]{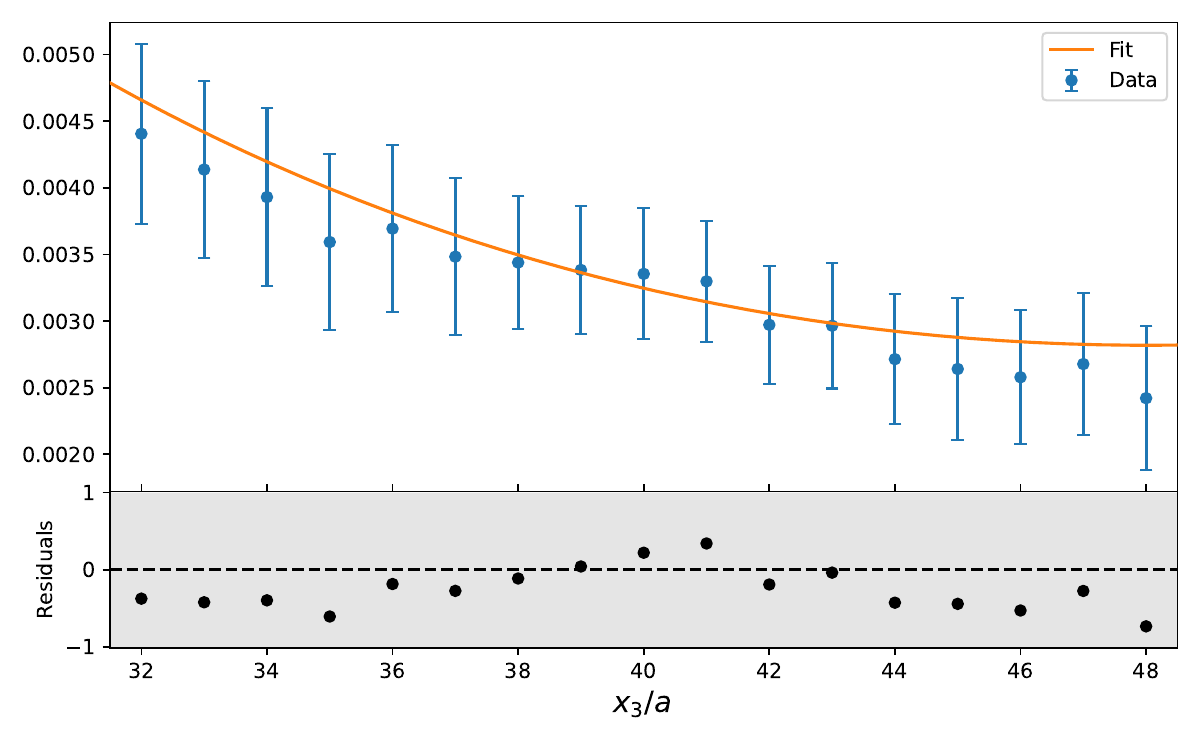}
	\includegraphics[scale=0.53]{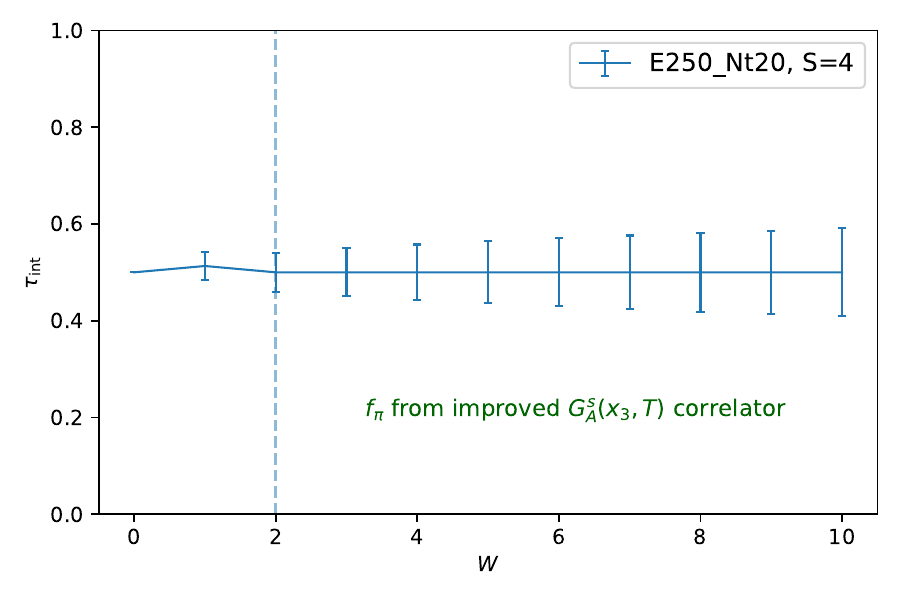}
	\includegraphics[scale=0.53]{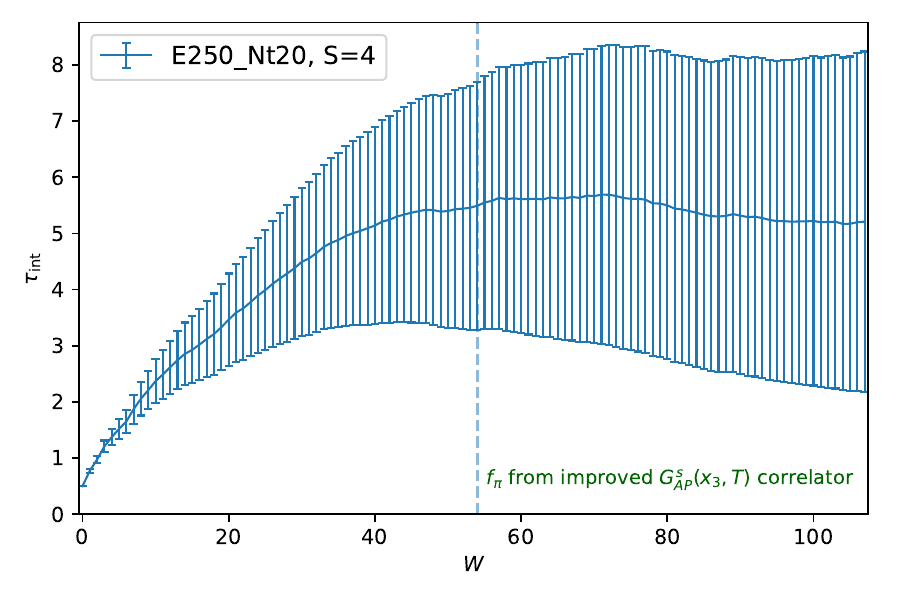}
\caption{
{\bfseries Upper panel}: Screening correlator $G_A^s(x_3,T=154\,\text{MeV})$ together with the corresponding fit.
{\bfseries Left panel}: Integrated autocorrelation time $\tau_{int}$ for $f_{\pi}$ determined as described in Sec.\,\ref{sec:extraction_screening_quant} {\bfseries Right panel }: Integrated autocorrelation time $\tau_{int}$ for $f_{\pi}$ determined from $G_A^s(x_3,T)$.
}
\label{fig:autocorr_fpi}
\end{figure}

We determine the screening pion decay constant on all three ensembles in Sec.\,\ref{sec:extraction_screening_quant}. The method described therein relies on a simultaneous fit ansatz of the screening pseudoscalar and axial-pseudoscalar correlator. As the latter correlator is an order parameter for chiral symmetry restoration, we are confronted with huge integrated autocorrelation times $\tau_{int}$ on the E250Nt20 ensemble which is generated at a temperature close to the pseudocritical temperature $T_{pc}$ [see Fig.\,\ref{fig:tau_int_Nt20and24}]. Alternatively, one could proceed as in Ref.\,{\cite{Ce:2022dax} and extract the pion decay constant simultaneously from the pseudoscalar and axial correlator. However, this fit ansatz relies on the PCAC mass, which itself needs to be extracted from the axial-pseudoscalar correlator. Thus, as a crosscheck, we determine the screening decay constant on the E250Nt20 ensemble solely from the noisy axial correlator $G_A^s$ [see top panel of Fig.\,\ref{fig:autocorr_fpi}]. The asymptotic form of this correlator defines the screening pion decay constant $f_{\pi}$ [see Eq.\,(\ref{eq:asymp})]. Consequently, using a cosh fit ansatz
\begin{equation}
    \label{eq:cosh_fit}
    G_A^s(x_3,T) \overset{\circ}{=} A_1^2\,\text{cosh}[m_1(x_3-L/2)]\,,
\end{equation}
the screening pion decay constant $f_{\pi}$ is obtained as
\begin{equation}
    \label{eq:f_pi_from_axial}
    f_{\pi}=A_1\sqrt{e^{m_1L/2}\,m_1}\,.
\end{equation}
We extract $f_{\pi}/T=0.179(24)$, compatible with our result $f_{\pi}/T=0.172(12)$ [see Table\,\ref{tab:results}].

\section{Free theory expression for spectral function of the axial charge}
\la{sec:freerhoA0}
The spectral representation for the axial-charge correlator reads
\be
\int \mathrm{d}^3x\; \<A_0^a(x) A_0^b(0)\> = \delta^{ab}\int_0^\infty \mathrm{d}\omega \;\frac{\cosh(\omega(\beta/2-x_0))}{\sinh(\omega\beta/2)}
\rho_A(\omega,\vec 0).
\ee
For non-interacting quarks of mass $m$, the spectral function reads
\be\la{eq:rhoAfree}
\rho_A(\omega,\vec 0) = \chi_A \,\omega\,\delta(\omega)
+ \frac{m^2 N_c}{4\pi^2}\,\theta(\omega^2-4m^2)\sqrt{1-\frac{4m^2}{\omega^2}}\,\tanh\left(\frac{\beta\omega}{4}\right),
\ee
with the axial-charge susceptibility reading
\be
\chi_A = 2\beta N_c\int \frac{d^3p}{(2\pi)^3}\,\frac{\vec p^2}{E_{\vec p}^2}\,f^F_{\vec p} (1-f^F_{\vec p}).
\ee
It is worth noting that the analogue of Eq.\ (\ref{eq:rhoAfree}) for the isospin charge,
\be
\int \mathrm{d}^3x\; \<V_0^a(x) V_0^b(0)\> = \delta^{ab}\int_0^\infty \mathrm{d}\omega \;\frac{\cosh(\omega(\beta/2-x_0))}{\sinh(\omega\beta/2)}
\rho_{00}(\omega,\vec 0),
\ee
reads
\be
\rho_{00}(\omega,\vec 0) = \chi_s \,\omega\,\delta(\omega)
\ee
with 
\be
\chi_s - \chi_A  = 2\beta N_c m^2\int \frac{\mathrm{d}^3p}{(2\pi)^3}\,\frac{1}{E_{\vec p}^2}\,f^F_{\vec p} (1-f^F_{\vec p}).
\ee

\section{Numerical values for temporal correlators}
\la{app:numerical_values}

In this appendix we list the means and errors of the (anti)symmetrized temporal correlators used in this work.
For convenience, the $PA_0$ and $V-A$ correlators are displayed in Figs.\,\ref{fig:ratio_A0P} and \ref{fig:thermal_vs_rec}, respectively.
The temporal $A_0A_0$ correlator at $T=128$\,MeV and its effective mass are displayed in Fig.\,\ref{fig:A0A0_E250Nt24_pred_fit}.

\begin{figure}[tp]
	\includegraphics[scale=0.5]{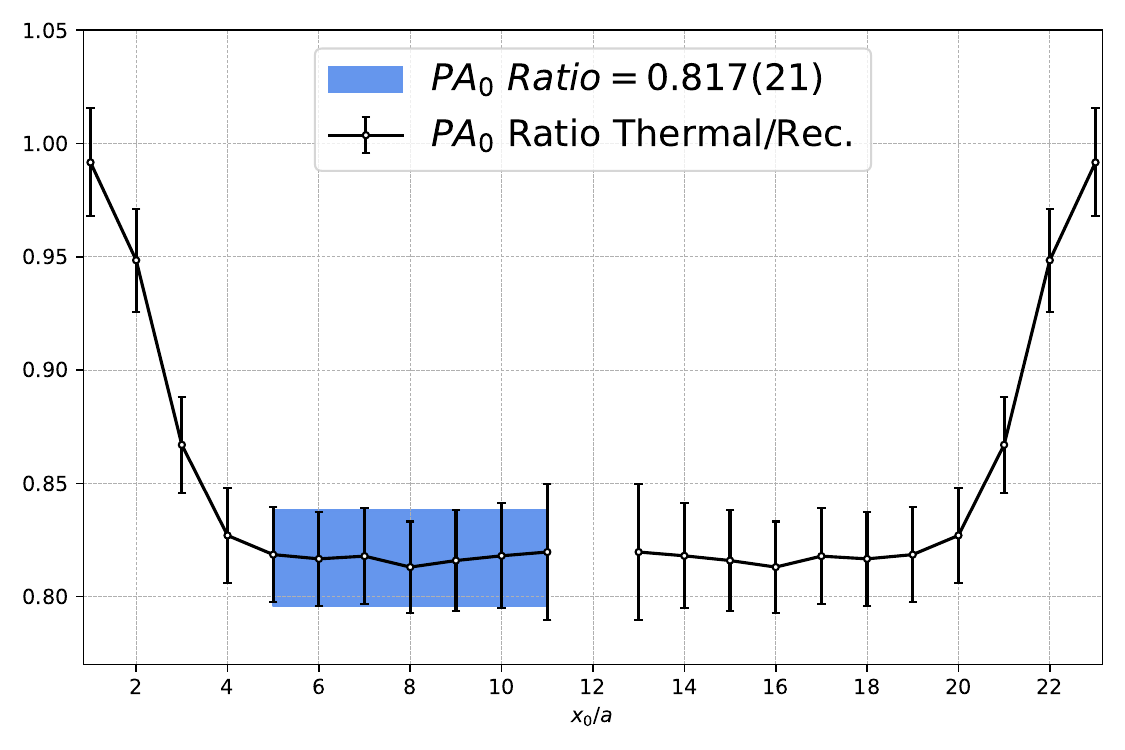}
        \includegraphics[scale=0.5]{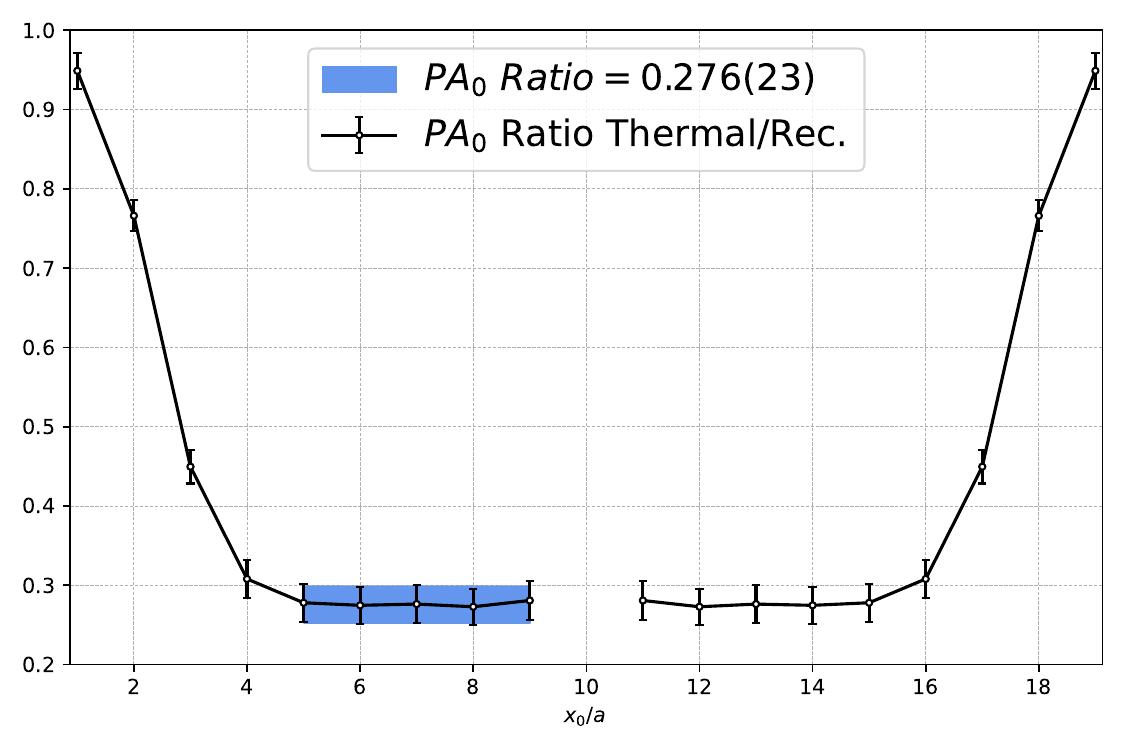}
        \includegraphics[scale=0.5]{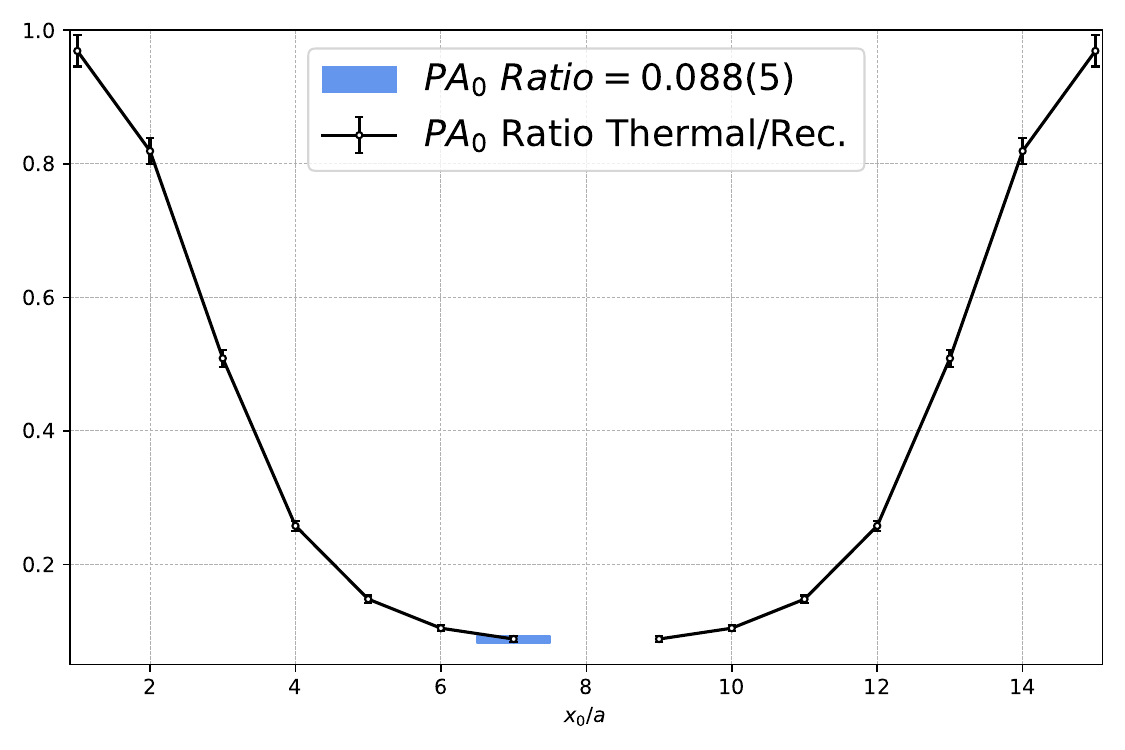}
	\caption{Ratios of the temporal thermal $(PA_0)$-correlators and  the reconstructed correlator $(PA_0)$-correlator. The errors on the ratio have been estimated using the $\Gamma$ method in the implementation of the pyerrors package introduced in Ref.\,\cite{Joswig:2022qfe}.
    {\bfseries Top and middle panel}: Result for the E250Nt24 and E250Nt20 ensemble respectively. The blue band shows the result from a correlated fit to the plateau. In both cases the $(PA_0)-$correlators have been $\mathcal{O}(a)$-improved.
    {\bfseries Bottom panel}: Result for the E250Nt16 ensemble. Since there is no clear plateau visible we quote the result for $x_0/a=7$. In the high-temperature phase the $(PA_0)-$correlators have not been $\mathcal{O}(a)$-improved.
	\label{fig:ratio_A0P}}
\end{figure}

\begin{figure}[tp]
        \includegraphics[scale=0.53]{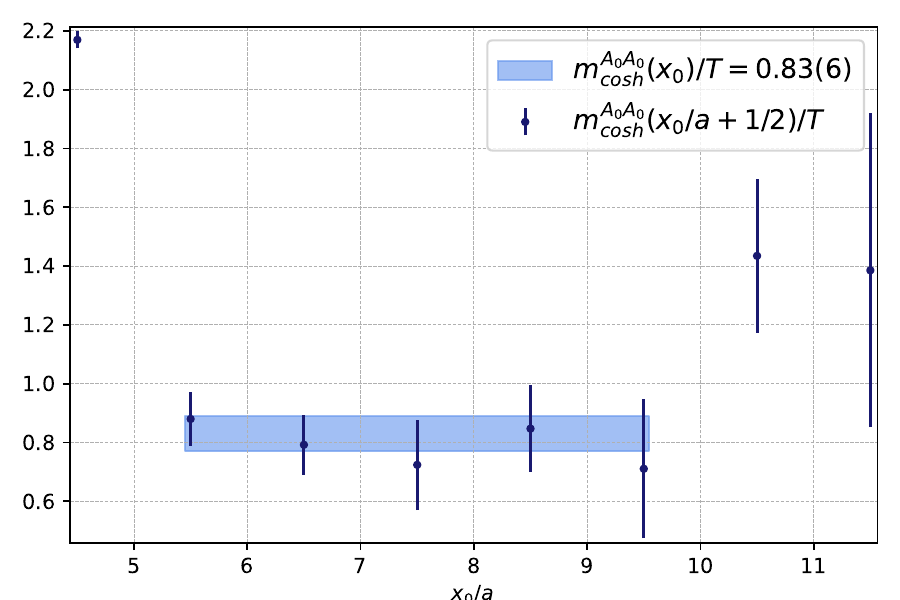}
	\includegraphics[scale=0.53]{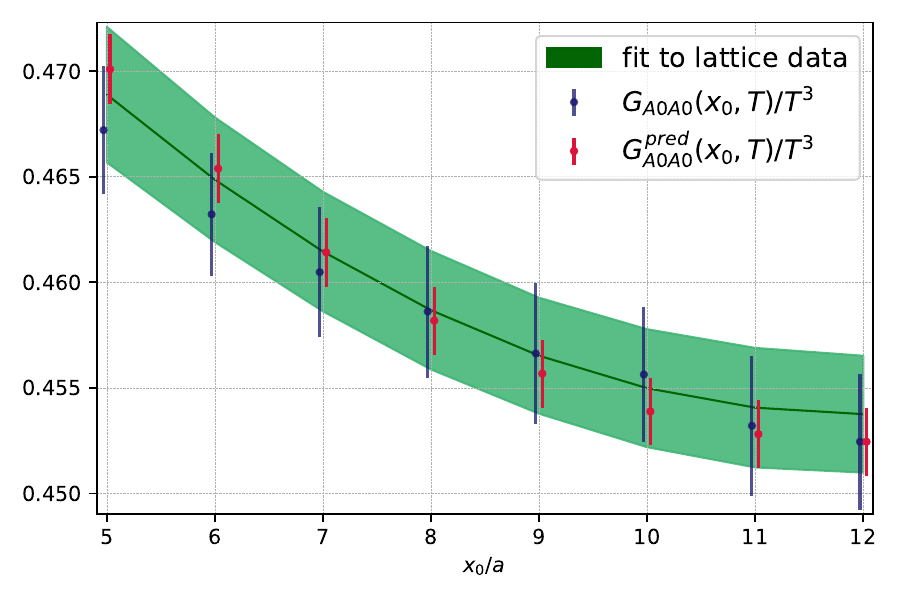}
\caption{Both plots show results from the hadronic E250Nt24 ensemble.
 {\bfseries Left panel}: Effective mass plot for the cosh mass $m_{\text{cosh}}(x_0)/T$ as a function of the $x_0$-coordinate in temperature units, obtained from the temporal axial correlation function at zero spatial momentum $G_{A_0}(x_0,T)$.
 {\bfseries Right panel}:
 The blue bars correspond to the lattice data of the temporal $G_{A_0}(x_0,T)$-correlator in the hadronic phase, while the red bars represent the prediction of the aforementioned correlator in terms of the screening quantities $f_{\pi}$ and $m_{\pi}$. The green band is the result of a direct one-state fit to the temporal correlator. From this fit we obtain $\omega_{\bold{0}}/T=0.91(9)$. 
}
\label{fig:A0A0_E250Nt24_pred_fit}
\end{figure}

\begin{figure}[tp]
        \includegraphics[scale=0.42]{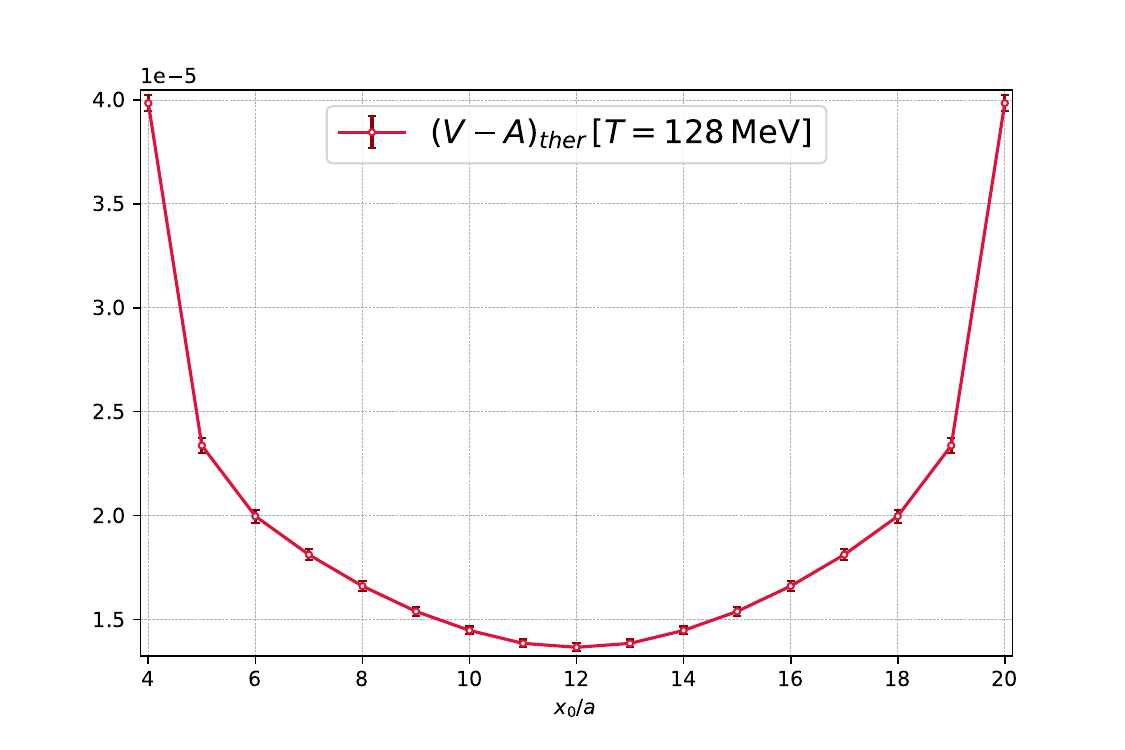}  
	\includegraphics[scale=0.42]{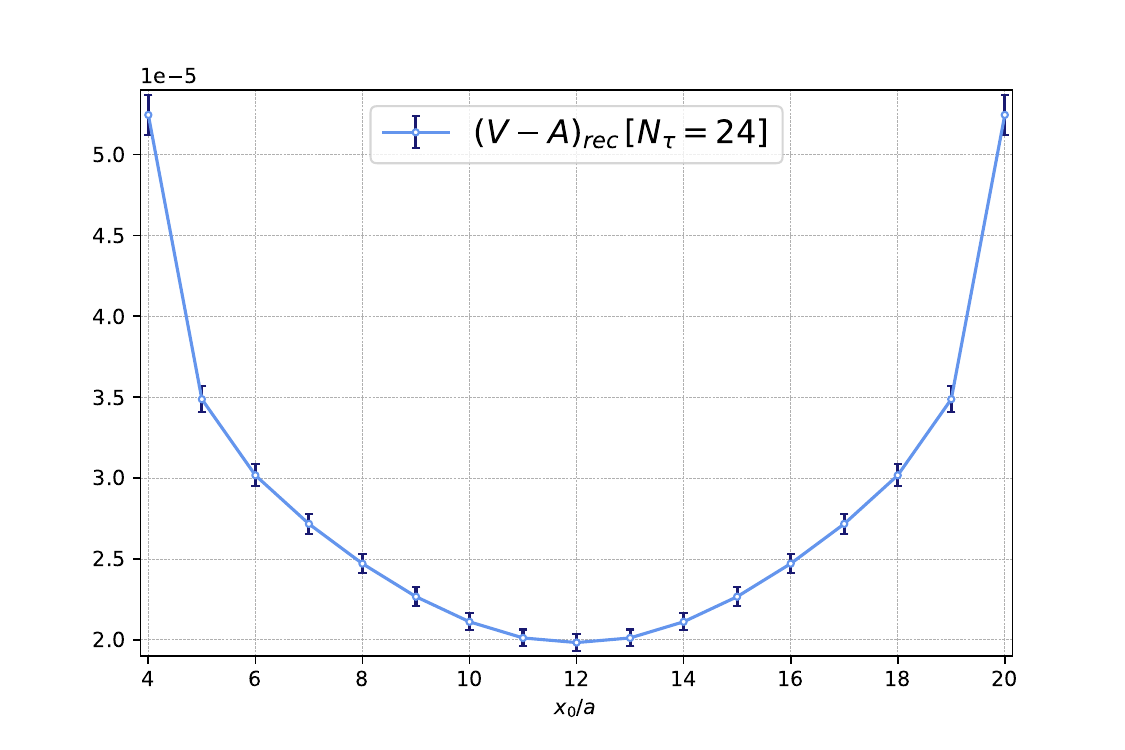}
        \includegraphics[scale=.42]{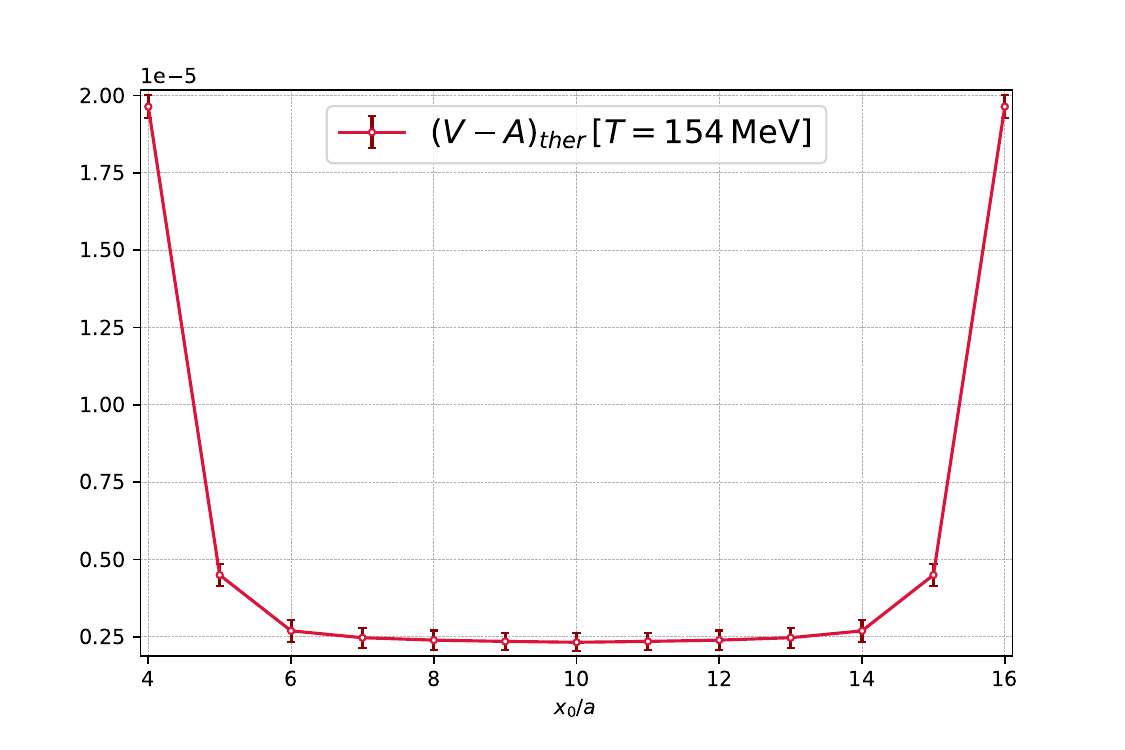}
        \includegraphics[scale=0.42]{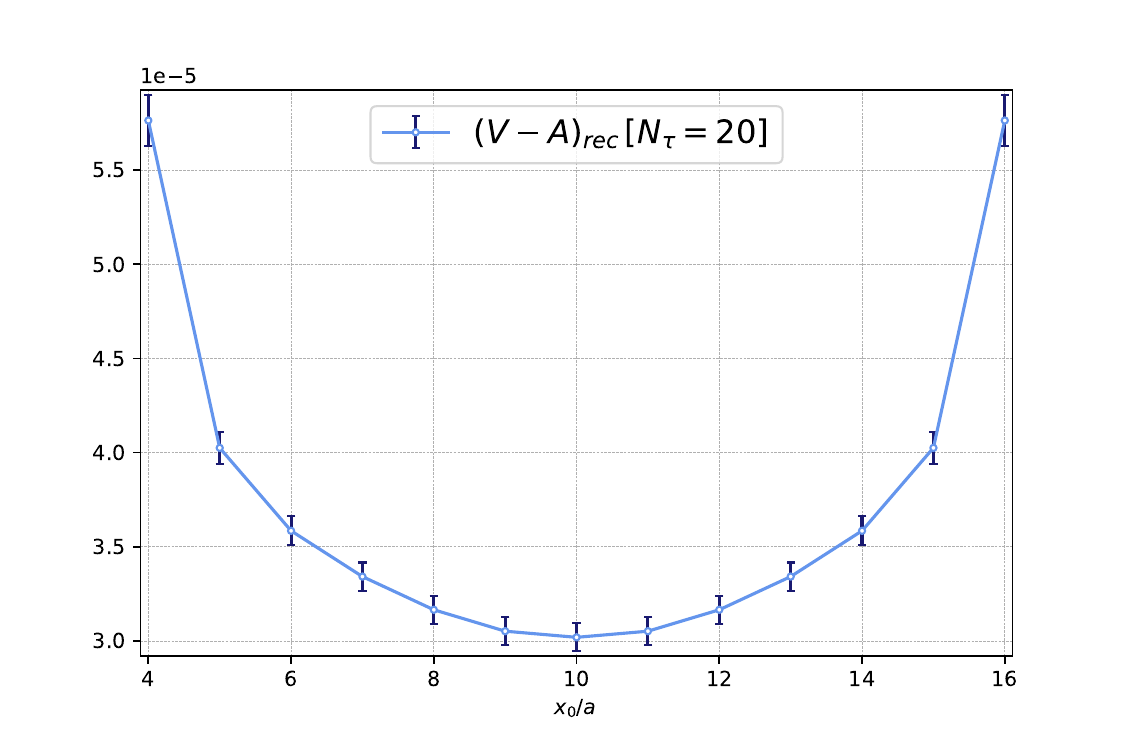}
        \includegraphics[scale=.42]{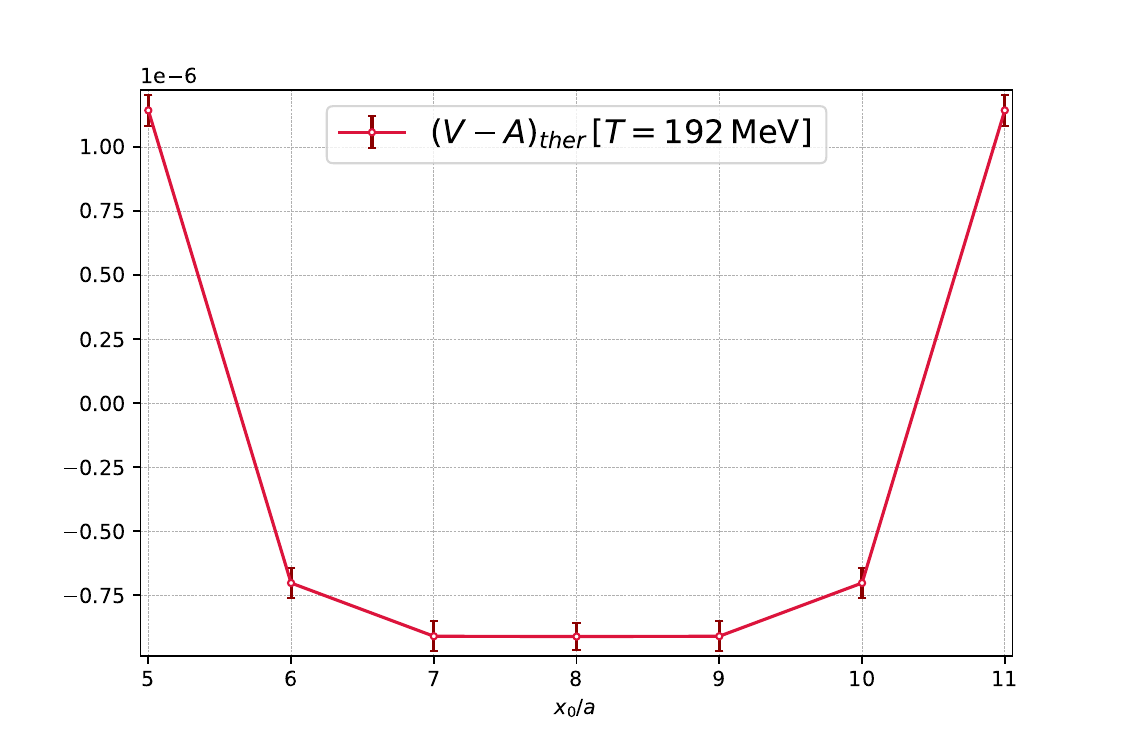}
        \includegraphics[scale=.42]{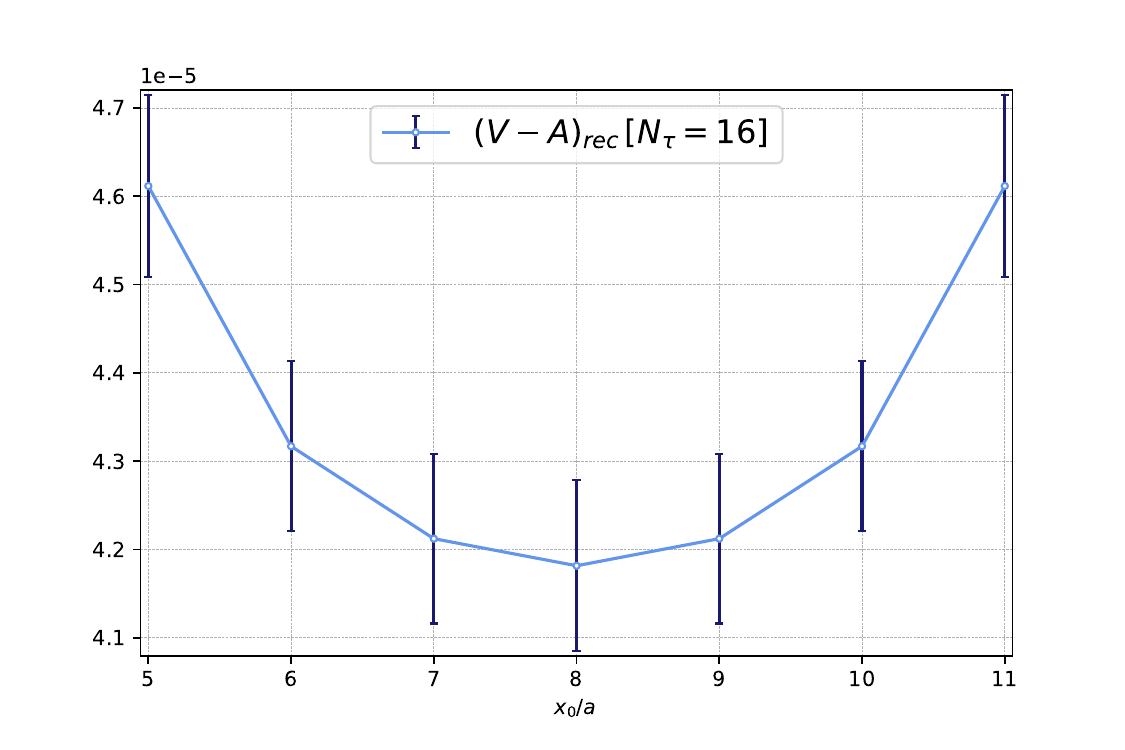}
\caption{{\bfseries L.h.s.}: The improved thermal correlator for the difference
“$(V-A)$”. The top panel shows the result in the hadronic phase, the middle panel at the finite temperature crossover and the bottom panel in the high-temperature phase. {\bfseries R.h.s.}: The corresponding reconstructed correlators [following Eq.\,(\ref{eq:reconstr})] for the $\mathcal{O}(a)-$improved difference “$(V-A)$”. All correlators are renormalized and the corresponding ratios are shown in Fig.\,\ref{fig:V-A}.
}
\label{fig:thermal_vs_rec}
\end{figure}

\begin{table}[tb]
\caption{Temporal symmetrized $V-A$ correlators projected to zero momentum. All errors quoted are purely statistical. The correlators are renormalized and quoted in units of temperature (i.e. scaled by a factor $N_{\tau}^3$).} 
\small
\begin{tabular}{c c c c c c c}
\hline
\hline
        & \multicolumn{2}{c}{$T=128$\,\text{MeV}} & \multicolumn{2}{c}{$T=154$\,\text{MeV}} & \multicolumn{2}{c}{$T=192$\,\text{MeV}} \\
$x_0/a$ & $\frac{G_{(V-A)}(x_0)}{T^3}$ & $\frac{G^\text{\,rec}_{(V-A)}(x_0)}{T^3}$ & $\frac{G_{(V-A)}(x_0)}{T^3}$ & $\frac{G^\text{\,rec}_{(V-A)}(x_0)}{T^3}$ & $\frac{G_{(V-A)}(x_0)}{T^3}$ & $\frac{G^\text{\,rec}_{(V-A)}(x_0)}{T^3}$ \\
\hline
3  & 2.7840(56) & 2.971(67)  & 1.4447(32) & 1.761(40)  & 0.73012(30)  & 0.917(21)  \\
4  & 0.5508(54) & 0.725(17)  & 0.1571(30) & 0.461(11)  & 0.06793(26)  & 0.2554(58) \\
5  & 0.3230(50) & 0.482(11)  & 0.0360(29) & 0.3220(68) & 0.00468(25)  & 0.1889(42) \\
6  & 0.2760(44) & 0.4170(95) & 0.0215(28) & 0.2868(62) & -0.00288(24) & 0.1768(39) \\
7  & 0.2504(36) & 0.3755(85) & 0.0197(25) & 0.2673(60) & -0.00373(21) & 0.1725(39) \\
8  & 0.2295(31) & 0.3415(81) & 0.0191(25) & 0.2532(59) & -0.00373(24) & 0.1713(40) \\
9  & 0.2127(28) & 0.3133(79) & 0.0188(23) & 0.2441(59) &              &            \\
10 & 0.2001(26) & 0.2918(72) & 0.0186(23) & 0.2415(59) &              &            \\
11 & 0.1914(26) & 0.2780(72) &            &            &              &            \\
12 & 0.1888(26) & 0.2741(74) &            &            &              &            \\
\hline
\hline
\end{tabular}
\label{tab:temporal_correlators}
\end{table}

\FloatBarrier



\bibliography{pion_quasiparticle.bib}

\providecommand{\href}[2]{#2}\begingroup\raggedright\begin{thebibliography}{10}

\bibitem{Aoki:2006we}
Y.~Aoki, G.~Endrodi, Z.~Fodor, S.~Katz and K.~Szabo, \emph{{The Order of the
  quantum chromodynamics transition predicted by the standard model of particle
  physics}}, \href{https://doi.org/10.1038/nature05120}{\emph{Nature}
  {\bfseries 443} (2006) 675}
  [\href{https://arxiv.org/abs/hep-lat/0611014}{{\ttfamily hep-lat/0611014}}].

\bibitem{HotQCD:2018}
{\scshape HotQCD} collaboration, \emph{{Chiral crossover in QCD at zero and
  non-zero chemical potentials}},
  \href{https://doi.org/10.1016/j.physletb.2019.05.013}{\emph{Phys. Lett. B}
  {\bfseries 795} (2019) 15}
  [\href{https://arxiv.org/abs/1812.08235}{{\ttfamily 1812.08235}}].

\bibitem{Borsanyi:2010bp}
{\scshape Wuppertal-Budapest} collaboration, \emph{{Is there still any Tc
  mystery in lattice QCD? Results with physical masses in the continuum limit
  III}}, \href{https://doi.org/10.1007/JHEP09(2010)073}{\emph{JHEP} {\bfseries
  09} (2010) 073} [\href{https://arxiv.org/abs/1005.3508}{{\ttfamily
  1005.3508}}].

\bibitem{HotQCD:2019}
{\scshape HotQCD} collaboration, \emph{{Chiral Phase Transition Temperature in
  ( 2+1 )-Flavor QCD}},
  \href{https://doi.org/10.1103/PhysRevLett.123.062002}{\emph{Phys. Rev. Lett.}
  {\bfseries 123} (2019) 062002}
  [\href{https://arxiv.org/abs/1903.04801}{{\ttfamily 1903.04801}}].

\bibitem{Shuryak:1990ie}
E.V.~Shuryak, \emph{{Physics of the pion liquid}},
  \href{https://doi.org/10.1103/PhysRevD.42.1764}{\emph{Phys. Rev. D}
  {\bfseries 42} (1990) 1764}.

\bibitem{Goity:1989gs}
J.~Goity and H.~Leutwyler, \emph{{On the Mean Free Path of Pions in Hot
  Matter}},
  \href{https://doi.org/10.1016/0370-2693(89)90985-4}{\emph{Phys.Lett.}
  {\bfseries B228} (1989) 517}.

\bibitem{Gasser:1987ah}
J.~Gasser and H.~Leutwyler, \emph{{Thermodynamics of Chiral Symmetry}},
  \href{https://doi.org/10.1016/0370-2693(87)91652-2}{\emph{Phys. Lett. B}
  {\bfseries 188} (1987) 477}.

\bibitem{Gerber:1988tt}
P.~Gerber and H.~Leutwyler, \emph{{Hadrons Below the Chiral Phase Transition}},
  \href{https://doi.org/10.1016/0550-3213(89)90349-0}{\emph{Nucl. Phys. B}
  {\bfseries 321} (1989) 387}.

\bibitem{Schenk:1993ru}
A.~Schenk, \emph{{Pion propagation at finite temperature}},
  \href{https://doi.org/10.1103/PhysRevD.47.5138}{\emph{Phys.Rev.} {\bfseries
  D47} (1993) 5138}.

\bibitem{Toublan:1997rr}
D.~Toublan, \emph{{Pion dynamics at finite temperature}},
  \href{https://doi.org/10.1103/PhysRevD.56.5629}{\emph{Phys.Rev.} {\bfseries
  D56} (1997) 5629} [\href{https://arxiv.org/abs/hep-ph/9706273}{{\ttfamily
  hep-ph/9706273}}].

\bibitem{Pisarski:1996mt}
R.D.~Pisarski and M.~Tytgat, \emph{{Propagation of cool pions}},
  \href{https://doi.org/10.1103/PhysRevD.54.R2989}{\emph{Phys.Rev.} {\bfseries
  D54} (1996) 2989} [\href{https://arxiv.org/abs/hep-ph/9604404}{{\ttfamily
  hep-ph/9604404}}].

\bibitem{Son:2001ff}
D.T.~Son and M.A.~Stephanov, \emph{{Pion propagation near the QCD chiral phase
  transition}},
  \href{https://doi.org/10.1103/PhysRevLett.88.202302}{\emph{Phys. Rev. Lett.}
  {\bfseries 88} (2002) 202302}
  [\href{https://arxiv.org/abs/hep-ph/0111100}{{\ttfamily hep-ph/0111100}}].

\bibitem{Son:2002ci}
D.T.~Son and M.A.~Stephanov, \emph{{Real time pion propagation in finite
  temperature QCD}},
  \href{https://doi.org/10.1103/PhysRevD.66.076011}{\emph{Phys. Rev. D}
  {\bfseries 66} (2002) 076011}
  [\href{https://arxiv.org/abs/hep-ph/0204226}{{\ttfamily hep-ph/0204226}}].

\bibitem{McLerran:1984ay}
L.D.~McLerran and T.~Toimela, \emph{{Photon and Dilepton Emission from the
  Quark - Gluon Plasma: Some General Considerations}},
  \href{https://doi.org/10.1103/PhysRevD.31.545}{\emph{Phys.Rev.} {\bfseries
  D31} (1985) 545}.

\bibitem{Braun-Munzinger:2015hba}
P.~Braun-Munzinger, V.~Koch, T.~Sch\"afer and J.~Stachel, \emph{{Properties of
  hot and dense matter from relativistic heavy ion collisions}},
  \href{https://doi.org/10.1016/j.physrep.2015.12.003}{\emph{Phys. Rept.}
  {\bfseries 621} (2016) 76}
  [\href{https://arxiv.org/abs/1510.00442}{{\ttfamily 1510.00442}}].

\bibitem{Rapp:1999ej}
R.~Rapp and J.~Wambach, \emph{{Chiral symmetry restoration and dileptons in
  relativistic heavy ion collisions}},
  \href{https://doi.org/10.1007/0-306-47101-9_1}{\emph{Adv. Nucl. Phys.}
  {\bfseries 25} (2000) 1}
  [\href{https://arxiv.org/abs/hep-ph/9909229}{{\ttfamily hep-ph/9909229}}].

\bibitem{Laine:2013vma}
M.~Laine, \emph{{NLO thermal dilepton rate at non-zero momentum}},
  \href{https://doi.org/10.1007/JHEP11(2013)120}{\emph{JHEP} {\bfseries 11}
  (2013) 120} [\href{https://arxiv.org/abs/1310.0164}{{\ttfamily 1310.0164}}].

\bibitem{Aarts:2015mma}
G.~Aarts, C.~Allton, S.~Hands, B.~J\"ager, C.~Praki and J.-I.~Skullerud,
  \emph{{Nucleons and parity doubling across the deconfinement transition}},
  \href{https://doi.org/10.1103/PhysRevD.92.014503}{\emph{Phys. Rev. D}
  {\bfseries 92} (2015) 014503}
  [\href{https://arxiv.org/abs/1502.03603}{{\ttfamily 1502.03603}}].

\bibitem{Aarts:2017rrl}
G.~Aarts, C.~Allton, D.~De~Boni, S.~Hands, B.~J\"ager, C.~Praki et~al.,
  \emph{{Light baryons below and above the deconfinement transition: medium
  effects and parity doubling}},
  \href{https://doi.org/10.1007/JHEP06(2017)034}{\emph{JHEP} {\bfseries 06}
  (2017) 034} [\href{https://arxiv.org/abs/1703.09246}{{\ttfamily
  1703.09246}}].

\bibitem{Brandt:2014qqa}
B.B.~Brandt, A.~Francis, H.B.~Meyer and D.~Robaina, \emph{{Chiral dynamics in
  the low-temperature phase of QCD}},
  \href{https://doi.org/10.1103/PhysRevD.90.054509}{\emph{Phys. Rev. D}
  {\bfseries 90} (2014) 054509}
  [\href{https://arxiv.org/abs/1406.5602}{{\ttfamily 1406.5602}}].

\bibitem{Brandt:2015sxa}
B.B.~Brandt, A.~Francis, H.B.~Meyer and D.~Robaina, \emph{{Pion quasiparticle
  in the low-temperature phase of QCD}},
  \href{https://doi.org/10.1103/PhysRevD.92.094510}{\emph{Phys. Rev. D}
  {\bfseries 92} (2015) 094510}
  [\href{https://arxiv.org/abs/1506.05732}{{\ttfamily 1506.05732}}].

\bibitem{Ce:2022dax}
M.~C\`e, T.~Harris, A.~Krasniqi, H.B.~Meyer and C.~T\"or\"ok, \emph{{Aspects of
  chiral symmetry in QCD at T=128\,\,MeV}},
  \href{https://doi.org/10.1103/PhysRevD.107.054509}{\emph{Phys. Rev. D}
  {\bfseries 107} (2023) 054509}
  [\href{https://arxiv.org/abs/2211.15558}{{\ttfamily 2211.15558}}].

\bibitem{Krasniqi:2022djb}
A.~Krasniqi, M.~C\`e, T.~Harris, H.~B.~Meyer, C.~T\"or\"ok and S.~Ruhl,
  \emph{{A $(2+1)$-flavor lattice study of the pion quasiparticle in the
  thermal hadronic phase at physical quark masses}},
  \href{https://doi.org/10.22323/1.430.0181}{\emph{PoS} {\bfseries Lattice2022}
  (2022) 0181}.

\bibitem{Luscher:1998pe}
M.~L{\"u}scher, \emph{{Advanced lattice QCD}},  in \emph{{Les Houches Summer
  School in Theoretical Physics, Session 68: Probing the Standard Model of
  Particle Interactions}}, pp.~229--280, 2, 1998
  [\href{https://arxiv.org/abs/hep-lat/9802029}{{\ttfamily hep-lat/9802029}}].

\bibitem{Meyer:2011gj}
H.B.~Meyer, \emph{{Transport Properties of the Quark-Gluon Plasma: A Lattice
  QCD Perspective}},
  \href{https://doi.org/10.1140/epja/i2011-11086-3}{\emph{Eur. Phys. J. A}
  {\bfseries 47} (2011) 86} [\href{https://arxiv.org/abs/1104.3708}{{\ttfamily
  1104.3708}}].

\bibitem{Sharpe:2006pu}
S.R.~Sharpe, \emph{{Applications of Chiral Perturbation theory to lattice
  QCD}},  in \emph{{Workshop on Perspectives in Lattice QCD}}, 7, 2006
  [\href{https://arxiv.org/abs/hep-lat/0607016}{{\ttfamily hep-lat/0607016}}].

\bibitem{Campos:2018ahf}
{\scshape ALPHA} collaboration, \emph{{Non-perturbative quark mass
  renormalisation and running in $N_f=3$ QCD}},
  \href{https://doi.org/10.1140/epjc/s10052-018-5870-5}{\emph{Eur. Phys. J. C}
  {\bfseries 78} (2018) 387}
  [\href{https://arxiv.org/abs/1802.05243}{{\ttfamily 1802.05243}}].

\bibitem{Bulava:2013cta}
J.~Bulava and S.~Schaefer, \emph{{Improvement of $N_f$ = 3 lattice QCD with
  Wilson fermions and tree-level improved gauge action}},
  \href{https://doi.org/10.1016/j.nuclphysb.2013.05.019}{\emph{Nucl. Phys. B}
  {\bfseries 874} (2013) 188}
  [\href{https://arxiv.org/abs/1304.7093}{{\ttfamily 1304.7093}}].

\bibitem{Bruno:2014jqa}
M.~Bruno et~al., \emph{{Simulation of QCD with N$_{f} =$ 2 $+$ 1 flavors of
  non-perturbatively improved Wilson fermions}},
  \href{https://doi.org/10.1007/JHEP02(2015)043}{\emph{JHEP} {\bfseries 02}
  (2015) 043} [\href{https://arxiv.org/abs/1411.3982}{{\ttfamily 1411.3982}}].

\bibitem{Mohler:2017wnb}
D.~Mohler, S.~Schaefer and J.~Simeth, \emph{{CLS 2+1 flavor simulations at
  physical light- and strange-quark masses}},
  \href{https://doi.org/10.1051/epjconf/201817502010}{\emph{EPJ Web Conf.}
  {\bfseries 175} (2018) 02010}
  [\href{https://arxiv.org/abs/1712.04884}{{\ttfamily 1712.04884}}].

\bibitem{Bruno:2016plf}
M.~Bruno, T.~Korzec and S.~Schaefer, \emph{{Setting the scale for the CLS $2 +
  1$ flavor ensembles}},
  \href{https://doi.org/10.1103/PhysRevD.95.074504}{\emph{Phys. Rev.}
  {\bfseries D95} (2017) 074504}
  [\href{https://arxiv.org/abs/1608.08900}{{\ttfamily 1608.08900}}].

\bibitem{Ce:2022kxy}
M.~C\`e et~al., \emph{{Window observable for the hadronic vacuum polarization
  contribution to the muon $g-2$ from lattice QCD}},
  \href{https://arxiv.org/abs/2206.06582}{{\ttfamily 2206.06582}}.

\bibitem{Luscher:2012av}
M.~L{\"u}scher and S.~Schaefer, \emph{{Lattice QCD with open boundary
  conditions and twisted-mass reweighting}},
  \href{https://doi.org/10.1016/j.cpc.2012.10.003}{\emph{Comput. Phys. Commun.}
  {\bfseries 184} (2013) 519}
  [\href{https://arxiv.org/abs/1206.2809}{{\ttfamily 1206.2809}}].

\bibitem{Clark:2006fx}
M.A.~Clark and A.D.~Kennedy, \emph{{Accelerating dynamical fermion computations
  using the rational hybrid Monte Carlo (RHMC) algorithm with multiple
  pseudofermion fields}},
  \href{https://doi.org/10.1103/PhysRevLett.98.051601}{\emph{Phys. Rev. Lett.}
  {\bfseries 98} (2007) 051601}
  [\href{https://arxiv.org/abs/hep-lat/0608015}{{\ttfamily hep-lat/0608015}}].

\bibitem{Mohler:2020txx}
D.~Mohler and S.~Schaefer, \emph{{Remarks on strange-quark simulations with
  Wilson fermions}},
  \href{https://doi.org/10.1103/PhysRevD.102.074506}{\emph{Phys. Rev. D}
  {\bfseries 102} (2020) 074506}
  [\href{https://arxiv.org/abs/2003.13359}{{\ttfamily 2003.13359}}].

\bibitem{Kuberski:2023zky}
S.~Kuberski, \emph{{Low-mode deflation for twisted-mass and RHMC reweighting in
  lattice QCD}}, \href{https://doi.org/10.1016/j.cpc.2024.109173}{\emph{Comput.
  Phys. Commun.} {\bfseries 300} (2024) 109173}
  [\href{https://arxiv.org/abs/2306.02385}{{\ttfamily 2306.02385}}].

\bibitem{Madras:1988ei}
N.~Madras and A.D.~Sokal, \emph{{The Pivot algorithm: a highly efficient Monte
  Carlo method for selfavoiding walk}},
  \href{https://doi.org/10.1007/BF01022990}{\emph{J. Statist. Phys.} {\bfseries
  50} (1988) 109}.

\bibitem{Wolff:2003sm}
{\scshape ALPHA} collaboration, \emph{{Monte Carlo errors with less errors}},
  \href{https://doi.org/10.1016/S0010-4655(03)00467-3}{\emph{Comput. Phys.
  Commun.} {\bfseries 156} (2004) 143}
  [\href{https://arxiv.org/abs/hep-lat/0306017}{{\ttfamily hep-lat/0306017}}].

\bibitem{Ramos:2018vgu}
A.~Ramos, \emph{{Automatic differentiation for error analysis of Monte Carlo
  data}}, \href{https://doi.org/10.1016/j.cpc.2018.12.020}{\emph{Comput. Phys.
  Commun.} {\bfseries 238} (2019) 19}
  [\href{https://arxiv.org/abs/1809.01289}{{\ttfamily 1809.01289}}].

\bibitem{Joswig:2022qfe}
F.~Joswig, S.~Kuberski, J.T.~Kuhlmann and J.~Neuendorf, \emph{{pyerrors: A
  python framework for error analysis of Monte Carlo data}},
  \href{https://doi.org/10.1016/j.cpc.2023.108750}{\emph{Comput. Phys. Commun.}
  {\bfseries 288} (2023) 108750}
  [\href{https://arxiv.org/abs/2209.14371}{{\ttfamily 2209.14371}}].

\bibitem{Foster:1998vw}
{\scshape UKQCD} collaboration, \emph{{Quark mass dependence of hadron masses
  from lattice QCD}},
  \href{https://doi.org/10.1103/PhysRevD.59.074503}{\emph{Phys. Rev. D}
  {\bfseries 59} (1999) 074503}
  [\href{https://arxiv.org/abs/hep-lat/9810021}{{\ttfamily hep-lat/9810021}}].

\bibitem{Boyle:2008rh}
P.A.~Boyle, A.~Juttner, C.~Kelly and R.D.~Kenway, \emph{{Use of stochastic
  sources for the lattice determination of light quark physics}},
  \href{https://doi.org/10.1088/1126-6708/2008/08/086}{\emph{JHEP} {\bfseries
  08} (2008) 086} [\href{https://arxiv.org/abs/0804.1501}{{\ttfamily
  0804.1501}}].

\bibitem{ETM:2008zte}
{\scshape ETM} collaboration, \emph{{Dynamical Twisted Mass Fermions with Light
  Quarks: Simulation and Analysis Details}},
  \href{https://doi.org/10.1016/j.cpc.2008.06.013}{\emph{Comput. Phys. Commun.}
  {\bfseries 179} (2008) 695}
  [\href{https://arxiv.org/abs/0803.0224}{{\ttfamily 0803.0224}}].

\bibitem{Bali:2009hu}
G.S.~Bali, S.~Collins and A.~Schafer, \emph{{Effective noise reduction
  techniques for disconnected loops in Lattice QCD}},
  \href{https://doi.org/10.1016/j.cpc.2010.05.008}{\emph{Comput. Phys. Commun.}
  {\bfseries 181} (2010) 1570}
  [\href{https://arxiv.org/abs/0910.3970}{{\ttfamily 0910.3970}}].

\bibitem{Backus:1968}
G.~Backus and F.~Gilbert, \emph{{The Resolving Power of Gross Earth Data}},
  \href{https://doi.org/10.1111/j.1365-246X.1968.tb00216.x}{\emph{Geophysical
  Journal International} {\bfseries 16} (1968) 169}
  [\href{https://arxiv.org/abs/https://academic.oup.com/gji/article-pdf/16/2/169/5891044/16-2-169.pdf}{{\ttfamily
  https://academic.oup.com/gji/article-pdf/16/2/169/5891044/16-2-169.pdf}}].

\bibitem{Wen:2024hgu}
N.~Wen, X.~Cao, J.~Chao and H.~Liu, \emph{{Neutral pion masses within a hot and
  magnetized medium in a lattice-improved soft-wall AdS/QCD model}},
  \href{https://arxiv.org/abs/2402.06239}{{\ttfamily 2402.06239}}.

\bibitem{Goderidze:2022vlm}
D.~Goderidze, A.~Friesen and Y.~Kalinovsky, \emph{{Pion damping width and pion
  spectral function in hot pion gas}},
  \href{https://doi.org/10.1142/S0217751X22501354}{\emph{Int. J. Mod. Phys. A}
  {\bfseries 37} (2022) 2250135}
  [\href{https://arxiv.org/abs/2205.11436}{{\ttfamily 2205.11436}}].

\bibitem{Bazavov:2019www}
A.~Bazavov et~al., \emph{{Meson screening masses in (2+1)-flavor QCD}},
  \href{https://doi.org/10.1103/PhysRevD.100.094510}{\emph{Phys. Rev. D}
  {\bfseries 100} (2019) 094510}
  [\href{https://arxiv.org/abs/1908.09552}{{\ttfamily 1908.09552}}].

\bibitem{Borsanyi:2011sw}
S.~Borsanyi, Z.~Fodor, S.D.~Katz, S.~Krieg, C.~Ratti and K.~Szabo,
  \emph{{Fluctuations of conserved charges at finite temperature from lattice
  QCD}}, \href{https://doi.org/10.1007/JHEP01(2012)138}{\emph{JHEP} {\bfseries
  01} (2012) 138} [\href{https://arxiv.org/abs/1112.4416}{{\ttfamily
  1112.4416}}].

\bibitem{Hagedorn:1984hz}
R.~Hagedorn, \emph{{How We Got to QCD Matter from the Hadron Side: 1984}},
  \href{https://doi.org/10.1007/978-3-319-17545-4_25}{\emph{Lect. Notes Phys.}
  {\bfseries 221} (1985) 53}.

\bibitem{Karsch:2003vd}
F.~Karsch, K.~Redlich and A.~Tawfik, \emph{{Hadron resonance mass spectrum and
  lattice QCD thermodynamics}},
  \href{https://doi.org/10.1140/epjc/s2003-01228-y}{\emph{Eur. Phys. J. C}
  {\bfseries 29} (2003) 549}
  [\href{https://arxiv.org/abs/hep-ph/0303108}{{\ttfamily hep-ph/0303108}}].

\bibitem{Brandt:2015aqk}
B.B.~Brandt, A.~Francis, B.~J\"ager and H.B.~Meyer, \emph{{Charge transport and
  vector meson dissociation across the thermal phase transition in lattice QCD
  with two light quark flavors}},
  \href{https://doi.org/10.1103/PhysRevD.93.054510}{\emph{Phys. Rev. D}
  {\bfseries 93} (2016) 054510}
  [\href{https://arxiv.org/abs/1512.07249}{{\ttfamily 1512.07249}}].

\bibitem{Davier:2005xq}
M.~Davier, A.~Hocker and Z.~Zhang, \emph{{The Physics of hadronic tau decays}},
  \href{https://doi.org/10.1103/RevModPhys.78.1043}{\emph{Rev.Mod.Phys.}
  {\bfseries 78} (2006) 1043}
  [\href{https://arxiv.org/abs/hep-ph/0507078}{{\ttfamily hep-ph/0507078}}].

\bibitem{Kapusta:1993hq}
J.I.~Kapusta and E.V.~Shuryak, \emph{{Weinberg type sum rules at zero and
  finite temperature}},
  \href{https://doi.org/10.1103/PhysRevD.49.4694}{\emph{Phys. Rev.} {\bfseries
  D49} (1994) 4694} [\href{https://arxiv.org/abs/hep-ph/9312245}{{\ttfamily
  hep-ph/9312245}}].

\bibitem{Hohler:2013eba}
P.M.~Hohler and R.~Rapp, \emph{{Is $\rho$-Meson Melting Compatible with Chiral
  Restoration?}},
  \href{https://doi.org/10.1016/j.physletb.2014.02.021}{\emph{Phys.Lett.}
  {\bfseries B731} (2014) 103}
  [\href{https://arxiv.org/abs/1311.2921}{{\ttfamily 1311.2921}}].

\bibitem{Dey:1990ba}
M.~Dey, V.L.~Eletsky and B.L.~Ioffe, \emph{{Mixing of vector and axial mesons
  at finite temperature: an Indication towards chiral symmetry restoration}},
  \href{https://doi.org/10.1016/0370-2693(90)90495-R}{\emph{Phys. Lett. B}
  {\bfseries 252} (1990) 620}.

\bibitem{Eletsky:1992ay}
V.L.~Eletsky and B.L.~Ioffe, \emph{{On the current correlators in QCD at finite
  temperature}}, \href{https://doi.org/10.1103/PhysRevD.47.3083}{\emph{Phys.
  Rev. D} {\bfseries 47} (1993) 3083}
  [\href{https://arxiv.org/abs/hep-ph/9302298}{{\ttfamily hep-ph/9302298}}].

\bibitem{Eletsky:1994rp}
V.L.~Eletsky and B.L.~Ioffe, \emph{{Next-to-leading order temperature
  corrections to correlators in QCD}},
  \href{https://doi.org/10.1103/PhysRevD.51.2371}{\emph{Phys. Rev. D}
  {\bfseries 51} (1995) 2371}
  [\href{https://arxiv.org/abs/hep-ph/9405371}{{\ttfamily hep-ph/9405371}}].

\bibitem{Meyer:2010}
H.B.~Meyer, \emph{{The bulk channel in thermal gauge theories}}, {\emph{JHEP 04
  (2010) 099} (2010) } [\href{https://arxiv.org/abs/arXiv:1002.3343}{{\ttfamily
  arXiv:1002.3343}}].

\bibitem{Gasser:1986vb}
J.~Gasser and H.~Leutwyler, \emph{{Light Quarks at Low Temperatures}},
  \href{https://doi.org/10.1016/0370-2693(87)90492-8}{\emph{Phys. Lett. B}
  {\bfseries 184} (1987) 83}.

\bibitem{Aarts:2020dda}
G.~Aarts and A.~Nikolaev, \emph{{Electrical conductivity of the quark-gluon
  plasma: perspective from lattice QCD}},
  \href{https://doi.org/10.1140/epja/s10050-021-00436-5}{\emph{Eur. Phys. J. A}
  {\bfseries 57} (2021) 118}
  [\href{https://arxiv.org/abs/2008.12326}{{\ttfamily 2008.12326}}].

\bibitem{Bernecker:2011gh}
D.~Bernecker and H.B.~Meyer, \emph{{Vector Correlators in Lattice QCD: Methods
  and applications}},
  \href{https://doi.org/10.1140/epja/i2011-11148-6}{\emph{Eur.Phys.J.}
  {\bfseries A47} (2011) 148}
  [\href{https://arxiv.org/abs/1107.4388}{{\ttfamily 1107.4388}}].

\bibitem{Caron-Huot:2009ypo}
S.~Caron-Huot, \emph{{Asymptotics of thermal spectral functions}},
  \href{https://doi.org/10.1103/PhysRevD.79.125009}{\emph{Phys. Rev. D}
  {\bfseries 79} (2009) 125009}
  [\href{https://arxiv.org/abs/0903.3958}{{\ttfamily 0903.3958}}].

\bibitem{Weinberg:1967kj}
S.~Weinberg, \emph{{Precise relations between the spectra of vector and axial
  vector mesons}},
  \href{https://doi.org/10.1103/PhysRevLett.18.507}{\emph{Phys. Rev. Lett.}
  {\bfseries 18} (1967) 507}.

\bibitem{Kuberski:2023qgx}
S.~Kuberski, \emph{{Muon $g-2$: Lattice calculations of the hadronic vacuum
  polarization}}, \href{https://doi.org/10.22323/1.453.0125}{\emph{PoS}
  {\bfseries LATTICE2023} (2024) 125}
  [\href{https://arxiv.org/abs/2312.13753}{{\ttfamily 2312.13753}}].

\bibitem{Galassi:2019}
M.~Galassi et~al., ``Gnu scientific library reference manual.''
  \url{https://www.gnu.org/software/gsl/}, 2019.

\bibitem{10.1145/1236463.1236468}
L.~Fousse, G.~Hanrot, V.~Lef\`{e}vre, P.~P\'{e}lissier and P.~Zimmermann,
  \emph{Mpfr: A multiple-precision binary floating-point library with correct
  rounding}, \href{https://doi.org/10.1145/1236463.1236468}{\emph{ACM Trans.
  Math. Softw.} {\bfseries 33} (2007) 13–es}.

\bibitem{Granlund:2020}
T.~Granlund and the GMP~development team. \url{https://gmplib.org/}, 2020.

\bibitem{Guagnelli:2000jw}
{\scshape ALPHA} collaboration, \emph{{Nonperturbative results for the
  coefficients b(m) and b(a) - b(P) in O(a) improved lattice QCD}},
  \href{https://doi.org/10.1016/S0550-3213(00)00675-1}{\emph{Nucl. Phys. B}
  {\bfseries 595} (2001) 44}
  [\href{https://arxiv.org/abs/hep-lat/0009021}{{\ttfamily hep-lat/0009021}}].

\bibitem{Bulava:2015bxa}
{\scshape ALPHA} collaboration, \emph{{Non-perturbative improvement of the
  axial current in $N_f$=3 lattice QCD with Wilson fermions and tree-level
  improved gauge action}},
  \href{https://doi.org/10.1016/j.nuclphysb.2015.05.003}{\emph{Nucl. Phys. B}
  {\bfseries 896} (2015) 555}
  [\href{https://arxiv.org/abs/1502.04999}{{\ttfamily 1502.04999}}].

\bibitem{Heitger:2020mkp}
{\scshape ALPHA} collaboration, \emph{{Ward identity determination of
  $Z_{\mathrm {S}}/Z_{\mathrm {P}}$ for $N_{\mathrm {f}}=3$ lattice QCD in a
  Schr\"odinger functional setup}},
  \href{https://doi.org/10.1140/epjc/s10052-020-8266-2}{\emph{Eur. Phys. J. C}
  {\bfseries 80} (2020) 765}
  [\href{https://arxiv.org/abs/2005.01352}{{\ttfamily 2005.01352}}].

\bibitem{Korcyl:2016ugy}
P.~Korcyl and G.S.~Bali, \emph{{Non-perturbative determination of improvement
  coefficients using coordinate space correlators in $N_f=2+1$ lattice QCD}},
  \href{https://doi.org/10.1103/PhysRevD.95.014505}{\emph{Phys. Rev. D}
  {\bfseries 95} (2017) 014505}
  [\href{https://arxiv.org/abs/1607.07090}{{\ttfamily 1607.07090}}].

\bibitem{DallaBrida:2018tpn}
M.~Dalla~Brida, T.~Korzec, S.~Sint and P.~Vilaseca, \emph{{High precision
  renormalization of the flavour non-singlet Noether currents in lattice QCD
  with Wilson quarks}},
  \href{https://doi.org/10.1140/epjc/s10052-018-6514-5}{\emph{Eur. Phys. J. C}
  {\bfseries 79} (2019) 23} [\href{https://arxiv.org/abs/1808.09236}{{\ttfamily
  1808.09236}}].

\bibitem{Bali:2023sdi}
{\scshape RQCD} collaboration, \emph{{Octet baryon isovector charges from
  Nf=2+1 lattice QCD}},
  \href{https://doi.org/10.1103/PhysRevD.108.034512}{\emph{Phys. Rev. D}
  {\bfseries 108} (2023) 034512}
  [\href{https://arxiv.org/abs/2305.04717}{{\ttfamily 2305.04717}}].

\bibitem{Gerardin:2018kpy}
A.~Gerardin, T.~Harris and H.B.~Meyer, \emph{{Nonperturbative renormalization
  and $O(a)$-improvement of the nonsinglet vector current with $N_f=2+1$ Wilson
  fermions and tree-level Symanzik improved gauge action}},
  \href{https://doi.org/10.1103/PhysRevD.99.014519}{\emph{Phys. Rev. D}
  {\bfseries 99} (2019) 014519}
  [\href{https://arxiv.org/abs/1811.08209}{{\ttfamily 1811.08209}}].

\bibitem{Ce:2022eix}
M.~C\`e, A.~G\'erardin, G.~von Hippel, H.B.~Meyer, K.~Miura, K.~Ottnad et~al.,
  \emph{{The hadronic running of the electromagnetic coupling and the
  electroweak mixing angle from lattice QCD}},
  \href{https://doi.org/10.1007/JHEP08(2022)220}{\emph{JHEP} {\bfseries 08}
  (2022) 220} [\href{https://arxiv.org/abs/2203.08676}{{\ttfamily
  2203.08676}}].

\bibitem{Agadjanov:2023efe}
A.~Agadjanov, D.~Djukanovic, G.~von Hippel, H.B.~Meyer, K.~Ottnad and
  H.~Wittig, \emph{{Nucleon Sigma Terms with Nf=2+1 Flavors of O(a)-Improved
  Wilson Fermions}},
  \href{https://doi.org/10.1103/PhysRevLett.131.261902}{\emph{Phys. Rev. Lett.}
  {\bfseries 131} (2023) 261902}
  [\href{https://arxiv.org/abs/2303.08741}{{\ttfamily 2303.08741}}].

\bibitem{Krasniqi:2024inr}
A.~Krasniqi, M.~C\`e, T.~Harris, R.J.~Hudspith, H.B.~Meyer and C.~T\"or\"ok,
  \emph{{The thermal photon emissivity at the QCD chiral crossover from
  imaginary momentum correlators}},
  \href{https://doi.org/10.22323/1.453.0180}{\emph{PoS} {\bfseries LATTICE2023}
  (2024) 180} [\href{https://arxiv.org/abs/2401.05951}{{\ttfamily
  2401.05951}}].

\bibitem{press2007numerical}
W.~Press, S.~Teukolsky, W.~Vetterling and B.~Flannery, \emph{Numerical Recipes:
  The Art of Scientific Computing}, Cambridge University Press, 3~ed. (2007).

\bibitem{Akaike:1998zah}
H.~Akaike, \emph{{Information Theory and an Extension of the Maximum Likelihood
  Principle}},  (New York), Springer Science+Business Media (1998),
  \href{https://doi.org/10.1007/978-1-4612-1694-0_15}{DOI}.

\bibitem{Jay:2020jkz}
W.I.~Jay and E.T.~Neil, \emph{{Bayesian model averaging for analysis of lattice
  field theory results}},
  \href{https://doi.org/10.1103/PhysRevD.103.114502}{\emph{Phys. Rev. D}
  {\bfseries 103} (2021) 114502}
  [\href{https://arxiv.org/abs/2008.01069}{{\ttfamily 2008.01069}}].

\end{thebibliography}\endgroup

\end{document}